\documentclass[11pt,preprintnumbers,pra,aps,
nofootinbib,tightenlines,floatfix, twocolumn,superscriptaddress,longbibliography]{revtex4-2}

\usepackage[colorlinks=true]{hyperref}
\hypersetup{
  linkcolor  = TealBlue!90!Black,
  citecolor  = YellowGreen!90!Black,
  urlcolor   = Peach!90!Black,
  colorlinks = true,
}
\usepackage[dvipsnames]{xcolor}
\usepackage[all]{hypcap}
\usepackage{amsmath,amssymb}
\usepackage{graphicx}
\usepackage{physics}
\usepackage{balance}
\usepackage{cancel}
\usepackage[utf8]{inputenc}
\usepackage{csquotes}
\usepackage{empheq}
\usepackage{longtable}
\usepackage{multirow}
\usepackage[section]{placeins}
\usepackage{comment}
\usepackage[customcolors]{hf-tikz}
\usepackage[top=1in,bottom=1in,left=1.8cm,right=1.6cm]{geometry}

\DeclareMathOperator{\diag}{diag}

\allowdisplaybreaks
\makeatletter
\renewcommand\onecolumngrid{
\do@columngrid{one}{\@ne}
\def\set@footnotewidth{\onecolumngrid}
\def\footnoterule{\kern-6pt\hrule width 1.5in\kern6pt}%
}
\makeatother
\DeclareMathOperator{\sign}{sgn}

\newcommand\scalemath[2]{\scalebox{#1}{\mbox{\ensuremath{\displaystyle #2}}}}

\catcode`,\active

\catcode`\,12

\def\Nslash{{ {\cal N}\hskip-0.55em /}}

\newcount\hour \newcount\hourminute \newcount\minute
\hour=\time \divide \hour by 60
\hourminute=\hour \multiply \hourminute by 60
\minute=\time \advance \minute by -\hourminute

\begin{document}

\title{Partial-transpose-guided entanglement classes and minimum noise filtering in many-body Gaussian quantum systems}

\author{Boyu Gao}
\email{boyu.gao@duke.edu}
\affiliation{{Duke Quantum Center and Department of Physics, Duke University, Durham, NC 27708, USA}}
\author{Natalie Klco}
\email{natalie.klco@duke.edu}
\affiliation{{Duke Quantum Center and Department of Physics, Duke University, Durham, NC 27708, USA}}
%%%%%%%%%%%%%%%%%

\begin{abstract}

The reduction and distortion of quantum correlations in the presence of classical noise leads to varied levels of inefficiency in the availability of entanglement as a resource for quantum information processing protocols. While generically minimizing required entanglement for mixed quantum states remains challenging, a class of many-body Gaussian quantum states ($\mathcal{N}$IC) is here identified that exhibits two-mode bipartite entanglement structure, resembling that of pure states, for which the logarithmic negativity entanglement measure remains invariant upon inclusion of the classical correlations and optimal entanglement resources can be clearly quantified. This subclass is found to be embedded within a broader class of many-body Gaussian states ($\mathcal{N}$-SOL) that retain two-mode entanglement structure for detection processes. These two entanglement classes are relevant in theoretical and experimental applications from the scalar field vacuum to the local axial motional modes of trapped ion chains. Utilizing the subspace that heralds inseparability in response to partial transposition, a minimum noise filtering process is designed to be necessary, sufficient, and computable for determining membership in these classes of entanglement structure. Application of this process to spacelike regions of the free scalar field vacuum is found to improve resource upper bounds, providing new understanding of the entanglement required for the quantum simulation of quantum fields as observed by arrays of local detectors.

\end{abstract}
\date{\today}
\maketitle

{
\scriptsize
\tableofcontents
}

\section{Introduction}

Beyond an intriguing feature of nature, the unique correlations of entanglement expressed by quantum particles~\cite{EPRoriginal,Bellineq,FreedmanClauser,Aspect1,Aspect2} and quantum fields~\cite{ReehSchlieder,summers1987maximal,Valentini,Reznik1,Reznik2,Witten} have been identified as a key resource in quantum information processing~\cite{feynman2018simulating,communicationandtele2,communicationandtele1,security1}, inspiring opportunities for exploring previously inaccessible phases of matter and dynamical properties of quantum many-body systems upon their incorporation into computational architectures~\cite{feynman2018simulating,coldatom,NKreview,hepquantumsimulation,bauer2023quantum}. For pure quantum states, in which all deviation from purity observed in a reduced density matrix, $\rho_A = \text{Tr}_B\left(\rho\right)$, arises from entanglement spanning the associated bipartition, entanglement can be directly quantified by the eigenspectrum of $\rho_A$~\cite{Srednicki:1993im,Concentratingpartialentanglement,ConditionsforaClass,MinimalConditions,Approximatetransformations,purestateent}. However, for mixed states, in which entropy is preexisting either due to partial measurements or noisy environments, perspectives of several entanglement measures are required to understand the quantum correlations, e.g., the Entanglement of Formation (EOF)~\cite{mixstateent}, logarithmic negativity~\cite{peresoriginalN,HORODECKIoriginalN,Simonreflection, Duanlocaltrans,computablemeasure,PlenioLogarithmic}, and distillable entanglement~\cite{mixstateent}, among others~\cite{mixstateent,quantifyent,Christandl_2004,PhysRevLett.101.140501,PhysRevLett.102.250503}, many of which are computationally formidable~\cite{Huang_2014,hiesmayr2021free}. The lack of degeneracy between such measures can be appreciated from a physical perspective by noting that classical correlations may obscure access to quantum correlations, and thus entanglement resources required for the creation of a mixed state are lower-bounded by, i.e., can exceed, the entanglement resource of the mixed state itself~\cite{originalboundent,anotherbound2,Gaussianboundent,anotherbound1}. Though entanglement in two-body and highly symmetric states is relatively well understood amid this complexity, challenges and open questions arise upon reductions of symmetry even in few-body contexts~\cite{Fourqubits,GiedkeEOF,PhysRevA.67.052311,WolfGEOF,It2,marian2008entanglement,TserkisandRalph,2019geof}. Focusing on general many-body bipartite Gaussian mixed states, this paper identifies two physically relevant and computable entanglement classes that have clear two-body entanglement structure in the crucial contexts of entanglement detection and state preparation.

Continuous variable (CV) quantum states~\cite{It2,It1,It3,BraunsteinGaussiareview,horodecki2009quantum,WeedbrookGaussiareview,It3,serafini2017quantum} are both experimentally accessible, e.g., Refs.~\cite{gaussianexample2,cavityqed,Cerf2007CVQIAL,Moran2014comb,gaussianphonon,Larsen2019cluster2d,royqs,kennethbrownqs}, and theoretically valuable, with the Gaussian subset commonly serving as a basis of physically motivated leading-order approximations~\cite{wolfextremality} that offer greater tractability for exploring many-body features of quantum correlations. In the many-body context, inspired in part by the ability to transform pure bipartite Gaussian states into a tensor-product series of $(1_A \times 1_B)$ two-mode entangled pairs with local unitaries~\cite{PhysRevA.67.052311,Giedkepurestatetrans}, several transformations have been devised to investigate properties of Gaussian mixed-state entanglement structure~\cite{wolfnotsonormal,NKcorehalo,Simonreflection,Duanlocaltrans,GiedkeEOF,WolfGEOF}. Among these, Ref.~\cite{NKcorehalo} identified a local unitary applicable to disjoint pairs of scalar field vacuum regions that produces a complete entanglement detection process through a tensor-product series of $(1_A \times 1_B)$ entangled mode pairs. In the present work, a deeper understanding of this transformation is developed and leveraged to generalize its performance to all possible many-body bipartite Gaussian states. The transformed state is found to be further conducive to the identification of the entanglement class of states composed entirely of optimal $(1_A \times 1_B)$ resources. These observations reinforce the practical value of complementary perspectives provided by local-unitary-based entanglement rearrangements in the understanding of quantum correlations in many-body mixed states.

\begin{figure}[t!]
\centering
\includegraphics[width=0.45\textwidth]{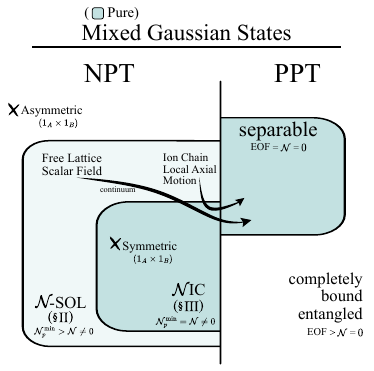}
\caption{Gaussian entanglement classifications, $\mathcal{N}$-SOL and $\mathcal{N}$IC, identified in the present work with trajectories depicting the change of entanglement structure for increasing separation between two regions in the latticed free scalar field vacuum~\cite{scalar1dextra,NKnegativitysphere,NKentsphere} and in the Gaussian approximation of local ion-chain axial motional modes symmetrically distributed in a quadratic trap~\cite{Retzker:2005dve,NKphononatural}. In order to quantify minimal resources for state formation, $\mathcal{N}_{p}^{\text{min}}$ is defined as the convex roof extension~\cite{horodecki2009quantum} of pure state logarithmic negativity. Classifications are also shown for the cases of symmetric~\cite{GiedkeEOF,WolfGEOF} and asymmetric~\cite{WolfGEOF,marian2008entanglement,TserkisandRalph,2019geof} two-mode Gaussian states.}
\label{fig:diagram}
\end{figure}

The landscape of entanglement in mixed Gaussian states, as shown in Fig.~\ref{fig:diagram}, can be divided into two sectors of Negative and Positive Partial Transpose (NPT and PPT) by the logarithmic negativity entanglement measure. For a subset of states, optimal convex decompositions exist in which the negativity remains invariant upon inclusion of the classical correlations ($\mathcal{N}$IC), i.e., can be constructed by the introduction of noise to a pure state with the same negativity. States in this entanglement class can be determined through a minimum noise filtering (MNF) procedure that is here designed to systematically construct an optimal underlying pure Gaussian state for $\mathcal{N}$IC-decomposable states by removing the minimum amount of noise that both \emph{aligns} the negativity\nobreakdash-contributing subspace ($\mathcal{V_{\mathcal{N}}}$) and maintains physicality. Separability in the remaining system after this $\mathcal{V}_{\mathcal{N}}$ alignment process is found to be a necessary and sufficient condition for determining the availability and explicit form of a $\mathcal{N}$IC-decomposition, simplifying the entirety of the entanglement into a tensor-product series of two-mode pairs. While asymmetric two-mode NPT Gaussian states belong to a broader class~\cite{WolfGEOF,marian2008entanglement,TserkisandRalph,2019geof}, their symmetric counterparts~\cite{GiedkeEOF,WolfGEOF} and regions of the scalar vacuum sensitive to the lattice at long distances are found to be $\mathcal{N}$IC-decomposable, and thus their optimal Gaussian decompositions calculable. For pure states, as the entanglement is completely captured by two mode structure~\cite{PhysRevA.67.052311, Giedkepurestatetrans} with a clear entanglement quantification, all states are either $\mathcal{N}$IC-decomposable or separable.

The $\mathcal{N}$IC entanglement class is found to be embedded within a broader class of multimode Gaussian states---those with local symplectic orthogonality in their $\mathcal{V_{\mathcal{N}}}$ ($\mathcal{N}$-SOL)---that generalizes applicability of a connection between negativity and separability structure originally observed in the specific context of disjoint regions in the free lattice scalar field vacuum~\cite{NKcorehalo}. The $\mathcal{N}$-SOL property is found to be both necessary and sufficient to consolidate the accessible entanglement into $(1_A \times 1_B)$ entangled pairs. Furthermore, the resulting separability among the entangled pairs and with the rest of the system can be analytically proven with the MNF procedure. From the perspective of two-mode Gaussian distillability~\cite{giedke2000inseparable,giedkedistill}, this entanglement structure indicates that $\mathcal{N}$-SOL states are particularly conducive to accessing available quantum correlations.

The organization of this article first establishes the $\mathcal{N}$-SOL and $\mathcal{N}$IC entanglement classes and techniques for their identification in Sections~\ref{sec:localtr} and~\ref{sec:NIC}, followed by application to the scalar field vacuum in Sec.~\ref{sec:scalarvac}. After defining the $\mathcal{N}$-SOL property from the PT symplectic structure, Section~\ref{sec:localon} presents a local transformation that consolidates such entanglement into a form conducive to two-body detection. The PT-guided MNF procedure is presented in Section~\ref{sec:postconsep} to establish the separability structure of post-consolidated $\mathcal{N}$-SOL states, and is subsequently utilized to identify $\mathcal{N}$IC as a subclass in Sec.~\ref{sec:NIC}. The pure state identification process provided by MNF is applied to spacelike separated regions of the free lattice scalar field vacuum in Sec.~\ref{sec:scalarvac}, identifying a stronger upper-bound of entanglement resources for state formation than previously established.

\section{Gaussian entanglement consolidation}
\label{sec:localtr}

The phase space of $n$ CV modes~\cite{It2,It1,It3,BraunsteinGaussiareview,horodecki2009quantum,WeedbrookGaussiareview,It3,serafini2017quantum} is spanned by $n$ pairs of position and momentum operators, $\hat{\boldsymbol{r}} \equiv \left( \hat{x}_1,\hat{p}_1,...,\hat{x}_n,\hat{p}_n\right)^{T}$, with canonical commutation relations (CCRs) forming a symplectic matrix, 
\begin{equation}
\Omega = \bigoplus\limits_{j = 1}^n\begin{pmatrix}
0 & 1 \\
-1 & 0 \\
\end{pmatrix} \ \ \ ,
\end{equation}
such that $[\hat{\boldsymbol{r}},\hat{\boldsymbol{r}}^{T}]=i\Omega$. Being unitary in the Hilbert space, symplectic transformations maintain the phase space CCRs, and thus satisfy $S\Omega S^{T}=\Omega$. Symplectic orthonormality,
\begin{equation}
   \bra{v_{i}}\Omega\ket{v_{j}}=\Omega_{ij} 
   \label{eq:symplecticorthonormality} \ \ \ ,
\end{equation}
holds for all row and column vectors, $\ket{v_{j}}$, of a valid symplectic transformation. 

Gaussian states can be fully described by first moments, $\bar{\boldsymbol{r}}$, of these phase space operators and their Covariance Matrix (CM),
\begin{equation}
\sigma=\mathrm{Tr} \left[ \rho_{G} \left\{ (\hat{\boldsymbol{r}} - \bar{\boldsymbol{r}}),(\hat{\boldsymbol{r}} - \bar{\boldsymbol{r}})^{T} \right\}  \right] \ \ \ ,
\end{equation}
where $\rho_G$ is the density matrix of the Gaussian state. Any CM can be diagonalized to Williamson normal form~\cite{williamsonnormalform} via global symplectic transformation, $S$,
\begin{equation}
    S \sigma S^{T} = \bigoplus\limits_{j = 1}^n\begin{pmatrix}
\nu_{j} & 0 \\
0 & \nu_{j} \\
\end{pmatrix}=D
\label{eq:onlys}\ \ \ ,
\end{equation}
where $\nu_{j}$ is the $j^{\text{th}}$ positive symplectic eigenvalue. For pure states, $\nu_{j} = 1$, while for mixed states $\nu_{j} \geq 1$. Methods for calculating $S$ from a CM can be found in, for example, Appendix~\ref{app:alignsub2} of this paper or Appendix D of Ref.~\cite{NKvolumemeasure}.

A necessary and sufficient condition for $\sigma$ to represent a physical Gaussian state is that it satisfies the uncertainty principle, expressed as the Positive Semidefinite (PSD) matrix property,
\begin{equation}
    \sigma_{\text{phys}} + i \Omega \geq 0 \ \ \ .
    \label{eq:phy}
\end{equation}
For bipartite Gaussian systems,
\begin{equation}
    \overline{\boldsymbol{r}}=\begin{pmatrix}
\overline{\boldsymbol{r}}_{A} \\
\overline{\boldsymbol{r}}_{B} \\
\end{pmatrix}, \quad \sigma=\begin{pmatrix}
\sigma_{A} & \sigma_{AB} \\
\sigma_{AB}^{T} & \sigma_{B} \\
\end{pmatrix} \ \ \ ,
\label{eq:gaussiantracing}
\end{equation}
the reduced density matrix $\text{Tr}_B(\rho_{AB})$ is also Gaussian~\cite{serafini2017quantum} with first moments $\overline{\boldsymbol{r}}_{A}$ and CM $\sigma_{A}$.

For Gaussian states, all information of quantum and classical correlations between modes resides in the CM. One informative quantifier of bipartite entanglement for mixed quantum states is the logarithmic negativity~\cite{peresoriginalN,computablemeasure, PlenioLogarithmic}, $\mathcal{N}=\log_{2} ||\Tilde{\rho}||_{1}$, where $\Tilde{\rho}$ is the partially transposed (PT) density matrix and $||\cdot||_{1}$ is the trace norm summing the absolute value of eigenvalues. In the phase space formalism, this translates to the sum of $n_{-}$ PT-symplectic eigenvalues less than one,
\begin{equation}
    \mathcal{N} = -\sum_{j=1}^{n_{-}}\log_{2}\Tilde{\nu}_{j} \ \ \ ,
    \label{eq:negg}
\end{equation}
where $\tilde{\nu}_j$ are the positive symplectic eigenvalues of the PT CM, $\tilde{\sigma} = \Lambda \sigma \Lambda$, and the PT operator $\Lambda$ is the local momentum reflection operator, $\hat{p}_{ \left\{n_b\right\} } \rightarrow -\hat{p}_{ \left\{n_b\right\} }$, in one party~\cite{Simonreflection},
\begin{equation}
    \Lambda = \Bigg( \bigoplus\limits_{j=1}^{n_{A}} \mathbb{I} \Bigg) \oplus \Bigg( \bigoplus\limits_{j=n_{A}+1}^{n_{A}+n_{B}} \sigma_{z}\Bigg) \ \ \ ,
    \label{eq:partialtrans}
\end{equation}
with $\mathbb{I}, \sigma_z$ the identity and Pauli-z matrix and $n_{A,B}$ the number of modes in space A,B of the A-B bipartition, $n = n_A + n_B$. As such, the logarithmic negativity heralds entanglement by identification of an invalid quantum state when locally transforming by a symmetry that is globally respected. Since Gaussianity is preserved under the PT operation~\cite{computablemeasure}, the transformation that diagonalizes the PT CM, $\tilde{S}$, is also symplectic,
\begin{equation}
    \Tilde{S} \Tilde{\sigma}  \Tilde{S}^{T} = \bigoplus\limits_{j = 1}^n\begin{pmatrix}
\Tilde{\nu}_{j}& 0 \\
0 & \Tilde{\nu}_{j} \\
\end{pmatrix} =\Tilde{D} \ \ \ ,
\label{eq:stilde}
\end{equation}
and provides a normal mode decomposition for the PT CM. This also gives a concrete definition of $\mathcal{V_{\mathcal{N}}}$ and the non-negativity\nobreakdash-contributing subspace ($\mathcal{V}_{\Nslash}$) as the subspaces with $\Tilde{\nu}_{j}<1$ and $\Tilde{\nu}_{j}\geq 1$, respectively, spanned by corresponding vectors of the symplectic transformation $\tilde{S}$.

Beyond its role in the negativity entanglement measure, the PT space will be here shown to carry further information of entanglement structure that informs broader properties of Gaussian quantum correlations and separability. The CM can be written in the PT basis as,
\begin{equation}
    \sigma =  \Lambda\Omega \Biggl\{ \sum_{j=1}^{2n} \Tilde{\nu}_{j} \ket{\Tilde{\nu}_{j}}\bra{\Tilde{\nu}_{j}}\Biggl\}\Omega^{T} \Lambda \ \ \ ,
    \label{eq:sigmatilde}
\end{equation}
with $\ket{\Tilde{\nu}_{j}}$ the corresponding row vectors of $\tilde{S}$, which also satisfy symplectic orthonormality, $\bra{\Tilde{\nu}_{i}}\Omega\ket{\Tilde{\nu}_{j}}=\Omega_{ij}$. As seen in Eq.~\eqref{eq:sigmatilde}, transformation by $\Lambda \Omega$ transfers the PT information of $\Tilde{S}$ to the CM structure. This transformation will be employed throughout the present work, allowing the PT information to guide entanglement reorganizations.

\subsection{\texorpdfstring{$\mathcal{N}$}{N}-SOL: symplectic orthogonality of local \texorpdfstring{$\mathcal{V}_\mathcal{N}$}{VN}}
\label{sec:localon}

For some physically relevant mixed Gaussian quantum states, there exists a local unitary transformation, $S_A \oplus S_B$, that consolidates bipartite negativity in multimode contexts into a tensor product series of $(1_A \times 1_B)$ pairs, each with a two-mode squeezed vacuum state (TMSVS) origin~\cite{NKcorehalo}. In the following, the property of $\mathcal{N}$-SOL---defined here as,
\begin{equation}
\bra{\Tilde{\nu}_{i,A}}\Omega\ket{\Tilde{\nu}_{j,A}}=\bra{\Tilde{\nu}_{i,B}}\Omega\ket{\Tilde{\nu}_{j,B}}=\Omega_{ij}/2 \ \ \ ,
\label{eq:nsolcondition}
\end{equation}
where $|\tilde{\nu}_{j,A}\rangle$ are the local $A$-space row or column vectors of $\tilde{S}$ in $\mathcal{V_{\mathcal{N}}}$---is identified as a necessary and sufficient condition for the availability of such a local unitary. As such, the $\mathcal{N}$-SOL criterion extends understanding of the multimode entanglement reorganization beyond the free scalar field.

Applying local symplectic transformations $S_{A,B}$ on a CM, $\sigma$, produces $\sigma'=(S_{A}\oplus S_{B})\sigma (S_{A}\oplus S_{B})^{T}$. Because local symplectics do not modify the PT-symplectic eigenvalues, $\tilde{D}' = \tilde{D}$. The PT diagonalizing operator after local symplectic operations, $\Tilde{S}^{'}$, can thus be related to the original $\tilde{S}$ through the equality $\Tilde{S}^{'} \Tilde{\sigma}' \Tilde{S}^{'T} = \Tilde{S} \Tilde{\sigma} \Tilde{S}^{T}$ to find
\begin{equation}
    \Tilde{S}^{'} = \Tilde{S} \Lambda(S_{A}\oplus S_{B})^{-1}\Lambda \ \ \ .
    \label{eq:sbar}
\end{equation}
Note that the symplecticity of $\Tilde{S}^{'}$ follows from $\Lambda(S_{A}\oplus S_{B})^{-1}\Lambda$ being symplectic for arbitrary $S_{A,B}$.

For an example of $\mathcal{N}$-SOL, consider symmetric two-mode Gaussian states.  
When calculating (G)EOF in this subset of quantum states, Ref.~\cite{GiedkeEOF} and~\cite{WolfGEOF} applied local squeezings to identify classical correlations that leave the pure-state negativity invariant. As will be discussed in Sec.~\ref{sec:NIC}, this is a defining characteristic that the $\mathcal{N}$IC entanglement class extends to multimode states, and thus identifies symmetric two-mode Gaussian states as members of $\mathcal{N}$-SOL (see Fig.~\ref{fig:diagram}). The following alternatively demonstrates the $\mathcal{N}$-SOL property from the perspective of the PT space, which is the perspective that will prove useful for subsequent multimode generalizations. Upon application of single-mode symplectic operations, any symmetric two-mode CM may be reduced to normal form~\cite{Simonreflection,Duanlocaltrans,Giedkesepflow,Gaussianboundent},
\begin{equation}
\sigma =  \begin{pmatrix}
n & 0 & k_{q} & 0  \\
0 & n & 0 & k_{p}  \\
k_{q} & 0 & n & 0  \\
0 & k_{p} & 0 & n   \\
\end{pmatrix} \ \ \ ,
\end{equation}
with $k_{q}\geq k_{p}$ without loss of generality. Additional constraints for this CM including physicality and entanglement conditions are discussed in Refs.~\cite{Simonreflection,Duanlocaltrans}~\footnote{Note conditions of Ref.~\cite{GiedkeEOF} deviate from Refs.~\cite{Simonreflection,Duanlocaltrans}.}. As a consequence of the physicality condition, $n \geq \abs{k_{q,p}}$ in order for $\sigma$ to be PSD. The $\mathcal{V_{\mathcal{N}}}$ subspace is characterized by symplectic eigenvalue $\Tilde{\nu}_{-}=\sqrt{(n+k_{p})(n-k_{q})}$ and corresponding $\Tilde{S}$ row vectors $\ket{\Tilde{\nu}_{-}}_{1} = \left(-(\frac{n+k_{p}}{4n-4k_{q}})^{1/4},0,(\frac{n+k_{p}}{4n-4k_{q}})^{1/4},0 \right)^{T}$ and $\ket{\Tilde{\nu}_{-}}_{2} = \left(0,-(\frac{n-k_{q}}{4n+4k_{p}})^{1/4},0,(\frac{n-k_{q}}{4n+4k_{p}})^{1/4} \right)^{T}$. For these $\tilde{S}$ vectors, the two-dimensional $A$- and $B$-space $\mathcal{N}$-SOL conditions, Eq.~\eqref{eq:nsolcondition}, can each be analytically confirmed. With Eq.~\eqref{eq:sbar}, applying a single mode squeezing operation, $S_\lambda = \diag\left( \lambda,\lambda^{-1} \right)$ with $\lambda = \left[(n+k_{p})/(n-k_{q})\right]^{1/4}$, to each CV mode transforms the $\mathcal{V_{\mathcal{N}}}$ subspace of $\Tilde{S}$ into that of a simple TMSVS, with two-mode squeezing parameter~\footnote{Note the square root updating Ref.~\cite{WolfGEOF}.}, $r = - \frac{1}{2} \ln{\Tilde{\nu}_{-}}$. Extending such \emph{alignment} of the $\mathcal{V_{\mathcal{N}}}$ subspace is an integral element in establishing the present entanglement classes for multimode $(n_{A,B} > 1)$ Gaussian states.

For bipartite Gaussian systems in $\mathcal{N}$-SOL, the local $\mathcal{V_{\mathcal{N}}}$ is spanned by $2 n_{-}$ symplectically orthonormal row vectors of $\Tilde{S}$. Symplectic transformations $S_{A}$ and $S_{B}$ can be designed from these row vectors with a symplectic Gram-Schmidt ($\text{GS}$) procedure as,
\begin{equation}
    S_{A} \oplus S_{B} = \Lambda (S_\pi \text{GS}[\Tilde{S}_{\mathcal{V_{\mathcal{N}}},A}] \oplus \text{GS}[\Tilde{S}_{\mathcal{V_{\mathcal{N}}},B}]) \Lambda \ \ \ .
    \label{eq:gseq}
\end{equation}
The notation $\text{GS}[\Tilde{S}_{\mathcal{V_{\mathcal{N}}},A}]$ indicates that the GS procedure starts with the $\mathcal{V}_{\mathcal{N}}$ subspace and completes the local symplectic with $2 \left( n_{A}-n_{-} \right) $ additional vectors, e.g., from $\mathcal{V}_{\Nslash}$, resulting in symplectic orthonormality, Eq.~\eqref{eq:symplecticorthonormality}. Details of this GS procedure can be found, for example, in Appendix A of Ref.~\cite{NKcorehalo}. Consistent with previous terminology~\cite{NKcorehalo}, each $(1_A \times 1_B)$ TMSVS space, governed by two vectors in $\mathcal{V_{\mathcal{N}}}$ spanning a degenerate $\tilde{\nu}_-$ subspace, will be referred as a core pair, while the complementary space of $\left( n-2n_- \right)$ CV modes will be referred to as the halo. Among the post-consolidation product of core pairs, if a negative squeezing pair---for which $\tilde{S}$ row vectors in $\mathcal{V}_{\mathcal{N}}\left(\mathcal{V}_{\Nslash}\right)$ are A-B symmetric(anti-symmetric), as shown in Appendix~\ref{app:corenoiseiden}---is found, $S_\pi=-\mathbb{I}_{2}$ applies a single-mode phase operation by angle $\pi$ to one of the modes in the core pair. The resulting structure of positive squeezing cores will be utilized throughout subsequent derivations. While the central component of Eq.~\eqref{eq:gseq} is consistent with procedures established in Ref.~\cite{NKcorehalo} from $\abs{i\Omega\Tilde{\sigma}}$, Eq.~\eqref{eq:gseq} generalizes applicability of the procedure to the entire $N$-SOL entanglement class, including when $x$-$p$ mixing is present in the CM~\footnote{For the special case of A-B symmetric PT CMs, $\tilde{\sigma}_A=\tilde{\sigma}_B$, with symmetric off-diagonal block, $\tilde{\sigma}_{AB}=\tilde{\sigma}_{AB}^{T}$, and vanishing $x$-$p$ matrix elements, relevant to physical applications of fields and trapped ions, a local consolidating transformation with $S_A = S_B$ may also be directly calculated through forthcoming techniques as the S that diagonalizes $\tilde{\sigma}_A - \tilde{\sigma}_{AB}$.}.

Because support of each post-consolidated $\Tilde{S}$ row vector in $\mathcal{V}_{\mathcal{N}}$ is isolated to the associated TMSVS, accessing individual core pairs upon tracing the rest of the system recovers all the PT symplectic eigenvalues in $\mathcal{V}_{\mathcal{N}}$. Therefore, after consolidation, the entirety of the negativity in the original mixed state is simplified to two-mode form. An example of entanglement consolidation for a general numerical CM can be found in Appendix~\ref{app:conso1}.

From Eq.~\eqref{eq:sbar}, note that the $\mathcal{N}$-SOL classification is preserved under local symplectic transformations. Because the TMSVS-structure in $\mathcal{V}_{\mathcal{N}}$ of the post-consolidation CM is a member of $\mathcal{N}$-SOL, the original state must be as well. Therefore, $\mathcal{N}$-SOL classification is a necessary condition for a Gaussian state to have $\mathcal{V}_{\mathcal{N}}$-consolidatable entanglement structure. Because the $\mathcal{N}$-SOL property provides sufficient linear independence to align each core TMSVS individually, it is also a sufficient condition for Gaussian entanglement consolidation. Thus, the $\mathcal{N}$-SOL classification serves as a necessary and sufficient condition that may be directly calculated to determine whether a mixed Gaussian state's negativity may be consolidated into a tensor-product series of TMSVS.

\subsection{Minimum noise filtering (MNF) and separability}
\label{sec:postconsep}

Though serving as a measure of mixed state entanglement, the logarithmic negativity is not generically a complete separability criterion when both parties in a bipartition are multimode CV systems. However, for states in the $\mathcal{N}$-SOL entanglement class, the PT-guided local unitary transformations of Sec.~\ref{sec:localon} simplify the many-body entanglement. In particular, the negativity-contributing subspace, $\mathcal{V}_{\mathcal{N}}$, can be organized into a set of cores that are both separable from each other and from the halo. This available separability structure for $\mathcal{N}$-SOL states will be presented in the following section, recovered systematically by first developing the MNF technique.

For mixed Gaussian states, pure Gaussian convex decompositions have non-vanishing support in the classical probability distribution only for pure states with correlations that decay faster than those of the mixed state~\cite{WolfGEOF}, i.e., $\sigma_{p} \leq \sigma_{m}$. Furthermore, a decomposition requiring the minimum entanglement resources may be composed of a single CM~\cite{WolfGEOF}, thus taking the form of Gaussian classical mixing,
\begin{equation}
    \sigma_{m} = \sigma_{p} + Y \ \ \ ,
    \label{eq:noisepmy}
\end{equation}
with Y a PSD matrix representing classical noise, e.g., an ensemble of Gaussian distributed first-moment displacements.

The first step of MNF is to strategically identify noise for each core. For any entangled core, there exists a Gaussian classical mixing decomposition with noise present solely in the $\mathcal{V}_{\Nslash}$ subspace. As discussed in Appendix~\ref{app:corenoiseiden}, noise isolated to $\mathcal{V}_{\Nslash}$ of a TMSVS will uniquely have the form, 
\begin{equation}
    Y_{c_{1}}=\frac{1}{2}\begin{pmatrix}
y_{11} & y_{12} & y_{11} & -y_{12}  \\
y_{12} & y_{22} & y_{12} & -y_{22}  \\
y_{11} & y_{12} & y_{11} & -y_{12}  \\
-y_{12} & -y_{22} & -y_{12} & y_{22}   \\
\end{pmatrix} \geq 0
\label{eq:Y11cform} \ \ \ ,
\end{equation}
where $y_{11} \geq 0$, $y_{22} \geq 0$ and $y_{11}y_{22} \geq y_{12}^{2}$. For the noise identified here, A-B symmetry is preserved up to local transformations.

The second step of MNF subtracts the minimum amount of noise in the rest of the system that maintains the core-halo structure given Eq.~\eqref{eq:Y11cform} identified in the first step. For ease of the following matrix notation, first permute modes of the consolidated CM, $\sigma'$, into core-halo ordering, $\hat{\mathbf{r}}_{ch} = \left( \hat{\mathbf{r}}_{c_1}, \hat{\mathbf{r}}_{c_2}, \ldots, \hat{\mathbf{r}}_{c_{n_-}}, \hat{\mathbf{r}}_{h} \right)$ where $\hat{\mathbf{r}}_{c_j} = \left( \hat{x}_A, \hat{p}_A, \hat{x}_B, \hat{p}_B\right)_{c_j}$ and $\hat{\mathbf{r}}_h$ captures the remaining halo modes. With the connection established in Eq.~\eqref{eq:sigmatilde}, the off-diagonal block of the CM between the first core and the rest of the system, $\sigma_{c_{1} r}$, can be expressed with the outer product of $\Tilde{S}^{'}$ row vectors from the corresponding subspaces. As an extension of Eq.~\eqref{eq:tmsvslike} to the entire core-halo system, entanglement consolidation with positive-squeezed underlying TMSVS respects A-B symmetric (anti-symmetric) $\Tilde{S}^{'}$ row vectors of $\mathcal{V}_{\Nslash}$ ($\mathcal{V}_{\mathcal{N}}$) in the core subspace. Due to the post-consolidation $\mathcal{V}_{\mathcal{N}}$ row vectors having support only in the corresponding core subspace, the A-B anti-symmetric vectors do not contribute to $\sigma_{c_{1} r}$. The $\Lambda\Omega$ transformation on the remaining A-B symmetric contributions produces A-B symmetry (anti-symmetry) in the $x$ ($p$) components, resulting in $\sigma_{c_{1}r}$ column vectors of the form, $\ket{l}\equiv \left( l_{i},-l_{j},l_{i},l_{j}\right)^{T}$.

In order to construct a matrix, $Y$, with $Y_{c_{1}}$, $Y_{r}$ diagonal blocks and $Y_{c_{1} r}$ off-diagonal block, $Y\geq 0$ is equivalent to~\cite{boyd2004convexschurcomp},
\begin{equation}
    (I-Y_{c_{1}}Y_{c_{1}}^{-1})Y_{c_{1} r}=0, \quad Y_{r}\geq Y_{c_{1} r}^{T}Y_{c_{1}}^{-1}Y_{c_{1} r} ,
     \label{eq:threeconditions}
\end{equation}
where $Y_{c_{1}}^{-1}$ denotes the pseudoinverse. The first condition requires every column vector of $Y_{c_{1} r}$ to be in the kernel of $I-Y_{c_{1}}Y_{c_{1}}^{-1}$. Consulting Eq.~\eqref{eq:Y11cform}, $Y_{c_1 r}$ is thus required to have the same $\ket{l}$ structure as $\sigma_{c_1 r}$. In order to both retain a physical CM and explicitly produce separability between the core and the rest of the system upon identification of $Y$, the saturation of $Y_{r}=Y_{c_{1} r}^{T}Y_{c_{1}}^{-1}Y_{c_{1} r}$ will be accompanied by $Y_{c_{1} r}= \sigma_{c_{1} r}$ to define the MNF process. To see that this filtration retains a physical CM, note that the expression of $\sigma_{c_{1}}$ established in Appendix~\ref{app:corenoiseiden} leads to $\bra{l} Y_{c_{1}}^{-1} \ket{l} =  \bra{l}(\sigma_{c_{1}} + i \Omega)^{-1}  \ket{l}$, independent of the squeezing parameter of $\sigma_{c_{1}}$. Therefore,
\begin{equation}
    Y_{c_{1} r}^{T} Y_{c_{1}}^{-1} Y_{c_{1} r} = Y_{c_{1} r}^{T} (\sigma_{c_{1}} + i \Omega)^{-1} Y_{c_{1} r} \ \ \ .
    \label{eq:minform}
\end{equation}
From the physicality, Eq.~\eqref{eq:phy}, of the post-consolidation CM,
\begin{equation}
\begin{pmatrix}
\sigma_{c_{1}} + i \Omega & \sigma_{c_{1} r} \\
\sigma_{c_{1} r}^{T} & \sigma_{r} + i \Omega \\
\end{pmatrix} \geq 0 \ \ \ .
\end{equation}
The Schur complement~\cite{boyd2004convexschurcomp} reads,
\begin{equation}
    \sigma_{r} - \sigma_{c_{1} r}^{T} (\sigma_{c_{1}} + i\Omega)^{-1} \sigma_{c_{1} r}+ i \Omega \geq 0 \ \ \ .
    \label{eq:minYsubtractedCM}
\end{equation}
With the MNF choices above, the second term is recognized as $Y_{r}$, indicating the filtered system remains physical. In summary, removing one core through the this filtering procedure will be defined as identifying $Y_{c_{1}}$ to be that isolated from a TMSVS with the same negativity as the dominant PT symplectic eigenvalue, $Y_{r} = Y_{c_{1} r}^{T} Y_{c_{1}}^{-1} Y_{c_{1} r}$ and $Y_{c_{1} r}= \sigma_{c_{1} r}$. The latter choice clearly reveals the separability between the core and the rest of the system.

The final step of the MNF algorithm is to iteratively repeat this noise-identifying filtration process. Such iteration is possible because the filtration does not affect the $\mathcal{V}_{\mathcal{N}}$ subspace of the remaining CM, relative to that of the original. To see this independence, consider the off-diagonal block of the CM between two cores, $\sigma_{c_{1} c_{2}}$, which has the structure of $\ket{l}$ embedded in both the columns and rows. Because $\sigma_{c_{1} c_{2}} \Lambda \Omega$ is composed of rows each with A-B symmetry, a vanishing result is found when acting on the two $\mathcal{V_{\mathcal{N}}}$ row vectors of the second core, which are A-B anti-symmetric. Furthermore, because the post-consolidation $\ket{\Tilde{\nu}_{-}}_{c_{2}}$ for the second core have support only in the two-mode $c_2$ subspace, from the connection between $\Tilde{S}$ and the CM structure established in Eq.~\eqref{eq:sigmatilde}, subtracting the minimum noise $Y_{c_{1} r}^{T} Y_{c_{1}}^{-1} Y_{c_{1} r}$ does not alter $\ket{\Tilde{\nu}_{-}}_{c_{2}}$,
\begin{equation}
   \bra{\Tilde{\nu}_{-}}_{c_{2}}  \Omega^{T} \Lambda Y_{c_{1} r}^{T} Y_{c_{1}}^{-1} Y_{c_{1} r} \Lambda\Omega \ket{\Tilde{\nu}_{-}}_{c_{2}} =0 \ \ \ .
   \label{eq:unalterN}
\end{equation}
In other words, the chosen $Y_{r}$, transformed via Eq.~\eqref{eq:sigmatilde} to impact the PT space of $\tilde{S}$, has no support in the remaining $\mathcal{V}_{\mathcal{N}}$ subspace of the original CM. 

Applying MNF sequentially for every core leads to post-consolidation separability both among the core pairs and with the halo,
\begin{equation}
    \sigma' = \Bigg( \bigoplus_{f} \sigma'_{c_{f}} \Bigg)  \oplus \sigma'_h + \sum_{f} Y_{f} \ \ \ ,
    \label{eq:MNF}
\end{equation}
with each filtration, $f$, identifying PSD noise $Y_{f}$. Through the iteration, the set $\left\{ \sigma'_{c_f} \right\}$ of TMSVS pure states can recover $n_-$ core pairs associated with the $\mathcal{V_{\mathcal{N}}}$ subspace of the CM, along with possible additional cores due to the subtraction of $Y_{r}$ in the rest of the system~\footnote{Note that choosing to identify separability with these additional cores can alter $\mathcal{V_{\mathcal{N}}}$ of the CM.}. The iteration stops when the identified $\sigma'_h$ is NPT without core structure or PPT, being either completely bound entangled~\cite{Gaussianboundent} or separable across the A-B bipartition, e.g., distinguished by the separability flow of Ref.~\cite{Giedkesepflow}. With the separability of Eq.~\eqref{eq:MNF} identified through MNF, operations on individual core pairs, $\sigma'_{c_{f}}$, do not alter the quantum correlations in the rest of the core-halo system.

\section{Gaussian negativity invariant classical correlations}
\label{sec:NIC}

The subclass of states for which $\sigma'_h$ is separable after $n_-$ iterations of the MNF procedure has physical properties of particular interest. For such states, the minimum negativity of the Gaussian-classical-mixing pure state decomposition, $\mathcal{N}_p^{\text{min}}$, saturates the basic lower bound, i.e., is equal to the negativity of the original mixed state, $\mathcal{N}_p^{\text{min}} = \mathcal{N} > 0$. Expressing this invariance of the negativity upon incorporation of classical correlations capable of relating the pure and mixed CMs, these states are here referred to as $\mathcal{N}$IC-decomposible. As seen from the separability structure, Eq.~\eqref{eq:MNF}, a collection of $n_-$ TMSVS states with squeezing parameters governed by the PT symplectic eigenvalues of $\sigma_m$ is capable of preparing such states. The entanglement of $\mathcal{N}$IC-decomposable states is thus completely captured by two-mode structure, a feature shared with all pure Gaussian states~\cite{PhysRevA.67.052311, Giedkepurestatetrans}. For this reason, the $\mathcal{N}$IC entanglement class may be regarded as a mixed-state generalization of pure-state entanglement.  

With the darker shading of Fig.~\ref{fig:diagram} expressing the simple pair of pure-state entanglement classes, the fact that $\mathcal{N}$IC decompositions are not available for all NPT mixed Gaussian states, $\mathcal{N}_p^{\text{min}} > \mathcal{N} > 0$, parallels the existence of bound entanglement in PPT mixed states. In general, the minimization of $\mathcal{N}_p^{\text{min}}$ remains a computationally challenging task. However, determining whether a state is $\mathcal{N}$IC-decomposable, and calculating a saturating $\sigma_p$ if so, can be systematically achieved through the developed MNF procedure. The following explores the entanglement structure of the $\mathcal{N}$IC entanglement class, showing that the framework of $\mathcal{V}_{\mathcal{N}}$ alignment and noise isolated to the $\mathcal{V}_{\Nslash}$ subspace is required for a state to be $\mathcal{N}$IC-decomposable. Thus, the designed MNF provides a complete and computable classifier in the identification of $\mathcal{N}$IC states and their optimal pure state decompositions.

\subsection{Structure and optimality of \texorpdfstring{$\mathcal{N}$IC}{NIC} decompositions}
\label{sec:Nalign}

Derived through Weyl's inequality~\cite{hornmatrixana} in Appendix~\ref{app:alignsub1}, negativity-invariant Gaussian classical mixing ($\mathcal{N}_{p}^{\text{min}} = \mathcal{N}$) is equivalent to four conditions,
\begin{equation}
     \text{spec}_{\mathcal{V_{\mathcal{N}}}}(-\Omega\Tilde{\sigma}_{p}\Omega\Tilde{\sigma}_{p})= \text{spec}_{\mathcal{V_{\mathcal{N}}}}(-\Omega\Tilde{\sigma}_{m}\Omega\Tilde{\sigma}_{m}) \ \ \ ,
    \label{eq:ppmm1}
\end{equation}
\begin{equation}
     \text{vec}_{\mathcal{V_{\mathcal{N}}}}(-\Omega\Tilde{\sigma}_{p}\Omega\Tilde{\sigma}_{p}) = \Omega\Tilde{\sigma}_{m} \text{vec}_{\mathcal{V_{\mathcal{N}}}}(-\Omega\Tilde{\sigma}_{m}\Omega\Tilde{\sigma}_{m}) \ \ \ ,
     \label{eq:ppmm2}
\end{equation}
\begin{equation}
     \text{spec}_{\mathcal{V_{\mathcal{N}}}}(-\Omega\Tilde{\sigma}_{p}\Omega\Tilde{\sigma}_{p}) =  \text{spec}_{\mathcal{V_{\mathcal{N}}}}(-\Omega\Tilde{\sigma}_{m}\Omega\Tilde{\sigma}_{p}) \ \ \ ,
     \label{eq:sqdof1}
\end{equation}
\begin{equation}
     \text{vec}_{\mathcal{V_{\mathcal{N}}}}(-\Omega\Tilde{\sigma}_{p}\Omega\Tilde{\sigma}_{p}) =  \text{vec}_{\mathcal{V_{\mathcal{N}}}}(-\Omega\Tilde{\sigma}_{m}\Omega\Tilde{\sigma}_{p}) \ \ \ ,
     \label{eq:sqdof2}
\end{equation}
where $\text{spec}_{\mathcal{V}_{\mathcal{N}}}(\cdot)$ and $\text{vec}_{\mathcal{V}_{\mathcal{N}}}(\cdot)$ are eigenvalues and eigenvectors in the $\mathcal{V}_{\mathcal{N}}$ subspace of the corresponding matrix. Because the negativity-contributing eigenvalues are smallest in the spectrum of a symmetric PSD matrix, noise can only be present in $\mathcal{V}_{\Nslash}$ to produce negativity-invariant PSD mixing.

Eigenvectors of $-\Omega\Tilde{\sigma}\Omega\Tilde{\sigma}$ can be constructed from linear combinations of pairs of $|i\Omega \tilde{\sigma}|$ eigenvectors, which are complex and arise in complex conjugate pairs with degenerate eigenvalues. As a result, $\Omega \tilde{\sigma}$ in Eq.~\eqref{eq:ppmm2} rotates eigenvectors within each two-dimensional degenerate subspace. As discussed in Appendix~\ref{app:alignsub2}, the conditions of Eq.~\eqref{eq:sqdof1} and Eq.~\eqref{eq:sqdof2} provide sensitivity to orientations in the single-mode subspaces that are otherwise degenerate from the $-\Omega\tilde{\sigma}\Omega\tilde{\sigma}$ perspective. After accounting for normalizations, Eqs.~\eqref{eq:ppmm1}--\eqref{eq:sqdof2} indicate that alignment of $\mathcal{V}_{\mathcal{N}}$ between the pure and mixed CMs is necessary for a multimode bipartite Gaussian state to have a $\mathcal{N}$IC decomposition. Hence, all Gaussian $\mathcal{N}$IC-decomposable states are members of $\mathcal{N}$-SOL.

As members of $\mathcal{N}$-SOL, post-consolidation $\mathcal{N}$IC states have a $\mathcal{V}_{\mathcal{N}}$ subspace organized in the tensor product of TMSVS form, i.e., described by a direct sum of subspaces of the form $\Tilde{S}_{\mathcal{V}_{\mathcal{N}}} = \frac{1}{\sqrt{2}}\begin{pmatrix}
-I_{2} & I_{2}
\end{pmatrix}$ in Eq.~\eqref{eq:a1}. The $\mathcal{V}_{\mathcal{N}}$ alignment required for a $\mathcal{N}$IC-decomposition extends to the pure-state normal-form symplectics, $\tilde{S}_{\mathcal{V}_\mathcal{N}, p} = \tilde{S}_{\mathcal{V}_\mathcal{N}, m}$. From Eq.~\eqref{eq:sbar} and the two-mode equivalence~\cite{PhysRevA.67.052311, Giedkepurestatetrans}, a general parameterization of the normal-form PT symplectic transformation for Gaussian pure states has the form,
\begin{equation}
\Tilde{S}_{p} = \frac{1}{\sqrt{2}}\begin{pmatrix}
S'_{A} &  S'_{B} \\
-S'_{A} & S'_{B} \\
\end{pmatrix}  \ \ \ .
\label{eq:purespparam} 
\end{equation}
As such, alignment also fixes each subspace of $\mathcal{V_{\Nslash}}$, via negation of row vectors in A, to be of the form $\tilde{S}_{\mathcal{V}_{\Nslash} \hspace{0.2em}, p} = \frac{1}{\sqrt{2}}\begin{pmatrix}
I_{2} & I_{2}
\end{pmatrix}$. Therefore, optimal underlying pure Gaussian states of post-consolidated $\mathcal{N}$IC states respect the tensor product of TMSVS form\footnote{Inspired by EOF results of Refs.~\cite{GiedkeEOF,marian2008entanglement}, the TMSVS structure derived here may provide opportunity for extension of this analysis to non-Gaussian entanglement resources.}.

Finally, because the negativity of the original CM is equal to that of the $n_-$ TMSVS core pairs, $\mathcal{V}_{\mathcal{N}}$ alignment and separability of the consolidated and filtered $\sigma'_h$ after $f_{max} = n_-$ iterations are determined to be necessary and sufficient conditions for the $\mathcal{N}$IC classification. Thus, all $\mathcal{N}$IC-decomposable states are identifiable through a three-step process: consolidating the entanglement through Eq.~\eqref{eq:gseq}, applying MNF to isolate the noise in $\mathcal{V}_{\Nslash}$, and checking the separability of $\sigma_h'$, e.g., with the techniques of Ref.~\cite{Giedkesepflow}. Beyond identifying all two-mode symmetric states to be in the $\mathcal{N}$IC entanglement class, consistent with the observations of Refs.~\cite{GiedkeEOF,WolfGEOF}, this procedure is applicable for Gaussian systems with any number of modes. A four-mode example from the free lattice scalar field vacuum is provided in Appendix~\ref{app:conso2}.

From Weyl's inequality~\cite{hornmatrixana,Giedkepurestatetrans}, noise identified solely in $\mathcal{V_{\Nslash}}$ maximizes $\mathcal{V_{\mathcal{N}}}$ symplectic eigenvalues, and hence minimizes entanglement for the TMSVS pairs of an underlying pure Gaussian state. Therefore, for Gaussian $\mathcal{N}$IC-decomposable states, alternate entanglement measures, e.g., GR2~\cite{adessoGR2} and GEOF~\cite{WolfGEOF}, will minimize to comparable pure states.

\section{MNF pure state identification}
\label{sec:scalarvac}

For mixed Gaussian states outside the $\mathcal{N}$IC entanglement class, the minimum negativity of the Gaussian-classical-mixing pure state decomposition is conclusively excluded from saturating the lower bound of the mixed-state negativity, $\mathcal{N}_p^{\text{min}}>\mathcal{N}>0$. Numerical searches designed to identify optimal decompositions above this bound are subject to non-trivial constraints with free parameters scaling quadratically with the number of CV modes~\cite{WolfGEOF,marian2008entanglement,2019geof}. Alternatively, the MNF process provides a systematic method for identifying an underlying pure state when $\sigma_h'$ of Eq.~\eqref{eq:MNF} is separable. Though no longer ensured to result in an optimal value of $\mathcal{N}_p^{\text{min}}$ (as it does within the $\mathcal{N}$IC entanglement class), the resulting upper bound is capable, due to Eq.~\eqref{eq:unalterN}, of identifying noise isolated to $\mathcal{V}_{\Nslash}$ for the $n_{-}$ core pairs of all $\mathcal{N}$-SOL mixed states. When applied to pairs of spatial regions in the free lattice scalar field vacuum, this feature allows the MNF process to produce an upper-bound to $\mathcal{N}_p^{\text{min}}$ that is lower than those of known procedures (e.g., canonical purification~\cite{gaussianpurification}, mixed-state normal-form symplectic transformation $\sigma_{p}=S_{m}S_{m}^{T}$~\cite{serafini2017quantum}, or volume measurement~\cite{NKvolumemeasure}), commonly by significant orders of magnitude at long distances where the $\mathcal{N}$IC entanglement structure arises.

\subsection{Application: lattice scalar field vacuum}
\label{sec:pureestimate}

The free lattice scalar field vacuum is a Gaussian state with CM elements that are analytic in the infinite volume limit. Upon isolating a pair of spatially separated regions (diameter $d$ and separation $\tilde{r}$) of the field---as might be observed by a pair of spacelike separated local detectors---calculations extrapolated to the continuum in up-to-three dimensions indicate exponentially decaying negativity with increasing separation~\cite{scalar1dextra,NKnegativitysphere,NKentsphere}, i.e., an exponentially decaying bound for the entanglement distillable from the vacuum, even for a massless field with polynomially and logarithmically decaying correlation functions. However, cutting through the noise of local detection by classically communicating measurements from the infinite volume has shown that the amount of entanglement fundamentally present between these regions of the massless field indeed scales as the poly/log two-point functions~\cite{NKvolumemeasure}. In the context of these observations, a fundamental question relevant to the quantum simulation of quantum fields becomes whether the required entanglement for representing a pair of free scalar field vacuum regions follows their negativity or their underlying entanglement, i.e., whether an exponential amount of entanglement must be suppressed when quantumly simulating observations of detector arrays, or whether there might be an exponentially more efficient design of quantum resources. In this section, employment of the MNF pure-state identification will conclusively answer this question at large separations where the lattice artifacts are strongest, and provide insight into the corresponding behavior toward the field continuum limit.

\begin{figure}[t!]
\centering
\includegraphics[width=0.48\textwidth]{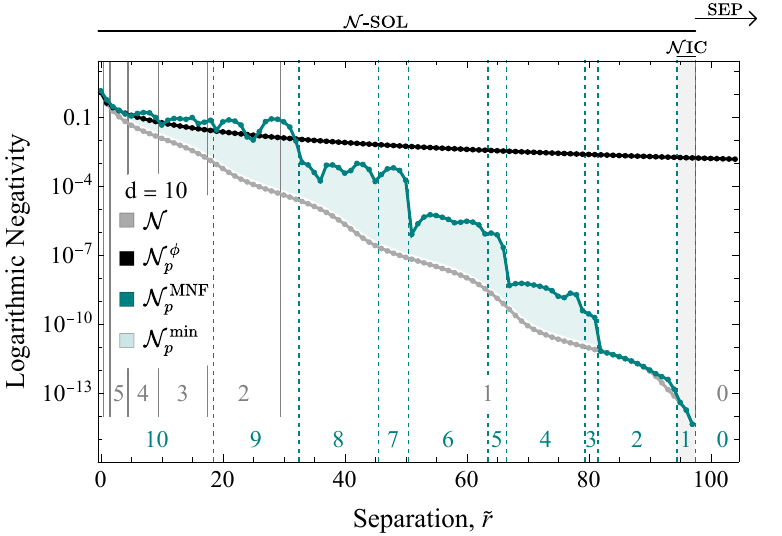}
\caption{Entanglement quantified by the logarithmic negativity $\mathcal{N}$ as a function of separation $\Tilde{r}$ between two regions ($d = 10$) of the infinite one-dimensional free lattice scalar field vacuum in the massless limit ($m = 10^{-10}$). Curves show the mixed-state negativity and the negativities of underlying pure states identified with field-basis volume measurement~\cite{NKvolumemeasure}, $\mathcal{N}_p^{\phi}$, or with MNF that creates TMSVS structure for dominant cores arising in the iteration, $\mathcal{N}_p^{\text{MNF}}$. Vertical lines indicate transitions between the labeled numbers of cores in the mixed state (solid, gray) and the MNF-identified pure state (dashed, teal). Transitions of the three entanglement classes---$\mathcal{N}$-SOL, $\mathcal{N}$IC, and separable---occur at $\Tilde{r}=95$ and $\Tilde{r}=98$.}
\label{fig:plot}
\end{figure}

Extending the observation of entanglement consolidation in Ref.~\cite{NKcorehalo}, regions of the scalar field vacuum are determined to be members of the $\mathcal{N}$-SOL entanglement class. In the latticized field at long distances near the vanishing negativity~\cite{vaneg2,scalarvacuumoriginal1,scalar1dextra,NKnegativitysphere,NKentsphere,vaneg3,vaneg4,vaneg5,vaneg6,vaneg7} transition to separability~\cite{NKentsphere}, the MNF process allows further identification of a small layer of $\mathcal{N}$IC-decomposable states~\footnote{It is observed that this transition from $\mathcal{N}$-SOL to $\mathcal{N}$IC entanglement structure coincides with the vanishing of negativity between the post-consolidated A- or B-space halo and the rest of the system.}. This layer appears as the grey shaded region of Fig.~\ref{fig:plot} for $d = 10$,  is demonstrated in Appendix~\ref{app:conso2} for the four-mode system of $(d,\tilde{r}) = (2,1)$, and has diminishing extent relative to $d$ for increasing system sizes toward the field continuum as documented in Table~\ref{tab:table2} of Appendix~\ref{app:num}. As such, the continuum limit of the scalar field vacuum is found to be outside the $\mathcal{N}$IC entanglement class, requiring $\mathcal{N}_p^{\text{min}} > \mathcal{N}$ as indicated by the teal shading in Fig.~\ref{fig:plot} for current possible values of $N_{p}^{\text{min}}$.

For entangled regions of the lattice scalar field vacuum outside the $\mathcal{N}$IC regime, the MNF pure-state identification provides a new upper bound to $\mathcal{N}_p^{\text{min}}$ as a function of $\tilde{r}$, shown by the teal line of Fig.~\ref{fig:plot}. Note that the developed MNF is a direct procedure, such that no optimization process is involved in the production of this upper bound. With $\mathcal{N}_p^{\text{MNF}} \sim \mathcal{N}$ at long distances, the $\mathcal{N}$IC entanglement structure is found to provide effective leading-order guidance for the decomposition of a range of $\mathcal{N}$-SOL states surrounding the $\mathcal{N}$IC regime. However, at reduced $\tilde{r}$ where impacts of incompatibility with the $\mathcal{N}$IC entanglement structure built into the MNF accumulate, $\mathcal{N}_p^{\text{MNF}}$ becomes comparable to the polynomially decaying pure-state negativities of previous techniques, e.g., the field-basis volume measurement~\cite{NKvolumemeasure} shown by the black line.

While the present results indicate that $\mathcal{N}_p^{\text{min}} \sim \mathcal{N}$ at long distances, the multi-core regime from which the continuum emerges remains to be determined. Beyond the MNF procedure, preliminary numerical optimizations of $\mathcal{N}_p^{\text{min}}$ intentionally deviating from the entanglement structure of Eq.~\eqref{eq:MNF} further support a conjecture that $\mathcal{N}_p^{\text{min}} \sim \mathcal{N}$, falling exponentially throughout the continuum regime of the field.

\section{Summary and Outlook}
\label{sec:discussion}

A complete understanding of mixed-state entanglement cannot be encapsulated in a single number, due to its multifaceted operational forms and varied relations to quantum information processing protocols. To highlight a few key properties, this work identifies two structural entanglement classes for which mixed-state Gaussian many-body entanglement may exhibit reduced two-body structure in the context of entanglement detection ($\mathcal{N}$-SOL) as well as in state preparation with clear entanglement resources that need not be increased by the present classical correlations ($\mathcal{N}$IC). The entanglement and separability organizations associated with such simplifications of the distributed quantum information are guided by the complete substructure of the partially transposed (PT) symplectic eigensystem, a frame usually synthesized to calculate the single negativity entanglement witness that imperfectly heralds the presence of many-body inseparability. While the $\mathcal{N}$-SOL classification is shown to be determined by inspection of the PT symplectic eigenvectors, development of a filtering procedure (MNF) that systematically isolates noise outside the negativity-contributing subspace was required in order to similarly provide a constructive technique for determining membership within the $\mathcal{N}$IC entanglement class. With $\mathcal{N}$-SOL entanglement structure connecting spacelike regions of the continuum free scalar field vaccum as well as between symmetric collections of local axial motional modes in trapped ion chains (transitioning to a NIC regime in the presence of strong lattice artifacts at large separations), early observations indicate these entanglement classes are prominent in fundamental physical quantum systems of both theoretical and experimental relevance.

Beyond future verification of the conjectured exponential decay of the Gaussian Entanglement of Formation with respect to the separation between spacelike regions of the free massless scalar field vacuum, the present paper supports several directions of advancing understanding and practical applications of quantum correlations. Expanding upon the current filtration process designed to produce optimal $\mathcal{N}$IC decompositions, alternate filtration strategies may be designed, e.g., governed by the substructure of other entanglement witnesses or the conclusive separability flow known for Gaussian quantum states~\cite{Giedkesepflow}, to identify improved pure-state decompositions of $\mathcal{N}$-SOL states not perturbatively close to the $\mathcal{N}$IC regime, e.g., those appearing between free scalar field vacuum regions in the continuum limit. Though the techniques and calculations of this paper have focused on the subset of Gaussian CV states, Gaussian approximations and extensions to the non-Gaussian scope will be important to address the entanglement properties of interacting quantum fields.

Mixed-state entanglement is central to both quantum information processing in noisy environments and quantum simulations of quantum fields observed by arrays of local detectors, as are frequently utilized in experimental nuclear and particle physics. As techniques emerge for incorporating entanglement structure of physical systems in associated quantum simulation design~\cite{NKdesign,NKfixpoint,simulationdesignalter,NKreview}, understanding of natural quantum information~\cite{entanglementinnature1,entanglementinnature3,entanglementinnature5,entanglementinnature2,entanglementinnature6,entanglementinnature7,entanglementinnature4,entanglementinnature8,entanglementinnature9,entanglementinnature10,entanglementinnature11,entanglementinnature13,entanglementinnature14,entanglementinnature12} becomes also a source of guidance in the creation of efficient quantum simulations.

\vspace{0.2cm}

\begin{acknowledgments}
We thank D.~H.~Beck and Ignacio Cirac for discussions of perspective at early stages of this research. Interactions with the latter were supported by the \emph{Quantum Computing Methods for High Energy Physics} program at the Munich Institute for Astro-, Particle and BioPhysics (MIAPbP), which is funded by the Deutsche Forschungsgemeinschaft (DFG, German Research Foundation) under Germany´s Excellence Strategy – EXC-2094 – 390783311. Presented calculations utilizing up to 1500 digits of precision were facilitated by the Mathematica 14.1 arbitrary precision libraries~\cite{mma}.
\end{acknowledgments}

\bibliography{biblio}

\begin{thebibliography}{107}%
\makeatletter
\providecommand \@ifxundefined [1]{%
 \@ifx{#1\undefined}
}%
\providecommand \@ifnum [1]{%
 \ifnum #1\expandafter \@firstoftwo
 \else \expandafter \@secondoftwo
 \fi
}%
\providecommand \@ifx [1]{%
 \ifx #1\expandafter \@firstoftwo
 \else \expandafter \@secondoftwo
 \fi
}%
\providecommand \natexlab [1]{#1}%
\providecommand \enquote  [1]{``#1''}%
\providecommand \bibnamefont  [1]{#1}%
\providecommand \bibfnamefont [1]{#1}%
\providecommand \citenamefont [1]{#1}%
\providecommand \href@noop [0]{\@secondoftwo}%
\providecommand \href [0]{\begingroup \@sanitize@url \@href}%
\providecommand \@href[1]{\@@startlink{#1}\@@href}%
\providecommand \@@href[1]{\endgroup#1\@@endlink}%
\providecommand \@sanitize@url [0]{\catcode `\\12\catcode `\$12\catcode
  `\&12\catcode `\#12\catcode `\^12\catcode `\_12\catcode `\%12\relax}%
\providecommand \@@startlink[1]{}%
\providecommand \@@endlink[0]{}%
\providecommand \url  [0]{\begingroup\@sanitize@url \@url }%
\providecommand \@url [1]{\endgroup\@href {#1}{\urlprefix }}%
\providecommand \urlprefix  [0]{URL }%
\providecommand \Eprint [0]{\href }%
\providecommand \doibase [0]{https://doi.org/}%
\providecommand \selectlanguage [0]{\@gobble}%
\providecommand \bibinfo  [0]{\@secondoftwo}%
\providecommand \bibfield  [0]{\@secondoftwo}%
\providecommand \translation [1]{[#1]}%
\providecommand \BibitemOpen [0]{}%
\providecommand \bibitemStop [0]{}%
\providecommand \bibitemNoStop [0]{.\EOS\space}%
\providecommand \EOS [0]{\spacefactor3000\relax}%
\providecommand \BibitemShut  [1]{\csname bibitem#1\endcsname}%
\let\auto@bib@innerbib\@empty
%</preamble>
\bibitem [{\citenamefont {Einstein}\ \emph {et~al.}(1935)\citenamefont
  {Einstein}, \citenamefont {Podolsky},\ and\ \citenamefont
  {Rosen}}]{EPRoriginal}%
  \BibitemOpen
  \bibfield  {author} {\bibinfo {author} {\bibfnamefont {A.}~\bibnamefont
  {Einstein}}, \bibinfo {author} {\bibfnamefont {B.}~\bibnamefont {Podolsky}},\
  and\ \bibinfo {author} {\bibfnamefont {N.}~\bibnamefont {Rosen}},\ }\bibfield
   {title} {\bibinfo {title} {Can quantum-mechanical description of physical
  reality be considered complete?},\ }\href
  {https://doi.org/10.1103/PhysRev.47.777} {\bibfield  {journal} {\bibinfo
  {journal} {Phys. Rev.}\ }\textbf {\bibinfo {volume} {47}},\ \bibinfo {pages}
  {777} (\bibinfo {year} {1935})}\BibitemShut {NoStop}%
\bibitem [{\citenamefont {Bell}(1964)}]{Bellineq}%
  \BibitemOpen
  \bibfield  {author} {\bibinfo {author} {\bibfnamefont {J.~S.}\ \bibnamefont
  {Bell}},\ }\bibfield  {title} {\bibinfo {title} {On the einstein podolsky
  rosen paradox},\ }\href {https://doi.org/10.1103/PhysicsPhysiqueFizika.1.195}
  {\bibfield  {journal} {\bibinfo  {journal} {Physics Physique Fizika}\
  }\textbf {\bibinfo {volume} {1}},\ \bibinfo {pages} {195} (\bibinfo {year}
  {1964})}\BibitemShut {NoStop}%
\bibitem [{\citenamefont {Freedman}\ and\ \citenamefont
  {Clauser}(1972)}]{FreedmanClauser}%
  \BibitemOpen
  \bibfield  {author} {\bibinfo {author} {\bibfnamefont {S.~J.}\ \bibnamefont
  {Freedman}}\ and\ \bibinfo {author} {\bibfnamefont {J.~F.}\ \bibnamefont
  {Clauser}},\ }\bibfield  {title} {\bibinfo {title} {Experimental test of
  local hidden-variable theories},\ }\href
  {https://doi.org/10.1103/PhysRevLett.28.938} {\bibfield  {journal} {\bibinfo
  {journal} {Phys. Rev. Lett.}\ }\textbf {\bibinfo {volume} {28}},\ \bibinfo
  {pages} {938} (\bibinfo {year} {1972})}\BibitemShut {NoStop}%
\bibitem [{\citenamefont {Aspect}\ \emph {et~al.}(1981)\citenamefont {Aspect},
  \citenamefont {Grangier},\ and\ \citenamefont {Roger}}]{Aspect1}%
  \BibitemOpen
  \bibfield  {author} {\bibinfo {author} {\bibfnamefont {A.}~\bibnamefont
  {Aspect}}, \bibinfo {author} {\bibfnamefont {P.}~\bibnamefont {Grangier}},\
  and\ \bibinfo {author} {\bibfnamefont {G.}~\bibnamefont {Roger}},\ }\bibfield
   {title} {\bibinfo {title} {Experimental tests of realistic local theories
  via bell's theorem},\ }\href {https://doi.org/10.1103/PhysRevLett.47.460}
  {\bibfield  {journal} {\bibinfo  {journal} {Phys. Rev. Lett.}\ }\textbf
  {\bibinfo {volume} {47}},\ \bibinfo {pages} {460} (\bibinfo {year}
  {1981})}\BibitemShut {NoStop}%
\bibitem [{\citenamefont {Aspect}\ \emph {et~al.}(1982)\citenamefont {Aspect},
  \citenamefont {Grangier},\ and\ \citenamefont {Roger}}]{Aspect2}%
  \BibitemOpen
  \bibfield  {author} {\bibinfo {author} {\bibfnamefont {A.}~\bibnamefont
  {Aspect}}, \bibinfo {author} {\bibfnamefont {P.}~\bibnamefont {Grangier}},\
  and\ \bibinfo {author} {\bibfnamefont {G.}~\bibnamefont {Roger}},\ }\bibfield
   {title} {\bibinfo {title} {Experimental realization of
  einstein-podolsky-rosen-bohm gedankenexperiment: A new violation of bell's
  inequalities},\ }\href {https://doi.org/10.1103/PhysRevLett.49.91} {\bibfield
   {journal} {\bibinfo  {journal} {Phys. Rev. Lett.}\ }\textbf {\bibinfo
  {volume} {49}},\ \bibinfo {pages} {91} (\bibinfo {year} {1982})}\BibitemShut
  {NoStop}%
\bibitem [{\citenamefont {Reeh}\ and\ \citenamefont
  {Schlieder}(1961)}]{ReehSchlieder}%
  \BibitemOpen
  \bibfield  {author} {\bibinfo {author} {\bibfnamefont {H.}~\bibnamefont
  {Reeh}}\ and\ \bibinfo {author} {\bibfnamefont {S.}~\bibnamefont
  {Schlieder}},\ }\bibfield  {title} {\bibinfo {title} {Bemerkungen zur
  unit{\"a}r{\"a}quivalenz von lorentzinvarianten feldern},\ }\href
  {https://doi.org/https://doi.org/10.1007/BF02787889} {\bibfield  {journal}
  {\bibinfo  {journal} {Il Nuovo Cimento (1955-1965)}\ }\textbf {\bibinfo
  {volume} {22}},\ \bibinfo {pages} {1051} (\bibinfo {year}
  {1961})}\BibitemShut {NoStop}%
\bibitem [{\citenamefont {Summers}\ and\ \citenamefont
  {Werner}(1987)}]{summers1987maximal}%
  \BibitemOpen
  \bibfield  {author} {\bibinfo {author} {\bibfnamefont {S.~J.}\ \bibnamefont
  {Summers}}\ and\ \bibinfo {author} {\bibfnamefont {R.}~\bibnamefont
  {Werner}},\ }\bibfield  {title} {\bibinfo {title} {Maximal violation of
  bell's inequalities is generic in quantum field theory},\ }\href
  {https://doi.org/https://doi.org/10.1007/BF01207366} {\bibfield  {journal}
  {\bibinfo  {journal} {Communications in Mathematical Physics}\ }\textbf
  {\bibinfo {volume} {110}},\ \bibinfo {pages} {247} (\bibinfo {year}
  {1987})}\BibitemShut {NoStop}%
\bibitem [{\citenamefont {Valentini}(1991)}]{Valentini}%
  \BibitemOpen
  \bibfield  {author} {\bibinfo {author} {\bibfnamefont {A.}~\bibnamefont
  {Valentini}},\ }\bibfield  {title} {\bibinfo {title} {Non-local correlations
  in quantum electrodynamics},\ }\href
  {https://doi.org/https://doi.org/10.1016/0375-9601(91)90952-5} {\bibfield
  {journal} {\bibinfo  {journal} {Physics Letters A}\ }\textbf {\bibinfo
  {volume} {153}},\ \bibinfo {pages} {321} (\bibinfo {year}
  {1991})}\BibitemShut {NoStop}%
\bibitem [{\citenamefont {Reznik}(2003)}]{Reznik1}%
  \BibitemOpen
  \bibfield  {author} {\bibinfo {author} {\bibfnamefont {B.}~\bibnamefont
  {Reznik}},\ }\bibfield  {title} {\bibinfo {title} {Entanglement from the
  vacuum},\ }\href {https://doi.org/https://doi.org/10.1023/A:1022875910744}
  {\bibfield  {journal} {\bibinfo  {journal} {Foundations of Physics}\ }\textbf
  {\bibinfo {volume} {33}},\ \bibinfo {pages} {167} (\bibinfo {year} {2003})},\
  \Eprint {https://arxiv.org/abs/quant-ph/0212044} {arXiv:quant-ph/0212044}
  \BibitemShut {NoStop}%
\bibitem [{\citenamefont {Reznik}\ \emph {et~al.}(2005)\citenamefont {Reznik},
  \citenamefont {Retzker},\ and\ \citenamefont {Silman}}]{Reznik2}%
  \BibitemOpen
  \bibfield  {author} {\bibinfo {author} {\bibfnamefont {B.}~\bibnamefont
  {Reznik}}, \bibinfo {author} {\bibfnamefont {A.}~\bibnamefont {Retzker}},\
  and\ \bibinfo {author} {\bibfnamefont {J.}~\bibnamefont {Silman}},\
  }\bibfield  {title} {\bibinfo {title} {Violating bell's inequalities in
  vacuum},\ }\href {https://doi.org/10.1103/PhysRevA.71.042104} {\bibfield
  {journal} {\bibinfo  {journal} {Phys. Rev. A}\ }\textbf {\bibinfo {volume}
  {71}},\ \bibinfo {pages} {042104} (\bibinfo {year} {2005})},\ \Eprint
  {https://arxiv.org/abs/quant-ph/0310058} {arXiv:quant-ph/0310058}
  \BibitemShut {NoStop}%
\bibitem [{\citenamefont {Witten}(2018)}]{Witten}%
  \BibitemOpen
  \bibfield  {author} {\bibinfo {author} {\bibfnamefont {E.}~\bibnamefont
  {Witten}},\ }\bibfield  {title} {\bibinfo {title} {Aps medal for exceptional
  achievement in research: Invited article on entanglement properties of
  quantum field theory},\ }\href {https://doi.org/10.1103/RevModPhys.90.045003}
  {\bibfield  {journal} {\bibinfo  {journal} {Rev. Mod. Phys.}\ }\textbf
  {\bibinfo {volume} {90}},\ \bibinfo {pages} {045003} (\bibinfo {year}
  {2018})},\ \Eprint {https://arxiv.org/abs/1803.04993} {arXiv:1803.04993
  [quant-ph]} \BibitemShut {NoStop}%
\bibitem [{\citenamefont {Feynman}(1982)}]{feynman2018simulating}%
  \BibitemOpen
  \bibfield  {author} {\bibinfo {author} {\bibfnamefont {R.~P.}\ \bibnamefont
  {Feynman}},\ }\bibfield  {title} {\bibinfo {title} {{Simulating physics with
  computers}},\ }\href {https://doi.org/10.1007/BF02650179} {\bibfield
  {journal} {\bibinfo  {journal} {Int. J. Theor. Phys.}\ }\textbf {\bibinfo
  {volume} {21}},\ \bibinfo {pages} {467} (\bibinfo {year} {1982})}\BibitemShut
  {NoStop}%
\bibitem [{\citenamefont {Bennett}\ and\ \citenamefont
  {Wiesner}(1992)}]{communicationandtele2}%
  \BibitemOpen
  \bibfield  {author} {\bibinfo {author} {\bibfnamefont {C.~H.}\ \bibnamefont
  {Bennett}}\ and\ \bibinfo {author} {\bibfnamefont {S.~J.}\ \bibnamefont
  {Wiesner}},\ }\bibfield  {title} {\bibinfo {title} {{Communication via one-
  and two-particle operators on Einstein-Podolsky-Rosen states}},\ }\href
  {https://doi.org/10.1103/PhysRevLett.69.2881} {\bibfield  {journal} {\bibinfo
   {journal} {Phys. Rev. Lett.}\ }\textbf {\bibinfo {volume} {69}},\ \bibinfo
  {pages} {2881} (\bibinfo {year} {1992})}\BibitemShut {NoStop}%
\bibitem [{\citenamefont {Bennett}\ \emph {et~al.}(1993)\citenamefont
  {Bennett}, \citenamefont {Brassard}, \citenamefont {Crepeau}, \citenamefont
  {Jozsa}, \citenamefont {Peres},\ and\ \citenamefont
  {Wootters}}]{communicationandtele1}%
  \BibitemOpen
  \bibfield  {author} {\bibinfo {author} {\bibfnamefont {C.~H.}\ \bibnamefont
  {Bennett}}, \bibinfo {author} {\bibfnamefont {G.}~\bibnamefont {Brassard}},
  \bibinfo {author} {\bibfnamefont {C.}~\bibnamefont {Crepeau}}, \bibinfo
  {author} {\bibfnamefont {R.}~\bibnamefont {Jozsa}}, \bibinfo {author}
  {\bibfnamefont {A.}~\bibnamefont {Peres}},\ and\ \bibinfo {author}
  {\bibfnamefont {W.~K.}\ \bibnamefont {Wootters}},\ }\bibfield  {title}
  {\bibinfo {title} {{Teleporting an unknown quantum state via dual classical
  and Einstein-Podolsky-Rosen channels}},\ }\href
  {https://doi.org/10.1103/PhysRevLett.70.1895} {\bibfield  {journal} {\bibinfo
   {journal} {Phys. Rev. Lett.}\ }\textbf {\bibinfo {volume} {70}},\ \bibinfo
  {pages} {1895} (\bibinfo {year} {1993})}\BibitemShut {NoStop}%
\bibitem [{\citenamefont {Bennett}\ and\ \citenamefont
  {Brassard}(2014)}]{security1}%
  \BibitemOpen
  \bibfield  {author} {\bibinfo {author} {\bibfnamefont {C.~H.}\ \bibnamefont
  {Bennett}}\ and\ \bibinfo {author} {\bibfnamefont {G.}~\bibnamefont
  {Brassard}},\ }\bibfield  {title} {\bibinfo {title} {{Quantum cryptography:
  Public key distribution and coin tossing}},\ }\href
  {https://doi.org/10.1016/j.tcs.2014.05.025} {\bibfield  {journal} {\bibinfo
  {journal} {Theor. Comput. Sci.}\ }\textbf {\bibinfo {volume} {560}},\
  \bibinfo {pages} {7} (\bibinfo {year} {2014})},\ \Eprint
  {https://arxiv.org/abs/2003.06557} {arXiv:2003.06557 [quant-ph]} \BibitemShut
  {NoStop}%
\bibitem [{\citenamefont {Aidelsburger}\ \emph {et~al.}(2022)\citenamefont
  {Aidelsburger} \emph {et~al.}}]{coldatom}%
  \BibitemOpen
  \bibfield  {author} {\bibinfo {author} {\bibfnamefont {M.}~\bibnamefont
  {Aidelsburger}} \emph {et~al.},\ }\bibfield  {title} {\bibinfo {title} {Cold
  atoms meet lattice gauge theory},\ }\href
  {https://doi.org/https://doi.org/10.1098/rsta.2021.0064} {\bibfield
  {journal} {\bibinfo  {journal} {Philosophical Transactions of the Royal
  Society A}\ }\textbf {\bibinfo {volume} {380}},\ \bibinfo {pages} {20210064}
  (\bibinfo {year} {2022})},\ \Eprint {https://arxiv.org/abs/2106.03063}
  {arXiv:2106.03063 [cond-mat.quant-gas]} \BibitemShut {NoStop}%
\bibitem [{\citenamefont {Klco}\ \emph {et~al.}(2022)\citenamefont {Klco},
  \citenamefont {Roggero},\ and\ \citenamefont {Savage}}]{NKreview}%
  \BibitemOpen
  \bibfield  {author} {\bibinfo {author} {\bibfnamefont {N.}~\bibnamefont
  {Klco}}, \bibinfo {author} {\bibfnamefont {A.}~\bibnamefont {Roggero}},\ and\
  \bibinfo {author} {\bibfnamefont {M.~J.}\ \bibnamefont {Savage}},\ }\bibfield
   {title} {\bibinfo {title} {{Standard model physics and the digital quantum
  revolution: thoughts about the interface}},\ }\href
  {https://doi.org/10.1088/1361-6633/ac58a4} {\bibfield  {journal} {\bibinfo
  {journal} {Rept. Prog. Phys.}\ }\textbf {\bibinfo {volume} {85}},\ \bibinfo
  {pages} {064301} (\bibinfo {year} {2022})},\ \Eprint
  {https://arxiv.org/abs/2107.04769} {arXiv:2107.04769 [quant-ph]} \BibitemShut
  {NoStop}%
\bibitem [{\citenamefont {Bauer}\ \emph
  {et~al.}(2023{\natexlab{a}})\citenamefont {Bauer} \emph
  {et~al.}}]{hepquantumsimulation}%
  \BibitemOpen
  \bibfield  {author} {\bibinfo {author} {\bibfnamefont {C.~W.}\ \bibnamefont
  {Bauer}} \emph {et~al.},\ }\bibfield  {title} {\bibinfo {title} {{Quantum
  Simulation for High-Energy Physics}},\ }\href
  {https://doi.org/10.1103/PRXQuantum.4.027001} {\bibfield  {journal} {\bibinfo
   {journal} {PRX Quantum}\ }\textbf {\bibinfo {volume} {4}},\ \bibinfo {pages}
  {027001} (\bibinfo {year} {2023}{\natexlab{a}})},\ \Eprint
  {https://arxiv.org/abs/2204.03381} {arXiv:2204.03381 [quant-ph]} \BibitemShut
  {NoStop}%
\bibitem [{\citenamefont {Bauer}\ \emph
  {et~al.}(2023{\natexlab{b}})\citenamefont {Bauer}, \citenamefont {Davoudi},
  \citenamefont {Klco},\ and\ \citenamefont {Savage}}]{bauer2023quantum}%
  \BibitemOpen
  \bibfield  {author} {\bibinfo {author} {\bibfnamefont {C.~W.}\ \bibnamefont
  {Bauer}}, \bibinfo {author} {\bibfnamefont {Z.}~\bibnamefont {Davoudi}},
  \bibinfo {author} {\bibfnamefont {N.}~\bibnamefont {Klco}},\ and\ \bibinfo
  {author} {\bibfnamefont {M.~J.}\ \bibnamefont {Savage}},\ }\bibfield  {title}
  {\bibinfo {title} {Quantum simulation of fundamental particles and forces},\
  }\href {https://doi.org/10.1038/s42254-023-00599-8} {\bibfield  {journal}
  {\bibinfo  {journal} {Nature Reviews Physics}\ }\textbf {\bibinfo {volume}
  {5}},\ \bibinfo {pages} {420} (\bibinfo {year} {2023}{\natexlab{b}})},\
  \Eprint {https://arxiv.org/abs/2404.06298} {arXiv:2404.06298 [hep-ph]}
  \BibitemShut {NoStop}%
\bibitem [{\citenamefont {Srednicki}(1993)}]{Srednicki:1993im}%
  \BibitemOpen
  \bibfield  {author} {\bibinfo {author} {\bibfnamefont {M.}~\bibnamefont
  {Srednicki}},\ }\bibfield  {title} {\bibinfo {title} {{Entropy and area}},\
  }\href {https://doi.org/10.1103/PhysRevLett.71.666} {\bibfield  {journal}
  {\bibinfo  {journal} {Phys. Rev. Lett.}\ }\textbf {\bibinfo {volume} {71}},\
  \bibinfo {pages} {666} (\bibinfo {year} {1993})},\ \Eprint
  {https://arxiv.org/abs/hep-th/9303048} {arXiv:hep-th/9303048} \BibitemShut
  {NoStop}%
\bibitem [{\citenamefont {Bennett}\ \emph
  {et~al.}(1996{\natexlab{a}})\citenamefont {Bennett}, \citenamefont
  {Bernstein}, \citenamefont {Popescu},\ and\ \citenamefont
  {Schumacher}}]{Concentratingpartialentanglement}%
  \BibitemOpen
  \bibfield  {author} {\bibinfo {author} {\bibfnamefont {C.~H.}\ \bibnamefont
  {Bennett}}, \bibinfo {author} {\bibfnamefont {H.~J.}\ \bibnamefont
  {Bernstein}}, \bibinfo {author} {\bibfnamefont {S.}~\bibnamefont {Popescu}},\
  and\ \bibinfo {author} {\bibfnamefont {B.}~\bibnamefont {Schumacher}},\
  }\bibfield  {title} {\bibinfo {title} {Concentrating partial entanglement by
  local operations},\ }\href {https://doi.org/10.1103/PhysRevA.53.2046}
  {\bibfield  {journal} {\bibinfo  {journal} {Phys. Rev. A}\ }\textbf {\bibinfo
  {volume} {53}},\ \bibinfo {pages} {2046} (\bibinfo {year}
  {1996}{\natexlab{a}})},\ \Eprint {https://arxiv.org/abs/quant-ph/9511030}
  {arXiv:quant-ph/9511030} \BibitemShut {NoStop}%
\bibitem [{\citenamefont {Nielsen}(1999)}]{ConditionsforaClass}%
  \BibitemOpen
  \bibfield  {author} {\bibinfo {author} {\bibfnamefont {M.~A.}\ \bibnamefont
  {Nielsen}},\ }\bibfield  {title} {\bibinfo {title} {Conditions for a class of
  entanglement transformations},\ }\href
  {https://doi.org/10.1103/PhysRevLett.83.436} {\bibfield  {journal} {\bibinfo
  {journal} {Phys. Rev. Lett.}\ }\textbf {\bibinfo {volume} {83}},\ \bibinfo
  {pages} {436} (\bibinfo {year} {1999})},\ \Eprint
  {https://arxiv.org/abs/quant-ph/9811053} {arXiv:quant-ph/9811053}
  \BibitemShut {NoStop}%
\bibitem [{\citenamefont {Jonathan}\ and\ \citenamefont
  {Plenio}(1999)}]{MinimalConditions}%
  \BibitemOpen
  \bibfield  {author} {\bibinfo {author} {\bibfnamefont {D.}~\bibnamefont
  {Jonathan}}\ and\ \bibinfo {author} {\bibfnamefont {M.~B.}\ \bibnamefont
  {Plenio}},\ }\bibfield  {title} {\bibinfo {title} {Minimal conditions for
  local pure-state entanglement manipulation},\ }\href
  {https://doi.org/10.1103/PhysRevLett.83.1455} {\bibfield  {journal} {\bibinfo
   {journal} {Phys. Rev. Lett.}\ }\textbf {\bibinfo {volume} {83}},\ \bibinfo
  {pages} {1455} (\bibinfo {year} {1999})},\ \Eprint
  {https://arxiv.org/abs/quant-ph/9903054} {arXiv:quant-ph/9903054}
  \BibitemShut {NoStop}%
\bibitem [{\citenamefont {Vidal}\ \emph {et~al.}(2000)\citenamefont {Vidal},
  \citenamefont {Jonathan},\ and\ \citenamefont
  {Nielsen}}]{Approximatetransformations}%
  \BibitemOpen
  \bibfield  {author} {\bibinfo {author} {\bibfnamefont {G.}~\bibnamefont
  {Vidal}}, \bibinfo {author} {\bibfnamefont {D.}~\bibnamefont {Jonathan}},\
  and\ \bibinfo {author} {\bibfnamefont {M.~A.}\ \bibnamefont {Nielsen}},\
  }\bibfield  {title} {\bibinfo {title} {Approximate transformations and robust
  manipulation of bipartite pure-state entanglement},\ }\href
  {https://doi.org/10.1103/PhysRevA.62.012304} {\bibfield  {journal} {\bibinfo
  {journal} {Phys. Rev. A}\ }\textbf {\bibinfo {volume} {62}},\ \bibinfo
  {pages} {012304} (\bibinfo {year} {2000})},\ \Eprint
  {https://arxiv.org/abs/quant-ph/9910099} {arXiv:quant-ph/9910099}
  \BibitemShut {NoStop}%
\bibitem [{\citenamefont {Vidal}(2000)}]{purestateent}%
  \BibitemOpen
  \bibfield  {author} {\bibinfo {author} {\bibfnamefont {G.}~\bibnamefont
  {Vidal}},\ }\bibfield  {title} {\bibinfo {title} {{Entanglement monotones}},\
  }\href {https://doi.org/10.1080/09500340008244048} {\bibfield  {journal}
  {\bibinfo  {journal} {J. Mod. Opt.}\ }\textbf {\bibinfo {volume} {47}},\
  \bibinfo {pages} {355} (\bibinfo {year} {2000})},\ \Eprint
  {https://arxiv.org/abs/quant-ph/9807077} {arXiv:quant-ph/9807077}
  \BibitemShut {NoStop}%
\bibitem [{\citenamefont {Bennett}\ \emph
  {et~al.}(1996{\natexlab{b}})\citenamefont {Bennett}, \citenamefont
  {DiVincenzo}, \citenamefont {Smolin},\ and\ \citenamefont
  {Wootters}}]{mixstateent}%
  \BibitemOpen
  \bibfield  {author} {\bibinfo {author} {\bibfnamefont {C.~H.}\ \bibnamefont
  {Bennett}}, \bibinfo {author} {\bibfnamefont {D.~P.}\ \bibnamefont
  {DiVincenzo}}, \bibinfo {author} {\bibfnamefont {J.~A.}\ \bibnamefont
  {Smolin}},\ and\ \bibinfo {author} {\bibfnamefont {W.~K.}\ \bibnamefont
  {Wootters}},\ }\bibfield  {title} {\bibinfo {title} {{Mixed state
  entanglement and quantum error correction}},\ }\href
  {https://doi.org/10.1103/PhysRevA.54.3824} {\bibfield  {journal} {\bibinfo
  {journal} {Phys. Rev. A}\ }\textbf {\bibinfo {volume} {54}},\ \bibinfo
  {pages} {3824} (\bibinfo {year} {1996}{\natexlab{b}})},\ \Eprint
  {https://arxiv.org/abs/quant-ph/9604024} {arXiv:quant-ph/9604024}
  \BibitemShut {NoStop}%
\bibitem [{\citenamefont {Peres}(1996)}]{peresoriginalN}%
  \BibitemOpen
  \bibfield  {author} {\bibinfo {author} {\bibfnamefont {A.}~\bibnamefont
  {Peres}},\ }\bibfield  {title} {\bibinfo {title} {{Separability criterion for
  density matrices}},\ }\href {https://doi.org/10.1103/PhysRevLett.77.1413}
  {\bibfield  {journal} {\bibinfo  {journal} {Phys. Rev. Lett.}\ }\textbf
  {\bibinfo {volume} {77}},\ \bibinfo {pages} {1413} (\bibinfo {year}
  {1996})},\ \Eprint {https://arxiv.org/abs/quant-ph/9604005}
  {arXiv:quant-ph/9604005} \BibitemShut {NoStop}%
\bibitem [{\citenamefont {Horodecki}\ \emph {et~al.}(1996)\citenamefont
  {Horodecki}, \citenamefont {Horodecki},\ and\ \citenamefont
  {Horodecki}}]{HORODECKIoriginalN}%
  \BibitemOpen
  \bibfield  {author} {\bibinfo {author} {\bibfnamefont {M.}~\bibnamefont
  {Horodecki}}, \bibinfo {author} {\bibfnamefont {P.}~\bibnamefont
  {Horodecki}},\ and\ \bibinfo {author} {\bibfnamefont {R.}~\bibnamefont
  {Horodecki}},\ }\bibfield  {title} {\bibinfo {title} {{Separability of mixed
  states: necessary and sufficient conditions}},\ }\href
  {https://doi.org/10.1016/S0375-9601(96)00706-2} {\bibfield  {journal}
  {\bibinfo  {journal} {Phys. Lett. A}\ }\textbf {\bibinfo {volume} {223}},\
  \bibinfo {pages} {1} (\bibinfo {year} {1996})},\ \Eprint
  {https://arxiv.org/abs/quant-ph/9605038} {arXiv:quant-ph/9605038}
  \BibitemShut {NoStop}%
\bibitem [{\citenamefont {Simon}(2000)}]{Simonreflection}%
  \BibitemOpen
  \bibfield  {author} {\bibinfo {author} {\bibfnamefont {R.}~\bibnamefont
  {Simon}},\ }\bibfield  {title} {\bibinfo {title} {{Peres-Horodecki
  Separability Criterion for Continuous Variable Systems}},\ }\href
  {https://doi.org/10.1103/PhysRevLett.84.2726} {\bibfield  {journal} {\bibinfo
   {journal} {Phys. Rev. Lett.}\ }\textbf {\bibinfo {volume} {84}},\ \bibinfo
  {pages} {2726} (\bibinfo {year} {2000})},\ \Eprint
  {https://arxiv.org/abs/quant-ph/9909044} {arXiv:quant-ph/9909044}
  \BibitemShut {NoStop}%
\bibitem [{\citenamefont {Duan}\ \emph
  {et~al.}(2000{\natexlab{a}})\citenamefont {Duan}, \citenamefont {Giedke},
  \citenamefont {Cirac},\ and\ \citenamefont {Zoller}}]{Duanlocaltrans}%
  \BibitemOpen
  \bibfield  {author} {\bibinfo {author} {\bibfnamefont {L.-M.}\ \bibnamefont
  {Duan}}, \bibinfo {author} {\bibfnamefont {G.}~\bibnamefont {Giedke}},
  \bibinfo {author} {\bibfnamefont {J.~I.}\ \bibnamefont {Cirac}},\ and\
  \bibinfo {author} {\bibfnamefont {P.}~\bibnamefont {Zoller}},\ }\bibfield
  {title} {\bibinfo {title} {{Inseparability Criterion for Continuous Variable
  Systems}},\ }\href {https://doi.org/10.1103/PhysRevLett.84.2722} {\bibfield
  {journal} {\bibinfo  {journal} {Phys. Rev. Lett.}\ }\textbf {\bibinfo
  {volume} {84}},\ \bibinfo {pages} {2722} (\bibinfo {year}
  {2000}{\natexlab{a}})},\ \Eprint {https://arxiv.org/abs/quant-ph/9908056}
  {arXiv:quant-ph/9908056} \BibitemShut {NoStop}%
\bibitem [{\citenamefont {Vidal}\ and\ \citenamefont
  {Werner}(2002)}]{computablemeasure}%
  \BibitemOpen
  \bibfield  {author} {\bibinfo {author} {\bibfnamefont {G.}~\bibnamefont
  {Vidal}}\ and\ \bibinfo {author} {\bibfnamefont {R.~F.}\ \bibnamefont
  {Werner}},\ }\bibfield  {title} {\bibinfo {title} {{Computable measure of
  entanglement}},\ }\href {https://doi.org/10.1103/PhysRevA.65.032314}
  {\bibfield  {journal} {\bibinfo  {journal} {Phys. Rev. A}\ }\textbf {\bibinfo
  {volume} {65}},\ \bibinfo {pages} {032314} (\bibinfo {year} {2002})},\
  \Eprint {https://arxiv.org/abs/quant-ph/0102117} {arXiv:quant-ph/0102117}
  \BibitemShut {NoStop}%
\bibitem [{\citenamefont {Plenio}(2005)}]{PlenioLogarithmic}%
  \BibitemOpen
  \bibfield  {author} {\bibinfo {author} {\bibfnamefont {M.~B.}\ \bibnamefont
  {Plenio}},\ }\bibfield  {title} {\bibinfo {title} {Logarithmic negativity: A
  full entanglement monotone that is not convex},\ }\href
  {https://doi.org/10.1103/PhysRevLett.95.090503} {\bibfield  {journal}
  {\bibinfo  {journal} {Phys. Rev. Lett.}\ }\textbf {\bibinfo {volume} {95}},\
  \bibinfo {pages} {090503} (\bibinfo {year} {2005})},\ \Eprint
  {https://arxiv.org/abs/quant-ph/0505071} {arXiv:quant-ph/0505071}
  \BibitemShut {NoStop}%
\bibitem [{\citenamefont {Vedral}\ \emph {et~al.}(1997)\citenamefont {Vedral},
  \citenamefont {Plenio}, \citenamefont {Rippin},\ and\ \citenamefont
  {Knight}}]{quantifyent}%
  \BibitemOpen
  \bibfield  {author} {\bibinfo {author} {\bibfnamefont {V.}~\bibnamefont
  {Vedral}}, \bibinfo {author} {\bibfnamefont {M.~B.}\ \bibnamefont {Plenio}},
  \bibinfo {author} {\bibfnamefont {M.~A.}\ \bibnamefont {Rippin}},\ and\
  \bibinfo {author} {\bibfnamefont {P.~L.}\ \bibnamefont {Knight}},\ }\bibfield
   {title} {\bibinfo {title} {Quantifying entanglement},\ }\href
  {https://doi.org/10.1103/PhysRevLett.78.2275} {\bibfield  {journal} {\bibinfo
   {journal} {Phys. Rev. Lett.}\ }\textbf {\bibinfo {volume} {78}},\ \bibinfo
  {pages} {2275} (\bibinfo {year} {1997})},\ \Eprint
  {https://arxiv.org/abs/quant-ph/9702027} {arXiv:quant-ph/9702027}
  \BibitemShut {NoStop}%
\bibitem [{\citenamefont {Christandl}\ and\ \citenamefont
  {Winter}(2004)}]{Christandl_2004}%
  \BibitemOpen
  \bibfield  {author} {\bibinfo {author} {\bibfnamefont {M.}~\bibnamefont
  {Christandl}}\ and\ \bibinfo {author} {\bibfnamefont {A.}~\bibnamefont
  {Winter}},\ }\bibfield  {title} {\bibinfo {title} {“squashed
  entanglement”: An additive entanglement measure},\ }\href
  {https://doi.org/10.1063/1.1643788} {\bibfield  {journal} {\bibinfo
  {journal} {Journal of Mathematical Physics}\ }\textbf {\bibinfo {volume}
  {45}},\ \bibinfo {pages} {829–840} (\bibinfo {year} {2004})},\ \Eprint
  {https://arxiv.org/abs/quant-ph/0308088} {arXiv:quant-ph/0308088}
  \BibitemShut {NoStop}%
\bibitem [{\citenamefont {Yang}\ \emph {et~al.}(2008)\citenamefont {Yang},
  \citenamefont {Horodecki},\ and\ \citenamefont
  {Wang}}]{PhysRevLett.101.140501}%
  \BibitemOpen
  \bibfield  {author} {\bibinfo {author} {\bibfnamefont {D.}~\bibnamefont
  {Yang}}, \bibinfo {author} {\bibfnamefont {M.}~\bibnamefont {Horodecki}},\
  and\ \bibinfo {author} {\bibfnamefont {Z.~D.}\ \bibnamefont {Wang}},\
  }\bibfield  {title} {\bibinfo {title} {An additive and operational
  entanglement measure: Conditional entanglement of mutual information},\
  }\href {https://doi.org/10.1103/PhysRevLett.101.140501} {\bibfield  {journal}
  {\bibinfo  {journal} {Phys. Rev. Lett.}\ }\textbf {\bibinfo {volume} {101}},\
  \bibinfo {pages} {140501} (\bibinfo {year} {2008})},\ \Eprint
  {https://arxiv.org/abs/0804.3683} {arXiv:0804.3683 [quant-ph]} \BibitemShut
  {NoStop}%
\bibitem [{\citenamefont {Piani}\ \emph {et~al.}(2009)\citenamefont {Piani},
  \citenamefont {Christandl}, \citenamefont {Mora},\ and\ \citenamefont
  {Horodecki}}]{PhysRevLett.102.250503}%
  \BibitemOpen
  \bibfield  {author} {\bibinfo {author} {\bibfnamefont {M.}~\bibnamefont
  {Piani}}, \bibinfo {author} {\bibfnamefont {M.}~\bibnamefont {Christandl}},
  \bibinfo {author} {\bibfnamefont {C.~E.}\ \bibnamefont {Mora}},\ and\
  \bibinfo {author} {\bibfnamefont {P.}~\bibnamefont {Horodecki}},\ }\bibfield
  {title} {\bibinfo {title} {Broadcast copies reveal the quantumness of
  correlations},\ }\href {https://doi.org/10.1103/PhysRevLett.102.250503}
  {\bibfield  {journal} {\bibinfo  {journal} {Phys. Rev. Lett.}\ }\textbf
  {\bibinfo {volume} {102}},\ \bibinfo {pages} {250503} (\bibinfo {year}
  {2009})},\ \Eprint {https://arxiv.org/abs/0901.1280} {arXiv:0901.1280
  [quant-ph]} \BibitemShut {NoStop}%
\bibitem [{\citenamefont {Huang}(2014)}]{Huang_2014}%
  \BibitemOpen
  \bibfield  {author} {\bibinfo {author} {\bibfnamefont {Y.}~\bibnamefont
  {Huang}},\ }\bibfield  {title} {\bibinfo {title} {Computing quantum discord
  is np-complete},\ }\href {https://doi.org/10.1088/1367-2630/16/3/033027}
  {\bibfield  {journal} {\bibinfo  {journal} {New Journal of Physics}\ }\textbf
  {\bibinfo {volume} {16}},\ \bibinfo {pages} {033027} (\bibinfo {year}
  {2014})},\ \Eprint {https://arxiv.org/abs/1305.5941} {arXiv:1305.5941
  [quant-ph]} \BibitemShut {NoStop}%
\bibitem [{\citenamefont {Hiesmayr}(2021)}]{hiesmayr2021free}%
  \BibitemOpen
  \bibfield  {author} {\bibinfo {author} {\bibfnamefont {B.~C.}\ \bibnamefont
  {Hiesmayr}},\ }\bibfield  {title} {\bibinfo {title} {Free versus bound
  entanglement, a np-hard problem tackled by machine learning},\ }\href
  {https://doi.org/https://doi.org/10.1038/s41598-021-98523-6} {\bibfield
  {journal} {\bibinfo  {journal} {Scientific Reports}\ }\textbf {\bibinfo
  {volume} {11}},\ \bibinfo {pages} {19739} (\bibinfo {year} {2021})},\ \Eprint
  {https://arxiv.org/abs/2106.03977} {arXiv:2106.03977 [quant-ph]} \BibitemShut
  {NoStop}%
\bibitem [{\citenamefont {Horodecki}\ \emph {et~al.}(1998)\citenamefont
  {Horodecki}, \citenamefont {Horodecki},\ and\ \citenamefont
  {Horodecki}}]{originalboundent}%
  \BibitemOpen
  \bibfield  {author} {\bibinfo {author} {\bibfnamefont {M.}~\bibnamefont
  {Horodecki}}, \bibinfo {author} {\bibfnamefont {P.}~\bibnamefont
  {Horodecki}},\ and\ \bibinfo {author} {\bibfnamefont {R.}~\bibnamefont
  {Horodecki}},\ }\bibfield  {title} {\bibinfo {title} {Mixed-state
  entanglement and distillation: Is there a ``bound'' entanglement in
  nature?},\ }\href {https://doi.org/10.1103/PhysRevLett.80.5239} {\bibfield
  {journal} {\bibinfo  {journal} {Phys. Rev. Lett.}\ }\textbf {\bibinfo
  {volume} {80}},\ \bibinfo {pages} {5239} (\bibinfo {year} {1998})},\ \Eprint
  {https://arxiv.org/abs/quant-ph/9801069} {arXiv:quant-ph/9801069}
  \BibitemShut {NoStop}%
\bibitem [{\citenamefont {Bennett}\ \emph {et~al.}(1999)\citenamefont
  {Bennett}, \citenamefont {DiVincenzo}, \citenamefont {Mor}, \citenamefont
  {Shor}, \citenamefont {Smolin},\ and\ \citenamefont
  {Terhal}}]{anotherbound2}%
  \BibitemOpen
  \bibfield  {author} {\bibinfo {author} {\bibfnamefont {C.~H.}\ \bibnamefont
  {Bennett}}, \bibinfo {author} {\bibfnamefont {D.~P.}\ \bibnamefont
  {DiVincenzo}}, \bibinfo {author} {\bibfnamefont {T.}~\bibnamefont {Mor}},
  \bibinfo {author} {\bibfnamefont {P.~W.}\ \bibnamefont {Shor}}, \bibinfo
  {author} {\bibfnamefont {J.~A.}\ \bibnamefont {Smolin}},\ and\ \bibinfo
  {author} {\bibfnamefont {B.~M.}\ \bibnamefont {Terhal}},\ }\bibfield  {title}
  {\bibinfo {title} {Unextendible product bases and bound entanglement},\
  }\href {https://doi.org/10.1103/PhysRevLett.82.5385} {\bibfield  {journal}
  {\bibinfo  {journal} {Phys. Rev. Lett.}\ }\textbf {\bibinfo {volume} {82}},\
  \bibinfo {pages} {5385} (\bibinfo {year} {1999})},\ \Eprint
  {https://arxiv.org/abs/quant-ph/9808030} {arXiv:quant-ph/9808030}
  \BibitemShut {NoStop}%
\bibitem [{\citenamefont {Werner}\ and\ \citenamefont
  {Wolf}(2001)}]{Gaussianboundent}%
  \BibitemOpen
  \bibfield  {author} {\bibinfo {author} {\bibfnamefont {R.~F.}\ \bibnamefont
  {Werner}}\ and\ \bibinfo {author} {\bibfnamefont {M.~M.}\ \bibnamefont
  {Wolf}},\ }\bibfield  {title} {\bibinfo {title} {Bound entangled gaussian
  states},\ }\href {https://doi.org/10.1103/PhysRevLett.86.3658} {\bibfield
  {journal} {\bibinfo  {journal} {Phys. Rev. Lett.}\ }\textbf {\bibinfo
  {volume} {86}},\ \bibinfo {pages} {3658} (\bibinfo {year} {2001})},\ \Eprint
  {https://arxiv.org/abs/quant-ph/0009118} {arXiv:quant-ph/0009118}
  \BibitemShut {NoStop}%
\bibitem [{\citenamefont {Breuer}(2006)}]{anotherbound1}%
  \BibitemOpen
  \bibfield  {author} {\bibinfo {author} {\bibfnamefont {H.-P.}\ \bibnamefont
  {Breuer}},\ }\bibfield  {title} {\bibinfo {title} {Optimal entanglement
  criterion for mixed quantum states},\ }\href
  {https://doi.org/10.1103/PhysRevLett.97.080501} {\bibfield  {journal}
  {\bibinfo  {journal} {Phys. Rev. Lett.}\ }\textbf {\bibinfo {volume} {97}},\
  \bibinfo {pages} {080501} (\bibinfo {year} {2006})},\ \Eprint
  {https://arxiv.org/abs/quant-ph/0605036} {arXiv:quant-ph/0605036}
  \BibitemShut {NoStop}%
\bibitem [{\citenamefont {Verstraete}\ \emph {et~al.}(2002)\citenamefont
  {Verstraete}, \citenamefont {Dehaene}, \citenamefont {De~Moor},\ and\
  \citenamefont {Verschelde}}]{Fourqubits}%
  \BibitemOpen
  \bibfield  {author} {\bibinfo {author} {\bibfnamefont {F.}~\bibnamefont
  {Verstraete}}, \bibinfo {author} {\bibfnamefont {J.}~\bibnamefont {Dehaene}},
  \bibinfo {author} {\bibfnamefont {B.}~\bibnamefont {De~Moor}},\ and\ \bibinfo
  {author} {\bibfnamefont {H.}~\bibnamefont {Verschelde}},\ }\bibfield  {title}
  {\bibinfo {title} {Four qubits can be entangled in nine different ways},\
  }\href {https://doi.org/10.1103/PhysRevA.65.052112} {\bibfield  {journal}
  {\bibinfo  {journal} {Phys. Rev. A}\ }\textbf {\bibinfo {volume} {65}},\
  \bibinfo {pages} {052112} (\bibinfo {year} {2002})},\ \Eprint
  {https://arxiv.org/abs/quant-ph/0109033} {arXiv:quant-ph/0109033}
  \BibitemShut {NoStop}%
\bibitem [{\citenamefont {Giedke}\ \emph
  {et~al.}(2003{\natexlab{a}})\citenamefont {Giedke}, \citenamefont {Wolf},
  \citenamefont {Krueger}, \citenamefont {Werner},\ and\ \citenamefont
  {Cirac}}]{GiedkeEOF}%
  \BibitemOpen
  \bibfield  {author} {\bibinfo {author} {\bibfnamefont {G.}~\bibnamefont
  {Giedke}}, \bibinfo {author} {\bibfnamefont {M.~M.}\ \bibnamefont {Wolf}},
  \bibinfo {author} {\bibfnamefont {O.}~\bibnamefont {Krueger}}, \bibinfo
  {author} {\bibfnamefont {R.~F.}\ \bibnamefont {Werner}},\ and\ \bibinfo
  {author} {\bibfnamefont {J.~I.}\ \bibnamefont {Cirac}},\ }\bibfield  {title}
  {\bibinfo {title} {{Entanglement of Formation for Symmetric Gaussian
  States}},\ }\href {https://doi.org/10.1103/PhysRevLett.91.107901} {\bibfield
  {journal} {\bibinfo  {journal} {Phys. Rev. Lett.}\ }\textbf {\bibinfo
  {volume} {91}},\ \bibinfo {pages} {107901} (\bibinfo {year}
  {2003}{\natexlab{a}})},\ \Eprint {https://arxiv.org/abs/quant-ph/0304042}
  {arXiv:quant-ph/0304042} \BibitemShut {NoStop}%
\bibitem [{\citenamefont {Botero}\ and\ \citenamefont
  {Reznik}(2003)}]{PhysRevA.67.052311}%
  \BibitemOpen
  \bibfield  {author} {\bibinfo {author} {\bibfnamefont {A.}~\bibnamefont
  {Botero}}\ and\ \bibinfo {author} {\bibfnamefont {B.}~\bibnamefont
  {Reznik}},\ }\bibfield  {title} {\bibinfo {title} {Modewise entanglement of
  gaussian states},\ }\href {https://doi.org/10.1103/PhysRevA.67.052311}
  {\bibfield  {journal} {\bibinfo  {journal} {Phys. Rev. A}\ }\textbf {\bibinfo
  {volume} {67}},\ \bibinfo {pages} {052311} (\bibinfo {year} {2003})},\
  \Eprint {https://arxiv.org/abs/quant-ph/0209026} {arXiv:quant-ph/0209026}
  \BibitemShut {NoStop}%
\bibitem [{\citenamefont {Wolf}\ \emph {et~al.}(2004)\citenamefont {Wolf},
  \citenamefont {Giedke}, \citenamefont {Krueger}, \citenamefont {Werner},\
  and\ \citenamefont {Cirac}}]{WolfGEOF}%
  \BibitemOpen
  \bibfield  {author} {\bibinfo {author} {\bibfnamefont {M.~M.}\ \bibnamefont
  {Wolf}}, \bibinfo {author} {\bibfnamefont {G.}~\bibnamefont {Giedke}},
  \bibinfo {author} {\bibfnamefont {O.}~\bibnamefont {Krueger}}, \bibinfo
  {author} {\bibfnamefont {R.~F.}\ \bibnamefont {Werner}},\ and\ \bibinfo
  {author} {\bibfnamefont {J.~I.}\ \bibnamefont {Cirac}},\ }\bibfield  {title}
  {\bibinfo {title} {{Gaussian entanglement of formation}},\ }\href
  {https://doi.org/10.1103/PhysRevA.69.052320} {\bibfield  {journal} {\bibinfo
  {journal} {Phys. Rev. A}\ }\textbf {\bibinfo {volume} {69}},\ \bibinfo
  {pages} {052320} (\bibinfo {year} {2004})},\ \Eprint
  {https://arxiv.org/abs/quant-ph/0306177} {arXiv:quant-ph/0306177}
  \BibitemShut {NoStop}%
\bibitem [{\citenamefont {Adesso}\ \emph {et~al.}(2006)\citenamefont {Adesso},
  \citenamefont {Serafini},\ and\ \citenamefont {Illuminati}}]{It2}%
  \BibitemOpen
  \bibfield  {author} {\bibinfo {author} {\bibfnamefont {G.}~\bibnamefont
  {Adesso}}, \bibinfo {author} {\bibfnamefont {A.}~\bibnamefont {Serafini}},\
  and\ \bibinfo {author} {\bibfnamefont {F.}~\bibnamefont {Illuminati}},\
  }\bibfield  {title} {\bibinfo {title} {Multipartite entanglement in
  three-mode gaussian states of continuous-variable systems: Quantification,
  sharing structure, and decoherence},\ }\href
  {https://doi.org/10.1103/PhysRevA.73.032345} {\bibfield  {journal} {\bibinfo
  {journal} {Phys. Rev. A}\ }\textbf {\bibinfo {volume} {73}},\ \bibinfo
  {pages} {032345} (\bibinfo {year} {2006})},\ \Eprint
  {https://arxiv.org/abs/quant-ph/0512124} {arXiv:quant-ph/0512124}
  \BibitemShut {NoStop}%
\bibitem [{\citenamefont {Marian}\ and\ \citenamefont
  {Marian}(2008)}]{marian2008entanglement}%
  \BibitemOpen
  \bibfield  {author} {\bibinfo {author} {\bibfnamefont {P.}~\bibnamefont
  {Marian}}\ and\ \bibinfo {author} {\bibfnamefont {T.~A.}\ \bibnamefont
  {Marian}},\ }\bibfield  {title} {\bibinfo {title} {{Entanglement of Formation
  for an Arbitrary Two-Mode Gaussian State}},\ }\href
  {https://doi.org/10.1103/PhysRevLett.101.220403} {\bibfield  {journal}
  {\bibinfo  {journal} {Phys. Rev. Lett.}\ }\textbf {\bibinfo {volume} {101}},\
  \bibinfo {pages} {220403} (\bibinfo {year} {2008})},\ \Eprint
  {https://arxiv.org/abs/0809.0321} {arXiv:0809.0321 [quant-ph]} \BibitemShut
  {NoStop}%
\bibitem [{\citenamefont {Tserkis}\ and\ \citenamefont
  {Ralph}(2017)}]{TserkisandRalph}%
  \BibitemOpen
  \bibfield  {author} {\bibinfo {author} {\bibfnamefont {S.}~\bibnamefont
  {Tserkis}}\ and\ \bibinfo {author} {\bibfnamefont {T.~C.}\ \bibnamefont
  {Ralph}},\ }\bibfield  {title} {\bibinfo {title} {Quantifying entanglement in
  two-mode gaussian states},\ }\href
  {https://doi.org/10.1103/PhysRevA.96.062338} {\bibfield  {journal} {\bibinfo
  {journal} {Phys. Rev. A}\ }\textbf {\bibinfo {volume} {96}},\ \bibinfo
  {pages} {062338} (\bibinfo {year} {2017})},\ \Eprint
  {https://arxiv.org/abs/1705.03612} {arXiv:1705.03612 [quant-ph]} \BibitemShut
  {NoStop}%
\bibitem [{\citenamefont {Tserkis}\ \emph {et~al.}(2019)\citenamefont
  {Tserkis}, \citenamefont {Onoe},\ and\ \citenamefont {Ralph}}]{2019geof}%
  \BibitemOpen
  \bibfield  {author} {\bibinfo {author} {\bibfnamefont {S.}~\bibnamefont
  {Tserkis}}, \bibinfo {author} {\bibfnamefont {S.}~\bibnamefont {Onoe}},\ and\
  \bibinfo {author} {\bibfnamefont {T.~C.}\ \bibnamefont {Ralph}},\ }\bibfield
  {title} {\bibinfo {title} {Quantifying entanglement of formation for two-mode
  gaussian states: Analytical expressions for upper and lower bounds and
  numerical estimation of its exact value},\ }\href
  {https://doi.org/10.1103/PhysRevA.99.052337} {\bibfield  {journal} {\bibinfo
  {journal} {Phys. Rev. A}\ }\textbf {\bibinfo {volume} {99}},\ \bibinfo
  {pages} {052337} (\bibinfo {year} {2019})},\ \Eprint
  {https://arxiv.org/abs/1903.09961} {arXiv:1903.09961 [quant-ph]} \BibitemShut
  {NoStop}%
\bibitem [{\citenamefont {Adesso}\ \emph {et~al.}(2004)\citenamefont {Adesso},
  \citenamefont {Serafini},\ and\ \citenamefont {Illuminati}}]{It1}%
  \BibitemOpen
  \bibfield  {author} {\bibinfo {author} {\bibfnamefont {G.}~\bibnamefont
  {Adesso}}, \bibinfo {author} {\bibfnamefont {A.}~\bibnamefont {Serafini}},\
  and\ \bibinfo {author} {\bibfnamefont {F.}~\bibnamefont {Illuminati}},\
  }\bibfield  {title} {\bibinfo {title} {{Extremal entanglement and mixedness
  in continuous variable systems}},\ }\href
  {https://doi.org/10.1103/PhysRevA.70.022318} {\bibfield  {journal} {\bibinfo
  {journal} {Phys. Rev. A}\ }\textbf {\bibinfo {volume} {70}},\ \bibinfo
  {pages} {022318} (\bibinfo {year} {2004})},\ \Eprint
  {https://arxiv.org/abs/quant-ph/0402124} {arXiv:quant-ph/0402124}
  \BibitemShut {NoStop}%
\bibitem [{\citenamefont {Adesso}\ and\ \citenamefont
  {Illuminati}(2005)}]{It3}%
  \BibitemOpen
  \bibfield  {author} {\bibinfo {author} {\bibfnamefont {G.}~\bibnamefont
  {Adesso}}\ and\ \bibinfo {author} {\bibfnamefont {F.}~\bibnamefont
  {Illuminati}},\ }\bibfield  {title} {\bibinfo {title} {Gaussian measures of
  entanglement versus negativities: Ordering of two-mode gaussian states},\
  }\href {https://doi.org/10.1103/PhysRevA.72.032334} {\bibfield  {journal}
  {\bibinfo  {journal} {Phys. Rev. A}\ }\textbf {\bibinfo {volume} {72}},\
  \bibinfo {pages} {032334} (\bibinfo {year} {2005})},\ \Eprint
  {https://arxiv.org/abs/quant-ph/0506124} {arXiv:quant-ph/0506124}
  \BibitemShut {NoStop}%
\bibitem [{\citenamefont {Braunstein}\ and\ \citenamefont {van
  Loock}(2005)}]{BraunsteinGaussiareview}%
  \BibitemOpen
  \bibfield  {author} {\bibinfo {author} {\bibfnamefont {S.~L.}\ \bibnamefont
  {Braunstein}}\ and\ \bibinfo {author} {\bibfnamefont {P.}~\bibnamefont {van
  Loock}},\ }\bibfield  {title} {\bibinfo {title} {Quantum information with
  continuous variables},\ }\href {https://doi.org/10.1103/RevModPhys.77.513}
  {\bibfield  {journal} {\bibinfo  {journal} {Rev. Mod. Phys.}\ }\textbf
  {\bibinfo {volume} {77}},\ \bibinfo {pages} {513} (\bibinfo {year} {2005})},\
  \Eprint {https://arxiv.org/abs/quant-ph/0410100} {arXiv:quant-ph/0410100}
  \BibitemShut {NoStop}%
\bibitem [{\citenamefont {Horodecki}\ \emph {et~al.}(2009)\citenamefont
  {Horodecki}, \citenamefont {Horodecki}, \citenamefont {Horodecki},\ and\
  \citenamefont {Horodecki}}]{horodecki2009quantum}%
  \BibitemOpen
  \bibfield  {author} {\bibinfo {author} {\bibfnamefont {R.}~\bibnamefont
  {Horodecki}}, \bibinfo {author} {\bibfnamefont {P.}~\bibnamefont
  {Horodecki}}, \bibinfo {author} {\bibfnamefont {M.}~\bibnamefont
  {Horodecki}},\ and\ \bibinfo {author} {\bibfnamefont {K.}~\bibnamefont
  {Horodecki}},\ }\bibfield  {title} {\bibinfo {title} {Quantum entanglement},\
  }\href {https://doi.org/10.1103/RevModPhys.81.865} {\bibfield  {journal}
  {\bibinfo  {journal} {Rev. Mod. Phys.}\ }\textbf {\bibinfo {volume} {81}},\
  \bibinfo {pages} {865} (\bibinfo {year} {2009})},\ \Eprint
  {https://arxiv.org/abs/quant-ph/0702225} {arXiv:quant-ph/0702225}
  \BibitemShut {NoStop}%
\bibitem [{\citenamefont {Weedbrook}\ \emph {et~al.}(2012)\citenamefont
  {Weedbrook}, \citenamefont {Pirandola}, \citenamefont {Garc\'{\i}a-Patr\'on},
  \citenamefont {Cerf}, \citenamefont {Ralph}, \citenamefont {Shapiro},\ and\
  \citenamefont {Lloyd}}]{WeedbrookGaussiareview}%
  \BibitemOpen
  \bibfield  {author} {\bibinfo {author} {\bibfnamefont {C.}~\bibnamefont
  {Weedbrook}}, \bibinfo {author} {\bibfnamefont {S.}~\bibnamefont
  {Pirandola}}, \bibinfo {author} {\bibfnamefont {R.}~\bibnamefont
  {Garc\'{\i}a-Patr\'on}}, \bibinfo {author} {\bibfnamefont {N.~J.}\
  \bibnamefont {Cerf}}, \bibinfo {author} {\bibfnamefont {T.~C.}\ \bibnamefont
  {Ralph}}, \bibinfo {author} {\bibfnamefont {J.~H.}\ \bibnamefont {Shapiro}},\
  and\ \bibinfo {author} {\bibfnamefont {S.}~\bibnamefont {Lloyd}},\ }\bibfield
   {title} {\bibinfo {title} {Gaussian quantum information},\ }\href
  {https://doi.org/10.1103/RevModPhys.84.621} {\bibfield  {journal} {\bibinfo
  {journal} {Rev. Mod. Phys.}\ }\textbf {\bibinfo {volume} {84}},\ \bibinfo
  {pages} {621} (\bibinfo {year} {2012})},\ \Eprint
  {https://arxiv.org/abs/1110.3234} {arXiv:1110.3234 [quant-ph]} \BibitemShut
  {NoStop}%
\bibitem [{\citenamefont {Serafini}(2017)}]{serafini2017quantum}%
  \BibitemOpen
  \bibfield  {author} {\bibinfo {author} {\bibfnamefont {A.}~\bibnamefont
  {Serafini}},\ }\href {https://doi.org/10.1201/9781315118727} {\emph {\bibinfo
  {title} {Quantum continuous variables: a primer of theoretical methods}}}\
  (\bibinfo  {publisher} {CRC press},\ \bibinfo {year} {2017})\BibitemShut
  {NoStop}%
\bibitem [{\citenamefont {Julsgaard}\ \emph {et~al.}(2000)\citenamefont
  {Julsgaard}, \citenamefont {Kozhekin},\ and\ \citenamefont
  {Polzik}}]{gaussianexample2}%
  \BibitemOpen
  \bibfield  {author} {\bibinfo {author} {\bibfnamefont {B.}~\bibnamefont
  {Julsgaard}}, \bibinfo {author} {\bibfnamefont {A.}~\bibnamefont
  {Kozhekin}},\ and\ \bibinfo {author} {\bibfnamefont {E.~S.}\ \bibnamefont
  {Polzik}},\ }\bibfield  {title} {\bibinfo {title} {Experimental long-lived
  entanglement of two macroscopic objects},\ }\href
  {https://doi.org/10.1038/35096524} {\bibfield  {journal} {\bibinfo  {journal}
  {Nature}\ }\textbf {\bibinfo {volume} {413}},\ \bibinfo {pages} {400}
  (\bibinfo {year} {2000})},\ \Eprint {https://arxiv.org/abs/quant-ph/0106057}
  {arXiv:quant-ph/0106057} \BibitemShut {NoStop}%
\bibitem [{\citenamefont {Blais}\ \emph {et~al.}(2004)\citenamefont {Blais},
  \citenamefont {Huang}, \citenamefont {Wallraff}, \citenamefont {Girvin},\
  and\ \citenamefont {Schoelkopf}}]{cavityqed}%
  \BibitemOpen
  \bibfield  {author} {\bibinfo {author} {\bibfnamefont {A.}~\bibnamefont
  {Blais}}, \bibinfo {author} {\bibfnamefont {R.-S.}\ \bibnamefont {Huang}},
  \bibinfo {author} {\bibfnamefont {A.}~\bibnamefont {Wallraff}}, \bibinfo
  {author} {\bibfnamefont {S.~M.}\ \bibnamefont {Girvin}},\ and\ \bibinfo
  {author} {\bibfnamefont {R.~J.}\ \bibnamefont {Schoelkopf}},\ }\bibfield
  {title} {\bibinfo {title} {Cavity quantum electrodynamics for superconducting
  electrical circuits: An architecture for quantum computation},\ }\href
  {https://doi.org/10.1103/PhysRevA.69.062320} {\bibfield  {journal} {\bibinfo
  {journal} {Phys. Rev. A}\ }\textbf {\bibinfo {volume} {69}},\ \bibinfo
  {pages} {062320} (\bibinfo {year} {2004})},\ \Eprint
  {https://arxiv.org/abs/cond-mat/0402216} {arXiv:cond-mat/0402216}
  \BibitemShut {NoStop}%
\bibitem [{\citenamefont {Cerf}\ \emph {et~al.}(2007)\citenamefont {Cerf},
  \citenamefont {Leuchs},\ and\ \citenamefont {Polzik}}]{Cerf2007CVQIAL}%
  \BibitemOpen
  \bibfield  {author} {\bibinfo {author} {\bibfnamefont {N.~J.}\ \bibnamefont
  {Cerf}}, \bibinfo {author} {\bibfnamefont {G.}~\bibnamefont {Leuchs}},\ and\
  \bibinfo {author} {\bibfnamefont {E.~S.}\ \bibnamefont {Polzik}},\ }\href
  {https://doi.org/10.1142/p489} {\emph {\bibinfo {title} {Quantum Information
  with Continuous Variables of Atoms and Light}}}\ (\bibinfo  {publisher}
  {Published by Imperial College Press and Distributed by World Scientific
  Publishing Co.},\ \bibinfo {year} {2007})\BibitemShut {NoStop}%
\bibitem [{\citenamefont {{Chen}}\ \emph {et~al.}(2014)\citenamefont {{Chen}},
  \citenamefont {{Menicucci}},\ and\ \citenamefont
  {{Pfister}}}]{Moran2014comb}%
  \BibitemOpen
  \bibfield  {author} {\bibinfo {author} {\bibfnamefont {M.}~\bibnamefont
  {{Chen}}}, \bibinfo {author} {\bibfnamefont {N.~C.}\ \bibnamefont
  {{Menicucci}}},\ and\ \bibinfo {author} {\bibfnamefont {O.}~\bibnamefont
  {{Pfister}}},\ }\bibfield  {title} {\bibinfo {title} {{Experimental
  Realization of Multipartite Entanglement of 60 Modes of a Quantum Optical
  Frequency Comb}},\ }\href {https://doi.org/10.1103/PhysRevLett.112.120505}
  {\bibfield  {journal} {\bibinfo  {journal} {Phys. Rev. Lett.}\ }\textbf
  {\bibinfo {volume} {112}},\ \bibinfo {eid} {120505} (\bibinfo {year}
  {2014})},\ \Eprint {https://arxiv.org/abs/1311.2957} {arXiv:1311.2957
  [quant-ph]} \BibitemShut {NoStop}%
\bibitem [{\citenamefont {Wittemer}\ \emph {et~al.}(2019)\citenamefont
  {Wittemer}, \citenamefont {Hakelberg}, \citenamefont {Kiefer}, \citenamefont
  {Schr\"oder}, \citenamefont {Fey}, \citenamefont {Sch\"utzhold},
  \citenamefont {Warring},\ and\ \citenamefont {Schaetz}}]{gaussianphonon}%
  \BibitemOpen
  \bibfield  {author} {\bibinfo {author} {\bibfnamefont {M.}~\bibnamefont
  {Wittemer}}, \bibinfo {author} {\bibfnamefont {F.}~\bibnamefont {Hakelberg}},
  \bibinfo {author} {\bibfnamefont {P.}~\bibnamefont {Kiefer}}, \bibinfo
  {author} {\bibfnamefont {J.-P.}\ \bibnamefont {Schr\"oder}}, \bibinfo
  {author} {\bibfnamefont {C.}~\bibnamefont {Fey}}, \bibinfo {author}
  {\bibfnamefont {R.}~\bibnamefont {Sch\"utzhold}}, \bibinfo {author}
  {\bibfnamefont {U.}~\bibnamefont {Warring}},\ and\ \bibinfo {author}
  {\bibfnamefont {T.}~\bibnamefont {Schaetz}},\ }\bibfield  {title} {\bibinfo
  {title} {{Phonon Pair Creation by Inflating Quantum Fluctuations in an Ion
  Trap}},\ }\href {https://doi.org/10.1103/PhysRevLett.123.180502} {\bibfield
  {journal} {\bibinfo  {journal} {Phys. Rev. Lett.}\ }\textbf {\bibinfo
  {volume} {123}},\ \bibinfo {pages} {180502} (\bibinfo {year} {2019})},\
  \Eprint {https://arxiv.org/abs/1903.05523} {arXiv:1903.05523 [quant-ph]}
  \BibitemShut {NoStop}%
\bibitem [{\citenamefont {{Larsen}}\ \emph {et~al.}(2019)\citenamefont
  {{Larsen}}, \citenamefont {{Guo}}, \citenamefont {{Breum}}, \citenamefont
  {{Neergaard-Nielsen}},\ and\ \citenamefont
  {{Andersen}}}]{Larsen2019cluster2d}%
  \BibitemOpen
  \bibfield  {author} {\bibinfo {author} {\bibfnamefont {M.~V.}\ \bibnamefont
  {{Larsen}}}, \bibinfo {author} {\bibfnamefont {X.}~\bibnamefont {{Guo}}},
  \bibinfo {author} {\bibfnamefont {C.~R.}\ \bibnamefont {{Breum}}}, \bibinfo
  {author} {\bibfnamefont {J.~S.}\ \bibnamefont {{Neergaard-Nielsen}}},\ and\
  \bibinfo {author} {\bibfnamefont {U.~L.}\ \bibnamefont {{Andersen}}},\
  }\bibfield  {title} {\bibinfo {title} {{Deterministic generation of a
  two-dimensional cluster state}},\ }\href
  {https://doi.org/10.1126/science.aay4354} {\bibfield  {journal} {\bibinfo
  {journal} {Science}\ }\textbf {\bibinfo {volume} {366}},\ \bibinfo {pages}
  {369} (\bibinfo {year} {2019})},\ \Eprint {https://arxiv.org/abs/1906.08709}
  {arXiv:1906.08709 [quant-ph]} \BibitemShut {NoStop}%
\bibitem [{\citenamefont {Roy}\ \emph {et~al.}(2021)\citenamefont {Roy},
  \citenamefont {Schuricht}, \citenamefont {Hauschild}, \citenamefont
  {Pollmann},\ and\ \citenamefont {Saleur}}]{royqs}%
  \BibitemOpen
  \bibfield  {author} {\bibinfo {author} {\bibfnamefont {A.}~\bibnamefont
  {Roy}}, \bibinfo {author} {\bibfnamefont {D.}~\bibnamefont {Schuricht}},
  \bibinfo {author} {\bibfnamefont {J.}~\bibnamefont {Hauschild}}, \bibinfo
  {author} {\bibfnamefont {F.}~\bibnamefont {Pollmann}},\ and\ \bibinfo
  {author} {\bibfnamefont {H.}~\bibnamefont {Saleur}},\ }\bibfield  {title}
  {\bibinfo {title} {The quantum sine-gordon model with quantum circuits},\
  }\href {https://doi.org/https://doi.org/10.1016/j.nuclphysb.2021.115445}
  {\bibfield  {journal} {\bibinfo  {journal} {Nuclear Physics B}\ }\textbf
  {\bibinfo {volume} {968}},\ \bibinfo {pages} {115445} (\bibinfo {year}
  {2021})},\ \Eprint {https://arxiv.org/abs/2007.06874} {arXiv:2007.06874
  [quant-ph]} \BibitemShut {NoStop}%
\bibitem [{\citenamefont {Whitlow}\ \emph {et~al.}(2023)\citenamefont
  {Whitlow}, \citenamefont {Jia}, \citenamefont {Wang}, \citenamefont {Fang},
  \citenamefont {Kim},\ and\ \citenamefont {Brown}}]{kennethbrownqs}%
  \BibitemOpen
  \bibfield  {author} {\bibinfo {author} {\bibfnamefont {J.}~\bibnamefont
  {Whitlow}}, \bibinfo {author} {\bibfnamefont {Z.}~\bibnamefont {Jia}},
  \bibinfo {author} {\bibfnamefont {Y.}~\bibnamefont {Wang}}, \bibinfo {author}
  {\bibfnamefont {C.}~\bibnamefont {Fang}}, \bibinfo {author} {\bibfnamefont
  {J.}~\bibnamefont {Kim}},\ and\ \bibinfo {author} {\bibfnamefont {K.~R.}\
  \bibnamefont {Brown}},\ }\bibfield  {title} {\bibinfo {title} {Quantum
  simulation of conical intersections using trapped ions},\ }\href
  {https://doi.org/https://doi.org/10.1038/s41557-023-01303-0} {\bibfield
  {journal} {\bibinfo  {journal} {Nature Chemistry}\ }\textbf {\bibinfo
  {volume} {15}},\ \bibinfo {pages} {1509} (\bibinfo {year} {2023})},\ \Eprint
  {https://arxiv.org/abs/2211.07319} {arXiv:2211.07319 [quant-ph]} \BibitemShut
  {NoStop}%
\bibitem [{\citenamefont {Wolf}\ \emph {et~al.}(2006)\citenamefont {Wolf},
  \citenamefont {Giedke},\ and\ \citenamefont {Cirac}}]{wolfextremality}%
  \BibitemOpen
  \bibfield  {author} {\bibinfo {author} {\bibfnamefont {M.~M.}\ \bibnamefont
  {Wolf}}, \bibinfo {author} {\bibfnamefont {G.}~\bibnamefont {Giedke}},\ and\
  \bibinfo {author} {\bibfnamefont {J.~I.}\ \bibnamefont {Cirac}},\ }\bibfield
  {title} {\bibinfo {title} {Extremality of gaussian quantum states},\ }\href
  {https://doi.org/10.1103/PhysRevLett.96.080502} {\bibfield  {journal}
  {\bibinfo  {journal} {Phys. Rev. Lett.}\ }\textbf {\bibinfo {volume} {96}},\
  \bibinfo {pages} {080502} (\bibinfo {year} {2006})},\ \Eprint
  {https://arxiv.org/abs/quant-ph/0509154} {arXiv:quant-ph/0509154}
  \BibitemShut {NoStop}%
\bibitem [{\citenamefont {Giedke}\ \emph
  {et~al.}(2003{\natexlab{b}})\citenamefont {Giedke}, \citenamefont {Eisert},
  \citenamefont {Cirac},\ and\ \citenamefont {Plenio}}]{Giedkepurestatetrans}%
  \BibitemOpen
  \bibfield  {author} {\bibinfo {author} {\bibfnamefont {G.}~\bibnamefont
  {Giedke}}, \bibinfo {author} {\bibfnamefont {J.}~\bibnamefont {Eisert}},
  \bibinfo {author} {\bibfnamefont {J.~I.}\ \bibnamefont {Cirac}},\ and\
  \bibinfo {author} {\bibfnamefont {M.~B.}\ \bibnamefont {Plenio}},\ }\bibfield
   {title} {\bibinfo {title} {{Entanglement transformations of pure Gaussian
  states}},\ }\href {https://doi.org/10.26421/QIC3.3-3} {\bibfield  {journal}
  {\bibinfo  {journal} {Quant. Inf. Comput.}\ }\textbf {\bibinfo {volume}
  {3}},\ \bibinfo {pages} {211} (\bibinfo {year} {2003}{\natexlab{b}})},\
  \Eprint {https://arxiv.org/abs/quant-ph/0301038} {arXiv:quant-ph/0301038}
  \BibitemShut {NoStop}%
\bibitem [{\citenamefont {Wolf}(2008)}]{wolfnotsonormal}%
  \BibitemOpen
  \bibfield  {author} {\bibinfo {author} {\bibfnamefont {M.~M.}\ \bibnamefont
  {Wolf}},\ }\bibfield  {title} {\bibinfo {title} {Not-so-normal mode
  decomposition},\ }\href {https://doi.org/10.1103/PhysRevLett.100.070505}
  {\bibfield  {journal} {\bibinfo  {journal} {Phys. Rev. Lett.}\ }\textbf
  {\bibinfo {volume} {100}},\ \bibinfo {pages} {070505} (\bibinfo {year}
  {2008})},\ \Eprint {https://arxiv.org/abs/0707.0604} {arXiv:0707.0604
  [quant-ph]} \BibitemShut {NoStop}%
\bibitem [{\citenamefont {Klco}\ \emph {et~al.}(2023)\citenamefont {Klco},
  \citenamefont {Beck},\ and\ \citenamefont {Savage}}]{NKcorehalo}%
  \BibitemOpen
  \bibfield  {author} {\bibinfo {author} {\bibfnamefont {N.}~\bibnamefont
  {Klco}}, \bibinfo {author} {\bibfnamefont {D.~H.}\ \bibnamefont {Beck}},\
  and\ \bibinfo {author} {\bibfnamefont {M.~J.}\ \bibnamefont {Savage}},\
  }\bibfield  {title} {\bibinfo {title} {{Entanglement structures in quantum
  field theories: Negativity cores and bound entanglement in the vacuum}},\
  }\href {https://doi.org/10.1103/PhysRevA.107.012415} {\bibfield  {journal}
  {\bibinfo  {journal} {Phys. Rev. A}\ }\textbf {\bibinfo {volume} {107}},\
  \bibinfo {pages} {012415} (\bibinfo {year} {2023})},\ \Eprint
  {https://arxiv.org/abs/2110.10736} {arXiv:2110.10736 [quant-ph]} \BibitemShut
  {NoStop}%
\bibitem [{\citenamefont {Marcovitch}\ \emph {et~al.}(2009)\citenamefont
  {Marcovitch}, \citenamefont {Retzker}, \citenamefont {Plenio},\ and\
  \citenamefont {Reznik}}]{scalar1dextra}%
  \BibitemOpen
  \bibfield  {author} {\bibinfo {author} {\bibfnamefont {S.}~\bibnamefont
  {Marcovitch}}, \bibinfo {author} {\bibfnamefont {A.}~\bibnamefont {Retzker}},
  \bibinfo {author} {\bibfnamefont {M.~B.}\ \bibnamefont {Plenio}},\ and\
  \bibinfo {author} {\bibfnamefont {B.}~\bibnamefont {Reznik}},\ }\bibfield
  {title} {\bibinfo {title} {{Critical and noncritical long-range entanglement
  in Klein-Gordon fields}},\ }\href
  {https://doi.org/10.1103/PhysRevA.80.012325} {\bibfield  {journal} {\bibinfo
  {journal} {Phys. Rev. A}\ }\textbf {\bibinfo {volume} {80}},\ \bibinfo
  {pages} {012325} (\bibinfo {year} {2009})},\ \Eprint
  {https://arxiv.org/abs/0811.1288} {arXiv:0811.1288 [quant-ph]} \BibitemShut
  {NoStop}%
\bibitem [{\citenamefont {Klco}\ and\ \citenamefont
  {Savage}(2021{\natexlab{a}})}]{NKnegativitysphere}%
  \BibitemOpen
  \bibfield  {author} {\bibinfo {author} {\bibfnamefont {N.}~\bibnamefont
  {Klco}}\ and\ \bibinfo {author} {\bibfnamefont {M.~J.}\ \bibnamefont
  {Savage}},\ }\bibfield  {title} {\bibinfo {title} {{Geometric quantum
  information structure in quantum fields and their lattice simulation}},\
  }\href {https://doi.org/10.1103/PhysRevD.103.065007} {\bibfield  {journal}
  {\bibinfo  {journal} {Phys. Rev. D}\ }\textbf {\bibinfo {volume} {103}},\
  \bibinfo {pages} {065007} (\bibinfo {year} {2021}{\natexlab{a}})},\ \Eprint
  {https://arxiv.org/abs/2008.03647} {arXiv:2008.03647 [quant-ph]} \BibitemShut
  {NoStop}%
\bibitem [{\citenamefont {Klco}\ and\ \citenamefont
  {Savage}(2021{\natexlab{b}})}]{NKentsphere}%
  \BibitemOpen
  \bibfield  {author} {\bibinfo {author} {\bibfnamefont {N.}~\bibnamefont
  {Klco}}\ and\ \bibinfo {author} {\bibfnamefont {M.~J.}\ \bibnamefont
  {Savage}},\ }\bibfield  {title} {\bibinfo {title} {{Entanglement Spheres and
  a UV-IR Connection in Effective Field Theories}},\ }\href
  {https://doi.org/10.1103/PhysRevLett.127.211602} {\bibfield  {journal}
  {\bibinfo  {journal} {Phys. Rev. Lett.}\ }\textbf {\bibinfo {volume} {127}},\
  \bibinfo {pages} {211602} (\bibinfo {year} {2021}{\natexlab{b}})},\ \Eprint
  {https://arxiv.org/abs/2103.14999} {arXiv:2103.14999 [hep-th]} \BibitemShut
  {NoStop}%
\bibitem [{\citenamefont {{Retzker}}\ \emph {et~al.}(2005)\citenamefont
  {{Retzker}}, \citenamefont {{Cirac}},\ and\ \citenamefont
  {{Reznik}}}]{Retzker:2005dve}%
  \BibitemOpen
  \bibfield  {author} {\bibinfo {author} {\bibfnamefont {A.}~\bibnamefont
  {{Retzker}}}, \bibinfo {author} {\bibfnamefont {J.~I.}\ \bibnamefont
  {{Cirac}}},\ and\ \bibinfo {author} {\bibfnamefont {B.}~\bibnamefont
  {{Reznik}}},\ }\bibfield  {title} {\bibinfo {title} {{Detecting Vacuum
  Entanglement in a Linear Ion Trap}},\ }\href
  {https://doi.org/10.1103/PhysRevLett.94.050504} {\bibfield  {journal}
  {\bibinfo  {journal} {Phys. Rev. Lett.}\ }\textbf {\bibinfo {volume} {94}},\
  \bibinfo {eid} {050504} (\bibinfo {year} {2005})},\ \Eprint
  {https://arxiv.org/abs/quant-ph/0408059} {arXiv:quant-ph/0408059 [quant-ph]}
  \BibitemShut {NoStop}%
\bibitem [{\citenamefont {Klco}\ and\ \citenamefont
  {Beck}(2024)}]{NKphononatural}%
  \BibitemOpen
  \bibfield  {author} {\bibinfo {author} {\bibfnamefont {N.}~\bibnamefont
  {Klco}}\ and\ \bibinfo {author} {\bibfnamefont {D.~H.}\ \bibnamefont
  {Beck}},\ }\bibfield  {title} {\bibinfo {title} {Identification of a natural
  fieldlike entanglement resource in trapped-ion chains},\ }\href
  {https://doi.org/10.1103/PhysRevA.109.062419} {\bibfield  {journal} {\bibinfo
   {journal} {Phys. Rev. A}\ }\textbf {\bibinfo {volume} {109}},\ \bibinfo
  {pages} {062419} (\bibinfo {year} {2024})},\ \Eprint
  {https://arxiv.org/abs/2311.08842} {arXiv:2311.08842 [quant-ph]} \BibitemShut
  {NoStop}%
\bibitem [{\citenamefont {Giedke}\ \emph {et~al.}(2000)\citenamefont {Giedke},
  \citenamefont {Duan}, \citenamefont {Cirac},\ and\ \citenamefont
  {Zoller}}]{giedke2000inseparable}%
  \BibitemOpen
  \bibfield  {author} {\bibinfo {author} {\bibfnamefont {G.}~\bibnamefont
  {Giedke}}, \bibinfo {author} {\bibfnamefont {L.-M.}\ \bibnamefont {Duan}},
  \bibinfo {author} {\bibfnamefont {J.~I.}\ \bibnamefont {Cirac}},\ and\
  \bibinfo {author} {\bibfnamefont {P.}~\bibnamefont {Zoller}},\ }\href@noop {}
  {\bibinfo {title} {All inseparable two-mode gaussian continuous variable
  states are distillable}} (\bibinfo {year} {2000}),\ \Eprint
  {https://arxiv.org/abs/quant-ph/0007061} {arXiv:quant-ph/0007061}
  \BibitemShut {NoStop}%
\bibitem [{\citenamefont {Giedke}\ \emph
  {et~al.}(2001{\natexlab{a}})\citenamefont {Giedke}, \citenamefont {Duan},
  \citenamefont {Cirac},\ and\ \citenamefont {Zoller}}]{giedkedistill}%
  \BibitemOpen
  \bibfield  {author} {\bibinfo {author} {\bibfnamefont {G.}~\bibnamefont
  {Giedke}}, \bibinfo {author} {\bibfnamefont {L.-M.}\ \bibnamefont {Duan}},
  \bibinfo {author} {\bibfnamefont {J.~I.}\ \bibnamefont {Cirac}},\ and\
  \bibinfo {author} {\bibfnamefont {P.}~\bibnamefont {Zoller}},\ }\bibfield
  {title} {\bibinfo {title} {{Distillability criterion for all bipartite
  Gaussian states}},\ }\href {https://doi.org/10.26421/QIC1.3-7} {\bibfield
  {journal} {\bibinfo  {journal} {Quant. Inf. Comput.}\ }\textbf {\bibinfo
  {volume} {1}},\ \bibinfo {pages} {79} (\bibinfo {year}
  {2001}{\natexlab{a}})},\ \Eprint {https://arxiv.org/abs/quant-ph/0104072}
  {arXiv:quant-ph/0104072} \BibitemShut {NoStop}%
\bibitem [{\citenamefont {Williamson}(1936)}]{williamsonnormalform}%
  \BibitemOpen
  \bibfield  {author} {\bibinfo {author} {\bibfnamefont {J.}~\bibnamefont
  {Williamson}},\ }\bibfield  {title} {\bibinfo {title} {On the algebraic
  problem concerning the normal forms of linear dynamical systems},\ }\href
  {http://www.jstor.org/stable/2371062} {\bibfield  {journal} {\bibinfo
  {journal} {American Journal of Mathematics}\ }\textbf {\bibinfo {volume}
  {58}},\ \bibinfo {pages} {141} (\bibinfo {year} {1936})}\BibitemShut
  {NoStop}%
\bibitem [{\citenamefont {Klco}\ and\ \citenamefont
  {Beck}(2023)}]{NKvolumemeasure}%
  \BibitemOpen
  \bibfield  {author} {\bibinfo {author} {\bibfnamefont {N.}~\bibnamefont
  {Klco}}\ and\ \bibinfo {author} {\bibfnamefont {D.~H.}\ \bibnamefont
  {Beck}},\ }\bibfield  {title} {\bibinfo {title} {{Entanglement structures in
  quantum field theories. II. Distortions of vacuum correlations through the
  lens of local observers}},\ }\href
  {https://doi.org/10.1103/PhysRevA.108.012429} {\bibfield  {journal} {\bibinfo
   {journal} {Phys. Rev. A}\ }\textbf {\bibinfo {volume} {108}},\ \bibinfo
  {pages} {012429} (\bibinfo {year} {2023})},\ \Eprint
  {https://arxiv.org/abs/2304.04143} {arXiv:2304.04143 [quant-ph]} \BibitemShut
  {NoStop}%
\bibitem [{\citenamefont {Giedke}\ \emph
  {et~al.}(2001{\natexlab{b}})\citenamefont {Giedke}, \citenamefont
  {Lewenstein}, \citenamefont {Cirac},\ and\ \citenamefont
  {Zoller}}]{Giedkesepflow}%
  \BibitemOpen
  \bibfield  {author} {\bibinfo {author} {\bibfnamefont {G.}~\bibnamefont
  {Giedke}}, \bibinfo {author} {\bibfnamefont {M.}~\bibnamefont {Lewenstein}},
  \bibinfo {author} {\bibfnamefont {J.~I.}\ \bibnamefont {Cirac}},\ and\
  \bibinfo {author} {\bibfnamefont {P.}~\bibnamefont {Zoller}},\ }\bibfield
  {title} {\bibinfo {title} {{Entanglement Criteria for All Bipartite Gaussian
  States}},\ }\href {https://doi.org/10.1103/PhysRevLett.87.167904} {\bibfield
  {journal} {\bibinfo  {journal} {Phys. Rev. Lett.}\ }\textbf {\bibinfo
  {volume} {87}},\ \bibinfo {pages} {167904} (\bibinfo {year}
  {2001}{\natexlab{b}})}\BibitemShut {NoStop}%
\bibitem [{\citenamefont {Boyd}\ and\ \citenamefont
  {Vandenberghe}(2004)}]{boyd2004convexschurcomp}%
  \BibitemOpen
  \bibfield  {author} {\bibinfo {author} {\bibfnamefont {S.}~\bibnamefont
  {Boyd}}\ and\ \bibinfo {author} {\bibfnamefont {L.}~\bibnamefont
  {Vandenberghe}},\ }\href
  {https://www.cambridge.org/highereducation/books/convex-optimization/17D2FAA54F641A2F62C7CCD01DFA97C4#overview}
  {\emph {\bibinfo {title} {Convex Optimization}}}\ (\bibinfo  {publisher}
  {Cambridge University Press},\ \bibinfo {year} {2004})\BibitemShut {NoStop}%
\bibitem [{\citenamefont {Horn}\ and\ \citenamefont
  {Johnson}(2012)}]{hornmatrixana}%
  \BibitemOpen
  \bibfield  {author} {\bibinfo {author} {\bibfnamefont {R.}~\bibnamefont
  {Horn}}\ and\ \bibinfo {author} {\bibfnamefont {C.}~\bibnamefont {Johnson}},\
  }\href
  {https://www.cambridge.org/us/universitypress/subjects/mathematics/algebra/matrix-analysis-2nd-edition?format=PB&isbn=9780521548236}
  {\emph {\bibinfo {title} {Matrix Analysis}}}\ (\bibinfo  {publisher}
  {Cambridge university press},\ \bibinfo {year} {2012})\BibitemShut {NoStop}%
\bibitem [{\citenamefont {Adesso}\ \emph {et~al.}(2012)\citenamefont {Adesso},
  \citenamefont {Girolami},\ and\ \citenamefont {Serafini}}]{adessoGR2}%
  \BibitemOpen
  \bibfield  {author} {\bibinfo {author} {\bibfnamefont {G.}~\bibnamefont
  {Adesso}}, \bibinfo {author} {\bibfnamefont {D.}~\bibnamefont {Girolami}},\
  and\ \bibinfo {author} {\bibfnamefont {A.}~\bibnamefont {Serafini}},\
  }\bibfield  {title} {\bibinfo {title} {Measuring gaussian quantum information
  and correlations using the r\'enyi entropy of order 2},\ }\href
  {https://doi.org/10.1103/PhysRevLett.109.190502} {\bibfield  {journal}
  {\bibinfo  {journal} {Phys. Rev. Lett.}\ }\textbf {\bibinfo {volume} {109}},\
  \bibinfo {pages} {190502} (\bibinfo {year} {2012})},\ \Eprint
  {https://arxiv.org/abs/1203.5116} {arXiv:1203.5116 [quant-ph]} \BibitemShut
  {NoStop}%
\bibitem [{\citenamefont {Duan}\ \emph
  {et~al.}(2000{\natexlab{b}})\citenamefont {Duan}, \citenamefont {Giedke},
  \citenamefont {Cirac},\ and\ \citenamefont {Zoller}}]{gaussianpurification}%
  \BibitemOpen
  \bibfield  {author} {\bibinfo {author} {\bibfnamefont {L.-M.}\ \bibnamefont
  {Duan}}, \bibinfo {author} {\bibfnamefont {G.}~\bibnamefont {Giedke}},
  \bibinfo {author} {\bibfnamefont {J.~I.}\ \bibnamefont {Cirac}},\ and\
  \bibinfo {author} {\bibfnamefont {P.}~\bibnamefont {Zoller}},\ }\bibfield
  {title} {\bibinfo {title} {Entanglement purification of gaussian continuous
  variable quantum states},\ }\href
  {https://doi.org/10.1103/PhysRevLett.84.4002} {\bibfield  {journal} {\bibinfo
   {journal} {Phys. Rev. Lett.}\ }\textbf {\bibinfo {volume} {84}},\ \bibinfo
  {pages} {4002} (\bibinfo {year} {2000}{\natexlab{b}})},\ \Eprint
  {https://arxiv.org/abs/quant-ph/9912017} {arXiv:quant-ph/9912017}
  \BibitemShut {NoStop}%
\bibitem [{\citenamefont {Audenaert}\ \emph {et~al.}(2002)\citenamefont
  {Audenaert}, \citenamefont {Eisert}, \citenamefont {Plenio},\ and\
  \citenamefont {Werner}}]{vaneg2}%
  \BibitemOpen
  \bibfield  {author} {\bibinfo {author} {\bibfnamefont {K.}~\bibnamefont
  {Audenaert}}, \bibinfo {author} {\bibfnamefont {J.}~\bibnamefont {Eisert}},
  \bibinfo {author} {\bibfnamefont {M.~B.}\ \bibnamefont {Plenio}},\ and\
  \bibinfo {author} {\bibfnamefont {R.~F.}\ \bibnamefont {Werner}},\ }\bibfield
   {title} {\bibinfo {title} {Entanglement properties of the harmonic chain},\
  }\href {https://doi.org/10.1103/PhysRevA.66.042327} {\bibfield  {journal}
  {\bibinfo  {journal} {Phys. Rev. A}\ }\textbf {\bibinfo {volume} {66}},\
  \bibinfo {pages} {042327} (\bibinfo {year} {2002})},\ \Eprint
  {https://arxiv.org/abs/quant-ph/0205025} {arXiv:quant-ph/0205025}
  \BibitemShut {NoStop}%
\bibitem [{\citenamefont {Botero}\ and\ \citenamefont
  {Reznik}(2004)}]{scalarvacuumoriginal1}%
  \BibitemOpen
  \bibfield  {author} {\bibinfo {author} {\bibfnamefont {A.}~\bibnamefont
  {Botero}}\ and\ \bibinfo {author} {\bibfnamefont {B.}~\bibnamefont
  {Reznik}},\ }\bibfield  {title} {\bibinfo {title} {Spatial structures and
  localization of vacuum entanglement in the linear harmonic chain},\ }\href
  {https://doi.org/10.1103/PhysRevA.70.052329} {\bibfield  {journal} {\bibinfo
  {journal} {Phys. Rev. A}\ }\textbf {\bibinfo {volume} {70}},\ \bibinfo
  {pages} {052329} (\bibinfo {year} {2004})},\ \Eprint
  {https://arxiv.org/abs/quant-ph/0403233} {arXiv:quant-ph/0403233}
  \BibitemShut {NoStop}%
\bibitem [{\citenamefont {Kofler}\ \emph {et~al.}(2006)\citenamefont {Kofler},
  \citenamefont {Vedral}, \citenamefont {Kim},\ and\ \citenamefont
  {Brukner}}]{vaneg3}%
  \BibitemOpen
  \bibfield  {author} {\bibinfo {author} {\bibfnamefont {J.}~\bibnamefont
  {Kofler}}, \bibinfo {author} {\bibfnamefont {V.}~\bibnamefont {Vedral}},
  \bibinfo {author} {\bibfnamefont {M.~S.}\ \bibnamefont {Kim}},\ and\ \bibinfo
  {author} {\bibfnamefont {i.~c.~v.}\ \bibnamefont {Brukner}},\ }\bibfield
  {title} {\bibinfo {title} {Entanglement between collective operators in a
  linear harmonic chain},\ }\href {https://doi.org/10.1103/PhysRevA.73.052107}
  {\bibfield  {journal} {\bibinfo  {journal} {Phys. Rev. A}\ }\textbf {\bibinfo
  {volume} {73}},\ \bibinfo {pages} {052107} (\bibinfo {year} {2006})},\
  \Eprint {https://arxiv.org/abs/quant-ph/0506236} {arXiv:quant-ph/0506236}
  \BibitemShut {NoStop}%
\bibitem [{\citenamefont {Calabrese}\ \emph {et~al.}(2009)\citenamefont
  {Calabrese}, \citenamefont {Cardy},\ and\ \citenamefont {Tonni}}]{vaneg4}%
  \BibitemOpen
  \bibfield  {author} {\bibinfo {author} {\bibfnamefont {P.}~\bibnamefont
  {Calabrese}}, \bibinfo {author} {\bibfnamefont {J.}~\bibnamefont {Cardy}},\
  and\ \bibinfo {author} {\bibfnamefont {E.}~\bibnamefont {Tonni}},\ }\bibfield
   {title} {\bibinfo {title} {Entanglement entropy of two disjoint intervals in
  conformal field theory},\ }\href
  {https://doi.org/10.1088/1742-5468/2009/11/P11001} {\bibfield  {journal}
  {\bibinfo  {journal} {Journal of Statistical Mechanics: Theory and
  Experiment}\ }\textbf {\bibinfo {volume} {2009}},\ \bibinfo {pages} {P11001}
  (\bibinfo {year} {2009})},\ \Eprint {https://arxiv.org/abs/0905.2069}
  {arXiv:0905.2069 [hep-th]} \BibitemShut {NoStop}%
\bibitem [{\citenamefont {Calabrese}\ \emph {et~al.}(2012)\citenamefont
  {Calabrese}, \citenamefont {Cardy},\ and\ \citenamefont {Tonni}}]{vaneg5}%
  \BibitemOpen
  \bibfield  {author} {\bibinfo {author} {\bibfnamefont {P.}~\bibnamefont
  {Calabrese}}, \bibinfo {author} {\bibfnamefont {J.}~\bibnamefont {Cardy}},\
  and\ \bibinfo {author} {\bibfnamefont {E.}~\bibnamefont {Tonni}},\ }\bibfield
   {title} {\bibinfo {title} {Entanglement negativity in quantum field
  theory},\ }\href {https://doi.org/10.1103/PhysRevLett.109.130502} {\bibfield
  {journal} {\bibinfo  {journal} {Phys. Rev. Lett.}\ }\textbf {\bibinfo
  {volume} {109}},\ \bibinfo {pages} {130502} (\bibinfo {year} {2012})},\
  \Eprint {https://arxiv.org/abs/1206.3092} {arXiv:1206.3092
  [cond-mat.stat-mech]} \BibitemShut {NoStop}%
\bibitem [{\citenamefont {Calabrese}\ \emph {et~al.}(2013)\citenamefont
  {Calabrese}, \citenamefont {Cardy},\ and\ \citenamefont {Tonni}}]{vaneg6}%
  \BibitemOpen
  \bibfield  {author} {\bibinfo {author} {\bibfnamefont {P.}~\bibnamefont
  {Calabrese}}, \bibinfo {author} {\bibfnamefont {J.}~\bibnamefont {Cardy}},\
  and\ \bibinfo {author} {\bibfnamefont {E.}~\bibnamefont {Tonni}},\ }\bibfield
   {title} {\bibinfo {title} {Entanglement negativity in extended systems: a
  field theoretical approach},\ }\href
  {https://doi.org/10.1088/1742-5468/2013/02/P02008} {\bibfield  {journal}
  {\bibinfo  {journal} {Journal of Statistical Mechanics: Theory and
  Experiment}\ }\textbf {\bibinfo {volume} {2013}},\ \bibinfo {pages} {P02008}
  (\bibinfo {year} {2013})},\ \Eprint {https://arxiv.org/abs/1210.5359}
  {arXiv:1210.5359 [cond-mat.stat-mech]} \BibitemShut {NoStop}%
\bibitem [{\citenamefont {Mohammadi~Mozaffar}\ and\ \citenamefont
  {Mollabashi}(2017)}]{vaneg7}%
  \BibitemOpen
  \bibfield  {author} {\bibinfo {author} {\bibfnamefont {M.~R.}\ \bibnamefont
  {Mohammadi~Mozaffar}}\ and\ \bibinfo {author} {\bibfnamefont
  {A.}~\bibnamefont {Mollabashi}},\ }\bibfield  {title} {\bibinfo {title}
  {Entanglement in lifshitz-type quantum field theories},\ }\href
  {https://doi.org/https://doi.org/10.1007/JHEP07(2017)120} {\bibfield
  {journal} {\bibinfo  {journal} {Journal of High Energy Physics}\ }\textbf
  {\bibinfo {volume} {2017}},\ \bibinfo {pages} {1} (\bibinfo {year} {2017})},\
  \Eprint {https://arxiv.org/abs/1705.00483} {arXiv:1705.00483 [hep-th]}
  \BibitemShut {NoStop}%
\bibitem [{\citenamefont {Klco}\ and\ \citenamefont
  {Savage}(2020{\natexlab{a}})}]{NKdesign}%
  \BibitemOpen
  \bibfield  {author} {\bibinfo {author} {\bibfnamefont {N.}~\bibnamefont
  {Klco}}\ and\ \bibinfo {author} {\bibfnamefont {M.~J.}\ \bibnamefont
  {Savage}},\ }\bibfield  {title} {\bibinfo {title} {{Systematically
  Localizable Operators for Quantum Simulations of Quantum Field Theories}},\
  }\href {https://doi.org/10.1103/PhysRevA.102.012619} {\bibfield  {journal}
  {\bibinfo  {journal} {Phys. Rev. A}\ }\textbf {\bibinfo {volume} {102}},\
  \bibinfo {pages} {012619} (\bibinfo {year} {2020}{\natexlab{a}})},\ \Eprint
  {https://arxiv.org/abs/1912.03577} {arXiv:1912.03577 [quant-ph]} \BibitemShut
  {NoStop}%
\bibitem [{\citenamefont {Klco}\ and\ \citenamefont
  {Savage}(2020{\natexlab{b}})}]{NKfixpoint}%
  \BibitemOpen
  \bibfield  {author} {\bibinfo {author} {\bibfnamefont {N.}~\bibnamefont
  {Klco}}\ and\ \bibinfo {author} {\bibfnamefont {M.~J.}\ \bibnamefont
  {Savage}},\ }\bibfield  {title} {\bibinfo {title} {Fixed-point quantum
  circuits for quantum field theories},\ }\href
  {https://doi.org/10.1103/PhysRevA.102.052422} {\bibfield  {journal} {\bibinfo
   {journal} {Phys. Rev. A}\ }\textbf {\bibinfo {volume} {102}},\ \bibinfo
  {pages} {052422} (\bibinfo {year} {2020}{\natexlab{b}})},\ \Eprint
  {https://arxiv.org/abs/2002.02018} {arXiv:2002.02018 [quant-ph]} \BibitemShut
  {NoStop}%
\bibitem [{\citenamefont {Ciavarella}\ \emph {et~al.}(2022)\citenamefont
  {Ciavarella}, \citenamefont {Klco},\ and\ \citenamefont
  {Savage}}]{simulationdesignalter}%
  \BibitemOpen
  \bibfield  {author} {\bibinfo {author} {\bibfnamefont {A.}~\bibnamefont
  {Ciavarella}}, \bibinfo {author} {\bibfnamefont {N.}~\bibnamefont {Klco}},\
  and\ \bibinfo {author} {\bibfnamefont {M.~J.}\ \bibnamefont {Savage}},\
  }\bibfield  {title} {\bibinfo {title} {{Some Conceptual Aspects of Operator
  Design for Quantum Simulations of Non-Abelian Lattice Gauge Theories}}\
  }(\bibinfo {year} {2022})\ \Eprint {https://arxiv.org/abs/2203.11988}
  {arXiv:2203.11988 [quant-ph]} \BibitemShut {NoStop}%
\bibitem [{\citenamefont {Cervera-Lierta}\ \emph {et~al.}(2017)\citenamefont
  {Cervera-Lierta}, \citenamefont {Latorre}, \citenamefont {Rojo},\ and\
  \citenamefont {Rottoli}}]{entanglementinnature1}%
  \BibitemOpen
  \bibfield  {author} {\bibinfo {author} {\bibfnamefont {A.}~\bibnamefont
  {Cervera-Lierta}}, \bibinfo {author} {\bibfnamefont {J.~I.}\ \bibnamefont
  {Latorre}}, \bibinfo {author} {\bibfnamefont {J.}~\bibnamefont {Rojo}},\ and\
  \bibinfo {author} {\bibfnamefont {L.}~\bibnamefont {Rottoli}},\ }\bibfield
  {title} {\bibinfo {title} {{Maximal Entanglement in High Energy Physics}},\
  }\href {https://doi.org/10.21468/SciPostPhys.3.5.036} {\bibfield  {journal}
  {\bibinfo  {journal} {SciPost Phys.}\ }\textbf {\bibinfo {volume} {3}},\
  \bibinfo {pages} {036} (\bibinfo {year} {2017})},\ \Eprint
  {https://arxiv.org/abs/1703.02989} {arXiv:1703.02989 [hep-th]} \BibitemShut
  {NoStop}%
\bibitem [{\citenamefont {Kharzeev}\ and\ \citenamefont
  {Levin}(2017)}]{entanglementinnature3}%
  \BibitemOpen
  \bibfield  {author} {\bibinfo {author} {\bibfnamefont {D.~E.}\ \bibnamefont
  {Kharzeev}}\ and\ \bibinfo {author} {\bibfnamefont {E.~M.}\ \bibnamefont
  {Levin}},\ }\bibfield  {title} {\bibinfo {title} {{Deep inelastic scattering
  as a probe of entanglement}},\ }\href
  {https://doi.org/10.1103/PhysRevD.95.114008} {\bibfield  {journal} {\bibinfo
  {journal} {Phys. Rev. D}\ }\textbf {\bibinfo {volume} {95}},\ \bibinfo
  {pages} {114008} (\bibinfo {year} {2017})},\ \Eprint
  {https://arxiv.org/abs/1702.03489} {arXiv:1702.03489 [hep-ph]} \BibitemShut
  {NoStop}%
\bibitem [{\citenamefont {Baker}\ and\ \citenamefont
  {Kharzeev}(2018)}]{entanglementinnature5}%
  \BibitemOpen
  \bibfield  {author} {\bibinfo {author} {\bibfnamefont {O.~K.}\ \bibnamefont
  {Baker}}\ and\ \bibinfo {author} {\bibfnamefont {D.~E.}\ \bibnamefont
  {Kharzeev}},\ }\bibfield  {title} {\bibinfo {title} {{Thermal radiation and
  entanglement in proton-proton collisions at energies available at the CERN
  Large Hadron Collider}},\ }\href {https://doi.org/10.1103/PhysRevD.98.054007}
  {\bibfield  {journal} {\bibinfo  {journal} {Phys. Rev. D}\ }\textbf {\bibinfo
  {volume} {98}},\ \bibinfo {pages} {054007} (\bibinfo {year} {2018})},\
  \Eprint {https://arxiv.org/abs/1712.04558} {arXiv:1712.04558 [hep-ph]}
  \BibitemShut {NoStop}%
\bibitem [{\citenamefont {Berges}\ \emph {et~al.}(2019)\citenamefont {Berges},
  \citenamefont {Floerchinger},\ and\ \citenamefont
  {Venugopalan}}]{entanglementinnature2}%
  \BibitemOpen
  \bibfield  {author} {\bibinfo {author} {\bibfnamefont {J.}~\bibnamefont
  {Berges}}, \bibinfo {author} {\bibfnamefont {S.}~\bibnamefont
  {Floerchinger}},\ and\ \bibinfo {author} {\bibfnamefont {R.}~\bibnamefont
  {Venugopalan}},\ }\bibfield  {title} {\bibinfo {title} {{Entanglement and
  thermalization}},\ }\href {https://doi.org/10.1016/j.nuclphysa.2018.12.008}
  {\bibfield  {journal} {\bibinfo  {journal} {Nucl. Phys. A}\ }\textbf
  {\bibinfo {volume} {982}},\ \bibinfo {pages} {819} (\bibinfo {year}
  {2019})},\ \Eprint {https://arxiv.org/abs/1812.08120} {arXiv:1812.08120
  [hep-th]} \BibitemShut {NoStop}%
\bibitem [{\citenamefont {Beane}\ and\ \citenamefont
  {Ehlers}(2019)}]{entanglementinnature6}%
  \BibitemOpen
  \bibfield  {author} {\bibinfo {author} {\bibfnamefont {S.~R.}\ \bibnamefont
  {Beane}}\ and\ \bibinfo {author} {\bibfnamefont {P.}~\bibnamefont {Ehlers}},\
  }\bibfield  {title} {\bibinfo {title} {{Chiral symmetry breaking,
  entanglement, and the nucleon spin decomposition}},\ }\href
  {https://doi.org/10.1142/S0217732320500480} {\bibfield  {journal} {\bibinfo
  {journal} {Mod. Phys. Lett. A}\ }\textbf {\bibinfo {volume} {35}},\ \bibinfo
  {pages} {2050048} (\bibinfo {year} {2019})},\ \Eprint
  {https://arxiv.org/abs/1905.03295} {arXiv:1905.03295 [hep-ph]} \BibitemShut
  {NoStop}%
\bibitem [{\citenamefont {Beane}\ \emph {et~al.}(2019)\citenamefont {Beane},
  \citenamefont {Kaplan}, \citenamefont {Klco},\ and\ \citenamefont
  {Savage}}]{entanglementinnature7}%
  \BibitemOpen
  \bibfield  {author} {\bibinfo {author} {\bibfnamefont {S.~R.}\ \bibnamefont
  {Beane}}, \bibinfo {author} {\bibfnamefont {D.~B.}\ \bibnamefont {Kaplan}},
  \bibinfo {author} {\bibfnamefont {N.}~\bibnamefont {Klco}},\ and\ \bibinfo
  {author} {\bibfnamefont {M.~J.}\ \bibnamefont {Savage}},\ }\bibfield  {title}
  {\bibinfo {title} {{Entanglement Suppression and Emergent Symmetries of
  Strong Interactions}},\ }\href
  {https://doi.org/10.1103/PhysRevLett.122.102001} {\bibfield  {journal}
  {\bibinfo  {journal} {Phys. Rev. Lett.}\ }\textbf {\bibinfo {volume} {122}},\
  \bibinfo {pages} {102001} (\bibinfo {year} {2019})},\ \Eprint
  {https://arxiv.org/abs/1812.03138} {arXiv:1812.03138 [nucl-th]} \BibitemShut
  {NoStop}%
\bibitem [{\citenamefont {Kharzeev}\ and\ \citenamefont
  {Levin}(2021)}]{entanglementinnature4}%
  \BibitemOpen
  \bibfield  {author} {\bibinfo {author} {\bibfnamefont {D.~E.}\ \bibnamefont
  {Kharzeev}}\ and\ \bibinfo {author} {\bibfnamefont {E.}~\bibnamefont
  {Levin}},\ }\bibfield  {title} {\bibinfo {title} {{Deep inelastic scattering
  as a probe of entanglement: Confronting experimental data}},\ }\href
  {https://doi.org/10.1103/PhysRevD.104.L031503} {\bibfield  {journal}
  {\bibinfo  {journal} {Phys. Rev. D}\ }\textbf {\bibinfo {volume} {104}},\
  \bibinfo {pages} {L031503} (\bibinfo {year} {2021})},\ \Eprint
  {https://arxiv.org/abs/2102.09773} {arXiv:2102.09773 [hep-ph]} \BibitemShut
  {NoStop}%
\bibitem [{\citenamefont {Beane}\ and\ \citenamefont
  {Farrell}(2021)}]{entanglementinnature8}%
  \BibitemOpen
  \bibfield  {author} {\bibinfo {author} {\bibfnamefont {S.~R.}\ \bibnamefont
  {Beane}}\ and\ \bibinfo {author} {\bibfnamefont {R.~C.}\ \bibnamefont
  {Farrell}},\ }\bibfield  {title} {\bibinfo {title} {{Geometry and
  entanglement in the scattering matrix}},\ }\href
  {https://doi.org/10.1016/j.aop.2021.168581} {\bibfield  {journal} {\bibinfo
  {journal} {Annals Phys.}\ }\textbf {\bibinfo {volume} {433}},\ \bibinfo
  {pages} {168581} (\bibinfo {year} {2021})},\ \Eprint
  {https://arxiv.org/abs/2011.01278} {arXiv:2011.01278 [hep-th]} \BibitemShut
  {NoStop}%
\bibitem [{\citenamefont {Beane}\ \emph {et~al.}(2021)\citenamefont {Beane},
  \citenamefont {Farrell},\ and\ \citenamefont
  {Varma}}]{entanglementinnature9}%
  \BibitemOpen
  \bibfield  {author} {\bibinfo {author} {\bibfnamefont {S.~R.}\ \bibnamefont
  {Beane}}, \bibinfo {author} {\bibfnamefont {R.~C.}\ \bibnamefont {Farrell}},\
  and\ \bibinfo {author} {\bibfnamefont {M.}~\bibnamefont {Varma}},\ }\bibfield
   {title} {\bibinfo {title} {{Entanglement minimization in hadronic scattering
  with pions}},\ }\href {https://doi.org/10.1142/S0217751X21502055} {\bibfield
  {journal} {\bibinfo  {journal} {Int. J. Mod. Phys. A}\ }\textbf {\bibinfo
  {volume} {36}},\ \bibinfo {pages} {2150205} (\bibinfo {year} {2021})},\
  \Eprint {https://arxiv.org/abs/2108.00646} {arXiv:2108.00646 [hep-ph]}
  \BibitemShut {NoStop}%
\bibitem [{\citenamefont {Robin}\ \emph {et~al.}(2021)\citenamefont {Robin},
  \citenamefont {Savage},\ and\ \citenamefont
  {Pillet}}]{entanglementinnature10}%
  \BibitemOpen
  \bibfield  {author} {\bibinfo {author} {\bibfnamefont {C.}~\bibnamefont
  {Robin}}, \bibinfo {author} {\bibfnamefont {M.~J.}\ \bibnamefont {Savage}},\
  and\ \bibinfo {author} {\bibfnamefont {N.}~\bibnamefont {Pillet}},\
  }\bibfield  {title} {\bibinfo {title} {{Entanglement Rearrangement in
  Self-Consistent Nuclear Structure Calculations}},\ }\href
  {https://doi.org/10.1103/PhysRevC.103.034325} {\bibfield  {journal} {\bibinfo
   {journal} {Phys. Rev. C}\ }\textbf {\bibinfo {volume} {103}},\ \bibinfo
  {pages} {034325} (\bibinfo {year} {2021})},\ \Eprint
  {https://arxiv.org/abs/2007.09157} {arXiv:2007.09157 [nucl-th]} \BibitemShut
  {NoStop}%
\bibitem [{\citenamefont {Low}\ and\ \citenamefont
  {Mehen}(2021)}]{entanglementinnature11}%
  \BibitemOpen
  \bibfield  {author} {\bibinfo {author} {\bibfnamefont {I.}~\bibnamefont
  {Low}}\ and\ \bibinfo {author} {\bibfnamefont {T.}~\bibnamefont {Mehen}},\
  }\bibfield  {title} {\bibinfo {title} {{Symmetry from entanglement
  suppression}},\ }\href {https://doi.org/10.1103/PhysRevD.104.074014}
  {\bibfield  {journal} {\bibinfo  {journal} {Phys. Rev. D}\ }\textbf {\bibinfo
  {volume} {104}},\ \bibinfo {pages} {074014} (\bibinfo {year} {2021})},\
  \Eprint {https://arxiv.org/abs/2104.10835} {arXiv:2104.10835 [hep-th]}
  \BibitemShut {NoStop}%
\bibitem [{\citenamefont {Roggero}(2021)}]{entanglementinnature13}%
  \BibitemOpen
  \bibfield  {author} {\bibinfo {author} {\bibfnamefont {A.}~\bibnamefont
  {Roggero}},\ }\bibfield  {title} {\bibinfo {title} {{Entanglement and
  many-body effects in collective neutrino oscillations}},\ }\href
  {https://doi.org/10.1103/PhysRevD.104.103016} {\bibfield  {journal} {\bibinfo
   {journal} {Phys. Rev. D}\ }\textbf {\bibinfo {volume} {104}},\ \bibinfo
  {pages} {103016} (\bibinfo {year} {2021})},\ \Eprint
  {https://arxiv.org/abs/2102.10188} {arXiv:2102.10188 [hep-ph]} \BibitemShut
  {NoStop}%
\bibitem [{\citenamefont {Mueller}\ \emph {et~al.}(2022)\citenamefont
  {Mueller}, \citenamefont {Zache},\ and\ \citenamefont
  {Ott}}]{entanglementinnature14}%
  \BibitemOpen
  \bibfield  {author} {\bibinfo {author} {\bibfnamefont {N.}~\bibnamefont
  {Mueller}}, \bibinfo {author} {\bibfnamefont {T.~V.}\ \bibnamefont {Zache}},\
  and\ \bibinfo {author} {\bibfnamefont {R.}~\bibnamefont {Ott}},\ }\bibfield
  {title} {\bibinfo {title} {{Thermalization of Gauge Theories from their
  Entanglement Spectrum}},\ }\href
  {https://doi.org/10.1103/PhysRevLett.129.011601} {\bibfield  {journal}
  {\bibinfo  {journal} {Phys. Rev. Lett.}\ }\textbf {\bibinfo {volume} {129}},\
  \bibinfo {pages} {011601} (\bibinfo {year} {2022})},\ \Eprint
  {https://arxiv.org/abs/2107.11416} {arXiv:2107.11416 [quant-ph]} \BibitemShut
  {NoStop}%
\bibitem [{\citenamefont {Liu}\ \emph {et~al.}(2023)\citenamefont {Liu},
  \citenamefont {Low},\ and\ \citenamefont {Mehen}}]{entanglementinnature12}%
  \BibitemOpen
  \bibfield  {author} {\bibinfo {author} {\bibfnamefont {Q.}~\bibnamefont
  {Liu}}, \bibinfo {author} {\bibfnamefont {I.}~\bibnamefont {Low}},\ and\
  \bibinfo {author} {\bibfnamefont {T.}~\bibnamefont {Mehen}},\ }\bibfield
  {title} {\bibinfo {title} {{Minimal entanglement and emergent symmetries in
  low-energy QCD}},\ }\href {https://doi.org/10.1103/PhysRevC.107.025204}
  {\bibfield  {journal} {\bibinfo  {journal} {Phys. Rev. C}\ }\textbf {\bibinfo
  {volume} {107}},\ \bibinfo {pages} {025204} (\bibinfo {year} {2023})},\
  \Eprint {https://arxiv.org/abs/2210.12085} {arXiv:2210.12085 [quant-ph]}
  \BibitemShut {NoStop}%
\bibitem [{\citenamefont {Inc.}(2024)}]{mma}%
  \BibitemOpen
  \bibfield  {author} {\bibinfo {author} {\bibfnamefont {W.~R.}\ \bibnamefont
  {Inc.}},\ }\href {https://www.wolfram.com/mathematica} {\bibinfo {title}
  {Mathematica, {V}ersion 14.0}} (\bibinfo {year} {2024}),\ \bibinfo {note}
  {champaign, IL, 2024}\BibitemShut {NoStop}%
\end{thebibliography}%

\onecolumngrid
\newpage
\appendix

\section{Gaussian classical mixing channel for a single core}
\label{app:corenoiseiden}

Section~\ref{sec:localon} established the $\mathcal{N}$-SOL entanglement class with symplectic transformation $\Tilde{S}$ that diagonalizes the PT CM. After the $\mathcal{V}_{\mathcal{N}}$ entanglement is consolidated in the core via the calculable local transformation, Eq.~\eqref{eq:gseq},  noise in $\mathcal{V}_{\Nslash}$ may be identified through a Gaussian classical mixing channel~\cite{WolfGEOF,GiedkeEOF}. This appendix provides the corresponding noise parameterization, and shows its equivalence to parameterization of $\Tilde{S}$ in the core. This noise parameterization is utilized in Sec.~\ref{sec:postconsep} to show post-consolidation separability between TMSVS pairs in $\mathcal{N}$-SOL states.

For TMSVS with squeezing parameter $r$, the PT CM can be diagonalized via,
\begin{equation}
    \Tilde{D}_{TMSVS} = \Tilde{S} \Tilde{\sigma}_{TMSVS} \Tilde{S}^{T} = \frac{1}{2}\begin{pmatrix}
\mathbb{I}_{2} & \mathbb{I}_{2} \\
-\mathbb{I}_{2} & \mathbb{I}_{2} \\
\end{pmatrix}\Tilde{\sigma}_{TMSVS}\begin{pmatrix}
\mathbb{I}_{2} & -\mathbb{I}_{2} \\
\mathbb{I}_{2} & \mathbb{I}_{2} \\
\end{pmatrix}=\begin{pmatrix}
e^{2r} & 0 & 0 & 0  \\
0 & e^{2r} & 0 & 0  \\
0 & 0 & e^{-2r} & 0  \\
0 & 0 & 0 & e^{-2r}   \\
\end{pmatrix} \ \ \ .
\label{eq:a1}
\end{equation}
In this basis with $r \geq 0$, the $\mathcal{V}_{\Nslash}$ and $\mathcal{V}_{\mathcal{N}}$ subspaces are organized into the first two and last two dimensions, respectively. Adding noise, $Y_{c_{1}}$, isolated to $\mathcal{V}_{\Nslash}$ through a Gaussian classical mixing channel in the form of $\sigma_{c_{1}}=\sigma_{TMSVS} + Y_{c_{1}}$ will result in a $\mathcal{N}$IC-decomposable CM,
\begin{equation}
    \sigma_{c_{1}} = \begin{pmatrix}
\cosh{2r} & 0 & \sinh{2r} & 0  \\
0 & \cosh{2r} & 0 & -\sinh{2r}  \\
\sinh{2r} & 0 & \cosh{2r} & 0  \\
0 & -\sinh{2r} & 0 & \cosh{2r}   \\
\end{pmatrix} +  \frac{1}{2} \Lambda \begin{pmatrix}
\mathbb{I}_{2} & -\mathbb{I}_{2} \\
\mathbb{I}_{2} & \mathbb{I}_{2} \\
\end{pmatrix} \begin{pmatrix}
y_{11} & y_{12} & 0 & 0  \\
y_{12} & y_{22} & 0 & 0  \\
0 & 0 & 0 & 0  \\
0 & 0 & 0 & 0   \\
\end{pmatrix} \begin{pmatrix}
\mathbb{I}_{2} & \mathbb{I}_{2} \\
-\mathbb{I}_{2} & \mathbb{I}_{2} \\
\end{pmatrix}  \Lambda \ \ \ .
\label{eq:c1exp}
\end{equation}
The first term in Eq.~\eqref{eq:c1exp} is the CM for TMSVS with squeezing parameter $r$ and the second term represents noise, $Y_{c_{1}}$, introduced via Gaussian classical mixing~\cite{serafini2017quantum}. The off-diagonal blocks of $Y_{c_{1}}$ in the diagonalized PT space are chosen to be zero in order to ensure all contributions in $\mathcal{V_{\mathcal{N}}}$ vanish. For the CM to remain physical, the noise must be PSD,
\begin{equation}
    \begin{pmatrix}
y_{11} & y_{12}  \\
y_{12} & y_{22}   \\
\end{pmatrix} \geq 0 \Rightarrow Y_{c_{1}} \geq 0 \ \ \ ,
\label{eq:a3}
\end{equation}
such that $y_{11} \geq 0$, $y_{22} \geq 0$ and $y_{11}y_{22} \geq y_{12}^{2}$.

In the following, the translation is illustrated between the framework of Eq.~\eqref{eq:c1exp} confining noise to $\mathcal{V}_{\Nslash}$ and the language of $\tilde{S}$ that is utilized in the main text to inform transformations and alignment of entanglement structure. From Sec.~\ref{sec:localon}, post-consolidated cores have $\mathcal{V_{\mathcal{N}}}$ subspaces aligned with that of TMSVS such that the general expression of $\Tilde{S}^{'}_{c_{1}}$ reads,
\begin{equation}
   \Tilde{S}^{'}_{c_{1}} = \begin{pmatrix}
a_{1} & a_{2} & a_{1} & a_{2} \\
a_{3} & a_{4} & a_{3} & a_{4} \\
-\frac{1}{\sqrt{2}} & 0 & \frac{1}{\sqrt{2}} & 0  \\
0 & -\frac{1}{\sqrt{2}} & 0 & \frac{1}{\sqrt{2}}  \\
\end{pmatrix} \ \ \ ,
\label{eq:tmsvslike}
\end{equation}
with real a's and $2a_{1}a_{4}-2a_{3}a_{2}=1$ for symplectic orthonormality, Eq.~\eqref{eq:symplecticorthonormality}. Because of symplectic orthogonality and positive squeezing assumed for the underlying TMSVS, $\Tilde{S}^{'}_{c_{1}}$ exhibits A-B symmetry in $\mathcal{V}_{\Nslash}$. With PT symplectic eigenvalues denoted by $\left\{ \Tilde{\nu}'_{+},\Tilde{\nu}'_{+},\Tilde{\nu}'_{-},\Tilde{\nu}'_{-}\right\}$, the CM of this single core can be constructed via Eq.~\eqref{eq:tmsvslike} and Eq.~\eqref{eq:sigmatilde}. Physicality of this CM leads to constraints $\Tilde{\nu}'_{+} \geq 1$, and the state is entangled if and only if $\Tilde{\nu}'_{-} < 1$~\cite{serafini2017quantum,computablemeasure,Simonreflection,Duanlocaltrans}. The CM associated with Eq.~\eqref{eq:tmsvslike} thus relates to that of Eq.~\eqref{eq:c1exp} by,
\begin{equation}
    y_{11} = 2\Tilde{\nu}'_{+}(a_{2}^{2}+a_{4}^{2})- \frac{1}{\Tilde{\nu}'_{-}} ,\quad y_{22} = 2\Tilde{\nu}'_{+}(a_{1}^{2}+a_{3}^{2})- \frac{1}{\Tilde{\nu}'_{-}} ,\quad y_{12} = -2\Tilde{\nu}'_{+}( a_{1}a_{2}+a_{3}a_{4} ),\quad r=-\frac{1}{2}\ln \Tilde{\nu}'_{-} \ \ \ .
\label{eq:eqparam}
\end{equation}
For a CM with PT symplectic transformation given by Eq.~\eqref{eq:tmsvslike}, there hence exists a decomposition in a Gaussian classical mixing channel specified by Eqs.~\eqref{eq:c1exp} and~\eqref{eq:eqparam}, where the PSD constraints of $Y_{c_{1}}$ follow from physicality of the CM. 

\section{Gaussian \texorpdfstring{$\mathcal{N}$IC}{NIC} decomposition derivations}
\label{app:align}

\subsection{Negativity invariant Gaussian classical mixing}
\label{app:alignsub1}

This appendix derives the four conditions of negativity-invariant Gaussian classical mixing discussed in Sec.~\ref{sec:Nalign}.

Mirroring their counterparts in the physical phase space, PT-symplectic eigenvalues, $\Tilde{\nu}$, of a CM, $\sigma$, may be calculated as eigenvalues of the product $|i \Omega \tilde{\sigma}|$. Given the invariance of eigenvalues under cyclic permutation of a matrix product, this expression can be expanded as the spectrum of a real positive symmetric matrix,
\begin{equation}
\sqrt{\text{spec}(-\Omega\Tilde{\sigma}\Omega\Tilde{\sigma})} = \sqrt{\text{spec}(\Tilde{\sigma}^{1/2}\Omega^{T}\Tilde{\sigma}\Omega\Tilde{\sigma}^{1/2})} \ \ \ ,
\label{eq:b1}
\end{equation}
where $-\Omega = \Omega^T$ and the square root is well-defined due to the physicality of CMs, Eq.~\eqref{eq:phy}.
After partial transposition, Gaussian classical mixing of Eq.~\eqref{eq:noisepmy} becomes,
\begin{equation}
   \Tilde{\sigma}_{m}=\Tilde{\sigma}_{p}+\Tilde{Y} \ \ \ .
\end{equation}
Because the positive semidefiniteness of Y is not altered by the PT coordinate transformation, $\tilde{\sigma}_p \leq \tilde{\sigma}_m$. Furthermore, as shown in Ref.~\cite{Giedkepurestatetrans} through Weyl's inequality~\cite{hornmatrixana}, each eigenvalue of the mixed-state PT-symplectic spectrum is greater than or equal to the sequentially corresponding one of the pure-state. These relations are expressed in the following inequalities:
\begin{equation}
    \Tilde{\sigma}_{p}^{1/2}\Omega^{T}\Tilde{\sigma}_{p}\Omega\Tilde{\sigma}_{p}^{1/2} \leq \Tilde{\sigma}_{p}^{1/2}\Omega^{T}\Tilde{\sigma}_{m}\Omega\Tilde{\sigma}_{p}^{1/2} \ \ \ ,
\end{equation}
\begin{equation}
    \Tilde{\sigma}_{m}^{1/2}\Omega^{T}\Tilde{\sigma}_{p}\Omega\Tilde{\sigma}_{m}^{1/2} \leq \Tilde{\sigma}_{m}^{1/2}\Omega^{T}\Tilde{\sigma}_{m}\Omega\Tilde{\sigma}_{m}^{1/2} \ \ \ ,
\end{equation}
\begin{equation}
    \text{spec}(\Tilde{\sigma}_{p}^{1/2}\Omega^{T}\Tilde{\sigma}_{p}\Omega\Tilde{\sigma}_{p}^{1/2}) \leq \text{spec}(\Tilde{\sigma}_{m}^{1/2}\Omega^{T}\Tilde{\sigma}_{m}\Omega\Tilde{\sigma}_{m}^{1/2}) \ \ \ .
    \label{eq:ptineqwithpsd}
\end{equation}
The latter relation, considered specifically for the $\mathcal{V}_{\mathcal{N}}$ subspace of PT-symplectic eigenvalues less than one, indicates that $\mathcal{N}$IC-decomposible mixed states ($\mathcal{N}_p^{\text{min}} = \mathcal{N}$) must saturate Eq.~\eqref{eq:ptineqwithpsd} in $\mathcal{V}_{\mathcal{N}}$, i.e., $\left\{ \tilde{\nu} \right\}_{\mathcal{V}_{\mathcal{N}}}$ must be invariant upon the the introduction of Y classical correlations.

For NIC-decomposable CMs, the invariance of eigenvalues in $\mathcal{V}_{\mathcal{N}}$ is accompanied by invariance also in the associated eigenvectors. To see this relationship, consider a real positive symmetric matrix, $M$, and positive semidefinite perturbation, $\delta M$, serving the role of $Y$. If the smallest eigenvalue (largest contribution to the negativity for NPT states) is constrained to be invariant, 
\begin{equation}
    \bra{m}_{0}M\ket{m}_{0} = \bra{m'}_{0}M\ket{m'}_{0} + \cancelto{0}{\bra{m'}_{0}\delta M\ket{m'}_{0}}  \ \ \ ,
    \label{eq:vectorunchange}
\end{equation}
where $\ket{m}_{0}$ and $\ket{m'}_{0}$ denote ground state before and after $\delta M$ perturbation. Due to the positive semidefiniteness of $\delta M$ and the fact that the expectation value is minimized on the left-hand side, the expectation value of $\delta M$ must vanish. Note that the symplectic eigenvalues are doubly degenerate, hence $\ket{m'}_{0}$ can be any vector in the degenerate subspace of the ground state. Applying Eq.~\eqref{eq:vectorunchange} a second time selects the vector orthogonal to the first one, and sequentially repeating the process gives eigenvectors that are equivalent to those of the original matrix, modulo the presence of local symplectic phase rotations aligning each mode. Thus, PT-symplectic eigenvectors of the $\mathcal{V}_{\mathcal{N}}$ subspace that are invariant between the pure and mixed CM can be constructed for $\mathcal{N}$IC-decomposable states, such that, 
\begin{multline}
     \text{vec}_{\mathcal{V_{\mathcal{N}}}}(\Tilde{\sigma}_{p}^{1/2}\Omega^{T}\Tilde{\sigma}_{p}\Omega\Tilde{\sigma}_{p}^{1/2})=\text{vec}_{\mathcal{V_{\mathcal{N}}}}(\Tilde{\sigma}_{p}^{1/2}\Omega^{T}\Tilde{\sigma}_{m}\Omega\Tilde{\sigma}_{p}^{1/2}) \\ =  \Tilde{\sigma}_{p}^{1/2}\Omega\Tilde{\sigma}_{m}^{1/2}\text{vec}_{\mathcal{V_{\mathcal{N}}}}(\Tilde{\sigma}_{m}^{1/2}\Omega^{T}\Tilde{\sigma}_{p}\Omega\Tilde{\sigma}_{m}^{1/2})=\Tilde{\sigma}_{p}^{1/2}\Omega\Tilde{\sigma}_{m}^{1/2}\text{vec}_{\mathcal{V_{\mathcal{N}}}}(\Tilde{\sigma}_{m}^{1/2}\Omega^{T}\Tilde{\sigma}_{m}\Omega\Tilde{\sigma}_{m}^{1/2}) \ \ \ ,
\end{multline}
with eigenvalue equalities, 
\begin{multline}
    \text{spec}_{\mathcal{V_{\mathcal{N}}}}(\Tilde{\sigma}_{p}^{1/2}\Omega^{T}\Tilde{\sigma}_{p}\Omega\Tilde{\sigma}_{p}^{1/2}) =  \text{spec}_{\mathcal{V_{\mathcal{N}}}}(\Tilde{\sigma}_{p}^{1/2}\Omega^{T}\Tilde{\sigma}_{m}\Omega\Tilde{\sigma}_{p}^{1/2}) \\=\text{spec}_{\mathcal{V_{\mathcal{N}}}}(\Tilde{\sigma}_{m}^{1/2}\Omega^{T}\Tilde{\sigma}_{p}\Omega\Tilde{\sigma}_{m}^{1/2}) =  \text{spec}_{\mathcal{V_{\mathcal{N}}}}(\Tilde{\sigma}_{m}^{1/2}\Omega^{T}\Tilde{\sigma}_{m}\Omega\Tilde{\sigma}_{m}^{1/2}) \ \ \ .
\end{multline}
By noting $\text{vec}_{k}(\Tilde{\sigma}^{1/2}\Omega^{T}\Tilde{\sigma}\Omega\Tilde{\sigma}^{1/2})=\Tilde{\sigma}^{1/2}\text{vec}_{k}(-\Omega\Tilde{\sigma}\Omega\Tilde{\sigma})$, this leads to Eqs.~\eqref{eq:ppmm1}--\eqref{eq:sqdof2}.

\subsection{Normal-form symplectic transformation}
\label{app:alignsub2}

This section presents an alternative construction (differing from those introduced in Refs.~\cite{serafini2017quantum,NKvolumemeasure}) of the normal-form symplectic transformation, $S$ of Eq.~\eqref{eq:onlys}, diagonalizing the CM. This will contribute to understanding of the four conditions of negativity-invariant Gaussian classical mixing derived in Appendix~\ref{app:alignsub1} and discussed in Sec.~\ref{sec:Nalign}. Because the procedures here are valid regardless whether a given CM satisfies the physicality condition, the CM $\sigma$ appearing throughout this section may be replaced by $\tilde{\sigma}$ for calculations of $\tilde{S}$ in Eq.~\eqref{eq:stilde}.

Because the CCRs are invariant under symplectic transformation, $S \Omega S^T = \Omega$, the form $-\Omega \sigma\Omega \sigma$ admits, 
\begin{equation}
    -\Omega\sigma\Omega\sigma = -S^{T}\Omega D \Omega D S^{-T}=-S^{T} \left[ \bigoplus_{i}\begin{pmatrix}
0 & d_{i} \\
-d_{i} & 0 \\
\end{pmatrix} \right] \left[ \bigoplus_{i}\begin{pmatrix}
0 & d_{i} \\
-d_{i} & 0 \\
\end{pmatrix} \right] S^{-T}=S^{T} \left[ \bigoplus_{i}\begin{pmatrix}
d_{i}^{2} & 0 \\
0 & d_{i}^{2} \\
\end{pmatrix} \right] S^{-T} ,
\label{eq:b9}
\end{equation}
hence $S^{T}$ diagonalizes $-\Omega\sigma\Omega\sigma$ by similarity. Reversely, denoting $L$ as the similarity transformation that diagonalizes $-\Omega \sigma \Omega \sigma$ (i.e., $-L\Omega\sigma\Omega\sigma L^{-1} = D^2$), methods for determining $L^{-T} = S$ do not generically satisfy the symplectic requirement. To enforce this property, the following procedure can be followed. 

Without mode-wise degeneracy of symplectic eigenvalues, symplectic orthogonality in Eq.~\eqref{eq:symplecticorthonormality} is automatically satisfied. This is because $L^{-T}$ is the stack of right eigenvectors of $-\Omega\sigma\Omega\sigma$, and $\Omega$ has block diagonal structure that only rotates eigenvectors in the degenerate subspace. If a degeneracy is present in the symplectic spectrum, linear combinations of the $L^{-T}$ row vectors within the degenerate subspace can be arranged to produce symplectic orthogonality. For the normalization, there exists the freedom, 
\begin{equation}
    L^{'} = D^{'} L \quad , \quad -L^{'}\Omega\sigma\Omega\sigma L^{'-1} = -L\Omega\sigma\Omega\sigma L^{-1} \ \ \ ,
\end{equation}
for arbitrary diagonal matrix $D^{'}$. Similar to procedures in Appendix D of Ref.~\cite{NKvolumemeasure}, the normalization can be achieved by tuning $D^{'-1}$ such that $L'^{-T} = D'^{-1}L^{-T}$ is a valid symplectic transformation,
\begin{equation}
L^{-T} \Omega L^{-1} = D' \Omega D' = \bigoplus_i \alpha_i \begin{pmatrix}
0 & 1 \\
-1 & 0
\end{pmatrix}
\quad , \quad
D' = \bigoplus_i  \begin{pmatrix}
\sqrt{|\alpha_i|} & 0 \\
0 & \sign \left( \alpha_i\right) \sqrt{|\alpha_i|}
\end{pmatrix} \ \ \ .
\label{eq:b11}
\end{equation}
Because the similarity transform of Eq.~\eqref{eq:b9} produces a matrix proportional to the identity for each mode, an additional freedom is present in the inclusion of arbitrary single-mode operations, $S^{(1)}_i$. Therefore, the final step adds single-mode operations such that $S$ completely diagonalizes the CM,
\begin{equation}
 S = \left[ \bigoplus_{i} S^{(1)}_i \right] D'^{-1} L^{-T} \quad , \quad S\sigma S^T = D \  \ \ . 
 \end{equation}
Note that the $S^{(1)}_i$ operations can be determined via standard single-mode normal form techniques, e.g., based on the symplectic eigenvectors of $i \Omega \sigma$, which retain a relative sign breaking the phase space degeneracy.

\section{Examples: four-mode Gaussian states}
\label{app:exp}

\subsection{General entanglement consolidation}
\label{app:conso1}

Section~\ref{sec:localon} indicates that membership in the $\mathcal{N}$-SOL subset of NPT states allows local symplectic consolidation of negativity into a tensor product TMSVS entanglement structure. This extends understanding of the availability of entanglement consolidation beyond regions of the scalar field vacuum. Here, this will be illustrated through an example with $x$-$p$ mixing, no reflection symmetry, and no connection to the scalar field. The following example is a selected 4-mode member of the $\mathcal{N}$-SOL entanglement class along the $(2_A \times 2_B)$ bipartition.

Consider a four-mode position-momentum correlated mixed Gaussian state with ordering ($A_1, A_2, B_1, B_2$),
\begin{equation}
\sigma=\begin{pmatrix}
2.89562& 1.86324& 1.20668& 3.63702& 0.388124& 0.199187& -1.61365& 0.752206\\
1.86324& 6.90690& 0.936903& -4.22060& 0.531462& -0.263280& -2.15507& 1.90189\\
1.20668& 0.936903& 1.23135& 2.42264& 1.21851& 2.45543& -0.213628& 0.763726\\
3.63702& -4.22060& 2.42264& 26.1167& 4.64899& 2.47557& 1.78513& 5.15944\\
0.388124& 0.531462& 1.21851& 4.64899& 3.26789& 5.95046& 2.06324& 1.95927\\
0.199187& -0.263280& 2.45543& 2.47557& 5.95046& 20.1834& 6.99689& 0.447081\\
-1.61365& -2.15507& -0.213628& 1.78513& 2.06324& 6.99689& 4.90273& 0.681893\\
0.752206& 1.90189& 0.763726& 5.15944& 1.95927& 0.447081& 0.681893& 4.20990\\
\end{pmatrix} \ \ \ ,
\end{equation}  
where the $\mathcal{V_{\mathcal{N}}}$ has one PT symplectic eigenvalue, $\tilde{\nu}_{-} = 0.213940$, hence the logarithmic negativity is $\mathcal{N}=2.22472$. The $\mathcal{V_{\mathcal{N}}}$ row vectors of $\Tilde{S}$ are governed by the symplectic orthonormality condition, ${}_1\langle \tilde{\nu}_-| \Omega | \tilde{\nu}_-\rangle_2 = 1$, and are calculated to be, 
\begin{equation}
    \ket{\Tilde{\nu}_{-}}_{1}=\begin{pmatrix}
0.325739& 0.00171586& 1.00356& -0.148923& 0.362151& 0.497660& 0.816754& 0.315735\\
\end{pmatrix}^{T} \ \ \ ,
\end{equation}
and,
\begin{equation}
    \ket{\Tilde{\nu}_{-}}_{2}=\begin{pmatrix}
-0.70301& 0.475514& 0& 0.342682& -0.729204& -0.249817& -0.181575& 0.208441\\
\end{pmatrix}^{T} \ \ \ .
\end{equation}
Confirming this CM is a member of $\mathcal{N}$-SOL via Eq.~\eqref{eq:nsolcondition}, a local transformation ($S_A \oplus S_B$) that consolidates the many-body negativity into TMSVS pairs can be designed through a symplectic GS procedure starting from the $\mathcal{V_{\mathcal{N}}}$ row vectors of $\tilde{S}$ as described in Eq.~\eqref{eq:gseq},
\begin{equation}
    S_{A}=\begin{pmatrix}
 1& 0& 0& 0\\
 0& 1& 0& 0\\
 0& 0& -1& 0\\
 0& 0& 0& -1\\
    \end{pmatrix}\begin{pmatrix}
 -0.906105& -0.588806& 2.46527& -0.176266\\
 0.581021& -0.381514& -0.0235654& 0.128324\\
 0.460665& 0.00242659& 1.41924& -0.210609\\
 -0.994206& 0.672478& 0& 0.484625\\
    \end{pmatrix} \ \ \ ,
\end{equation}
\begin{equation}
    S_{B}= (\sigma_{z}\oplus\sigma_{z}) \begin{pmatrix}
 -0.851396& 0.0275469& -0.489418& -0.720178\\
 -1.78259& -0.476916& 0.219893& -0.789690\\
 0.512159& 0.703799& 1.15507& 0.446517\\
 -1.03125& -0.353295& -0.256786& 0.294780\\
    \end{pmatrix} (\sigma_{z}\oplus\sigma_{z}) \ \ \ .
\end{equation}
The entanglement consolidated CM, $\sigma' = \left( S_A \oplus S_B \right) \sigma \left( S_A\oplus S_B \right)^T $, rearranged in core-halo form with ordering $ \left( c_A,c_B,h_A,h_B \right)$ is,
\begin{equation}
\sigma'=\begin{pmatrix}
3.6920& -0.38369& 3.4781& 0.38369& -0.60432& -0.054119& -0.073460& -0.24042 \\
-0.38369&  3.3723& -0.38369& -3.1583& 0.16689& -0.57620&  0.23274& -0.95387 \\
3.4781& -0.38369& 3.6920& 0.38369& -0.60432& -0.054119& -0.073460& -0.24042\\
0.38369& -3.1583& 0.38369& 3.3723& -0.16689& 0.57620& -0.23274&  0.95387 \\ 
-0.60432& 0.16689& -0.60432& -0.16689& 5.1245& 0.27367& -2.9693& 0.28105\\
-0.054119& -0.57620& -0.054119& 0.57620& 0.27367& 2.5123& -0.31802& 0.51894\\
-0.073459& 0.23274& -0.073460& -0.23274& -2.9693& -0.31802& 5.0281& -0.33108\\
-0.24042& -0.95387& -0.24042& 0.95387& 0.28105& 0.51894& -0.33108& 2.6272 \\
\end{pmatrix} \ \ \ ,
\end{equation}
where the upper diagonal $4 \times 4$ block, $\sigma_{c_1}$, captures all the negativity and available entanglement.

\subsection{Scalar field \texorpdfstring{$\mathcal{N}$IC}{NIC} decomposition}
\label{app:conso2}

Section~\ref{sec:Nalign} identifies entanglement consolidation as the first step in determining a $\mathcal{N}$IC decomposition, followed by MNF and a separability check~\cite{Giedkesepflow} of the minimum noise subtracted halo. This process is illustrated by the following example for the free lattice scalar field vacuum discussed in Sec.~\ref{sec:pureestimate}.

For two regions of the lattice scalar field vacuum in the massless regime characterized by $\left( d, \tilde{r}, m\right) = \left( 2, 1, 10^{-10}\right)$, the post-consolidation CM with mode ordering $\left( c_A, c_B, h_A, h_B\right)$ is,
\begin{equation}
\sigma'_{d=2,r=1}=\begin{pmatrix}
7.65761& 0& 6.72931& 0& 5.22809& 0& 5.22809& 0\\
0& 1.02441& 0& -0.096102& 0& 0.047747& 0& -0.047747\\
6.72931& 0& 7.65761& 0& 5.22809& 0& 5.22809& 0\\
0& -0.096102& 0& 1.02441& 0& -0.047747& 0& 0.047747\\
5.22809& 0& 5.22809& 0& 5.18827& 0& 4.10771& 0\\
0& 0.047747& 0& -0.047747& 0& 1.11078& 0& -0.026938\\
5.22809& 0& 5.22809& 0& 4.10771& 0& 5.18827& 0\\
0& -0.047747& 0& 0.047747& 0& -0.026938& 0&1.11078\\
\end{pmatrix} \ \ \ .
\end{equation}
From Eq.~\eqref{eq:Y11cform}, noise isolated to $\mathcal{V}_{\Nslash}$ of the core may be identified as, 
\begin{equation}
  Y_{c_{1}}=\begin{pmatrix}
6.65484& 0& 6.65484& 0\\
0& 0.021637& 0& -0.021637\\
6.65484& 0& 6.65484& 0\\
0& -0.021637& 0& 0.021637\\
\end{pmatrix} \ \ \ .
\end{equation}
The minimum noise subtracted halo is identified, as discussed in Sec.~\ref{sec:postconsep}, to be $\sigma_{h}- Y_{c_1 r}^{T}Y_{c_{1}}^{-1} Y_{c_1 r}$, with $Y_{c_1 r} = \sigma_{c_1 r}$. Hence, an underlying state with the same negativity as the mixed state organized in pure TMSVS form is,
\begin{multline}
    \sigma'_{d=2,r=1}-\begin{pmatrix}
 Y_{c_{1}} &  Y_{c_1 r}\\
  Y_{c_1 r}^{T}&  Y_{c_1 r}^{T}Y_{c_{1}}^{-1} Y_{c_1 r}\\
  \end{pmatrix} =\\ \scalemath{0.925}{ \begin{pmatrix}
1.00277& 0& 0.074465& 0& 0& 0& 0& 0\\
0& 1.00277& 0& -0.074465& 0& 0& 0& 0\\
0.074465& 0& 1.00277& 0& 0& 0& 0& 0\\
0& -0.074465& 0& 1.00277& 0& 0& 0& 0\\
0& 0& 0& 0& 1.08269& 0& 0.000490& 0\\
0& 0& 0& 0& 0& 1.00389& & 0.078307\\
0& 0& 0& 0& 0.000490& 0& 1.08269& 0\\
0& 0& 0& 0& 0& 0.078307& 0&1.00389\\
\end{pmatrix} } \ \ \ .
\end{multline}
Because the minimum noise subtracted halo (lower right block) is two-mode PPT, and thus separable~\cite{Simonreflection,Duanlocaltrans}, $\sigma_{d = 2, r = 1}$ is Gaussian $\mathcal{N}$IC-decomposable.

\section{Numerical tables}
\label{app:num}

In this appendix, numerical values are provided for $\mathcal{N}$IC transitions in the free massless scalar field vacuum and for the calculations presented in Fig.~\ref{fig:plot}.

\begin{table}[b!]
\tiny
\begin{tabular}{c|ccc}
\hline
\hline
$d$ & $\Tilde{r}_{\mathcal{N}IC}$ & $\Tilde{r}_{sep}$ & $\Tilde{r}_{sep}/\Tilde{r}_{\mathcal{N}IC}$ \\
\hline
\hline
$1$ & 0 & 1 & N/A\\
$2$ & 1 & 2 & 2.000\\
$3$ & 4 & 9 & 2.250\\
$4$ & 11 & 13 & 1.182\\
\hline
\hline
\end{tabular}
\begin{tabular}{c|ccc}
\hline
\hline
$d$ & $\Tilde{r}_{\mathcal{N}IC}$ & $\Tilde{r}_{sep}$ & $\Tilde{r}_{sep}/\Tilde{r}_{\mathcal{N}IC}$ \\
\hline
\hline
$5$ & 19 & 26 & 1.368\\
$6$ & 30 & 32 & 1.067\\
$7$ & 42 & 52 & 1.238\\
$8$ & 58 & 61 & 1.052\\
\hline
\hline
\end{tabular}
\begin{tabular}{c|ccc}
\hline
\hline
$d$ & $\Tilde{r}_{\mathcal{N}IC}$ & $\Tilde{r}_{sep}$ & $\Tilde{r}_{sep}/\Tilde{r}_{\mathcal{N}IC}$ \\
\hline
\hline
$9$ & 74 & 87 & 1.176\\
$10$ & 95 & 98 &1.032 \\
$11$ & 116 & 131 & 1.129\\
$12$ & 140 & 144 & 1.029\\
\hline
\hline
\end{tabular}
\begin{tabular}{c|ccc}
\hline
\hline
$d$ & $\Tilde{r}_{\mathcal{N}IC}$ & $\Tilde{r}_{sep}$ & $\Tilde{r}_{sep}/\Tilde{r}_{\mathcal{N}IC}$ \\
\hline
\hline
$13$ & 166 & 184 & 1.108\\
$14$ & 195 & 199 & 1.021\\
$15$ & 225 & 245 & 1.089\\
$16$ & 259 & 264 & 1.019\\
\hline
\hline
\end{tabular}
\begin{tabular}{c|ccc}
\hline
\hline
$d$ & $\Tilde{r}_{\mathcal{N}IC}$ & $\Tilde{r}_{sep}$ & $\Tilde{r}_{sep}/\Tilde{r}_{\mathcal{N}IC}$ \\
\hline
\hline
$17$ & 293 & 316 &1.078\\
$18$ & 331 & 337 &1.018\\
$19$ & 370 & 395 &1.068\\
$20$ & 413 & 419 & 1.015\\
\hline
\hline
\end{tabular}
\caption{Entanglement structure transitions from $\mathcal{N}$-SOL to $\mathcal{N}$IC to separable for $\tilde{r}$-separated regions of the one-dimensional free lattice scalar field vacuum in the massless limit ($m = 10^{-10}$) with a nearest-neighbor (leading order) lattice action. The left and right columns, $\Tilde{r}_{\mathcal{N}IC}$ and $\Tilde{r}_{sep}$, report the smallest $\tilde{r}$ for which the lattice vacuum regions are members of the $\mathcal{N}$IC entanglement class or separable, respectively. }
\label{tab:table2}
\end{table}

\begin{table}[b!]
\tiny
\begin{tabular}{c|ccc}
\hline
\hline
$\tilde{r}$ & $\mathcal{N}$ & $\mathcal{N}_p^{\phi}$ & $\mathcal{N}_p^{\text{MNF}}$ \\
\hline
\hline
$0$ & $1.214$ & $1.252$ & $1.464 $ \\
$1$ & $3.915\times 10^{-1}$ & $4.297\times 10^{-1}$ & $5.912\times 10^{-1} $ \\
$2$ & $1.907\times 10^{-1}$ & $2.715\times 10^{-1}$ & $2.950\times 10^{-1} $ \\
$3$ & $1.079\times 10^{-1}$ & $1.977\times 10^{-1}$ & $2.082\times 10^{-1} $ \\
$4$ & $6.725\times 10^{-2}$ & $1.540\times 10^{-1}$ & $1.434\times 10^{-1} $ \\
$5$ & $4.597\times 10^{-2}$ & $1.249\times 10^{-1}$ & $1.174\times 10^{-1} $ \\
$6$ & $3.353\times 10^{-2}$ & $1.040\times 10^{-1}$ & $1.503\times 10^{-1} $ \\
$7$ & $2.546\times 10^{-2}$ & $8.845\times 10^{-2}$ & $1.652\times 10^{-1} $ \\
$8$ & $1.981\times 10^{-2}$ & $7.637\times 10^{-2}$ & $1.600\times 10^{-1} $ \\
$9$ & $1.563\times 10^{-2}$ & $6.676\times 10^{-2}$ & $1.007\times 10^{-1} $ \\
$10$ & $1.243\times 10^{-2}$ & $5.896\times 10^{-2}$ & $4.696\times 10^{-2} $ \\
$11$ & $9.889\times 10^{-3}$ & $5.252\times 10^{-2}$ & $7.663\times 10^{-2} $ \\
$12$ & $7.837\times 10^{-3}$ & $4.712\times 10^{-2}$ & $9.036\times 10^{-2} $ \\
$13$ & $6.154\times 10^{-3}$ & $4.255\times 10^{-2}$ & $8.883\times 10^{-2} $ \\
$14$ & $4.766\times 10^{-3}$ & $3.864\times 10^{-2}$ & $8.370\times 10^{-2} $ \\
$15$ & $3.619\times 10^{-3}$ & $3.527\times 10^{-2}$ & $1.000\times 10^{-1} $ \\
$16$ & $2.680\times 10^{-3}$ & $3.233\times 10^{-2}$ & $5.441\times 10^{-2} $ \\
$17$ & $1.926\times 10^{-3}$ & $2.975\times 10^{-2}$ & $6.248\times 10^{-2} $ \\
$18$ & $1.340\times 10^{-3}$ & $2.748\times 10^{-2}$ & $7.596\times 10^{-2} $ \\
$19$ & $9.046\times 10^{-4}$ & $2.547\times 10^{-2}$ & $2.858\times 10^{-2} $ \\
$20$ & $6.024\times 10^{-4}$ & $2.367\times 10^{-2}$ & $6.465\times 10^{-2} $ \\
$21$ & $4.061\times 10^{-4}$ & $2.206\times 10^{-2}$ & $7.834\times 10^{-2} $ \\
$22$ & $2.836\times 10^{-4}$ & $2.062\times 10^{-2}$ & $6.978\times 10^{-2} $ \\
$23$ & $2.065\times 10^{-4}$ & $1.931\times 10^{-2}$ & $4.716\times 10^{-2} $ \\
$24$ & $1.560\times 10^{-4}$ & $1.813\times 10^{-2}$ & $1.428\times 10^{-2} $ \\
$25$ & $1.213\times 10^{-4}$ & $1.705\times 10^{-2}$ & $1.044\times 10^{-2} $ \\
$26$ & $9.624\times 10^{-5}$ & $1.607\times 10^{-2}$ & $2.434\times 10^{-2} $ \\
$27$ & $7.746\times 10^{-5}$ & $1.517\times 10^{-2}$ & $6.606\times 10^{-2} $ \\
$28$ & $6.292\times 10^{-5}$ & $1.435\times 10^{-2}$ & $8.724\times 10^{-2} $ \\
$29$ & $5.134\times 10^{-5}$ & $1.359\times 10^{-2}$ & $8.413\times 10^{-2} $ \\
$30$ & $4.193\times 10^{-5}$ & $1.289\times 10^{-2}$ & $6.499\times 10^{-2} $ \\
$31$ & $3.415\times 10^{-5}$ & $1.225\times 10^{-2}$ & $3.827\times 10^{-2} $ \\
$32$ & $2.763\times 10^{-5}$ & $1.165\times 10^{-2}$ & $9.426\times 10^{-3} $ \\
$33$ & $2.212\times 10^{-5}$ & $1.110\times 10^{-2}$ & $1.105\times 10^{-3} $ \\
$34$ & $1.744\times 10^{-5}$ & $1.058\times 10^{-2}$ & $8.656\times 10^{-4} $ \\
\end{tabular}
\begin{tabular}{c|ccc}
$35$ & $1.346\times 10^{-5}$ & $1.010\times 10^{-2}$ & $4.049\times 10^{-4} $ \\
$36$ & $1.011\times 10^{-5}$ & $9.655\times 10^{-3}$ & $1.730\times 10^{-4} $ \\
$37$ & $7.349\times 10^{-6}$ & $9.237\times 10^{-3}$ & $8.390\times 10^{-4} $ \\
$38$ & $5.142\times 10^{-6}$ & $8.846\times 10^{-3}$ & $8.614\times 10^{-4} $ \\
$39$ & $3.458\times 10^{-6}$ & $8.479\times 10^{-3}$ & $6.552\times 10^{-4} $ \\
$40$ & $2.241\times 10^{-6}$ & $8.134\times 10^{-3}$ & $3.774\times 10^{-4} $ \\
$41$ & $1.408\times 10^{-6}$ & $7.811\times 10^{-3}$ & $4.855\times 10^{-4} $ \\
$42$ & $8.743\times 10^{-7}$ & $7.506\times 10^{-3}$ & $9.986\times 10^{-4} $ \\
$43$ & $5.563\times 10^{-7}$ & $7.220\times 10^{-3}$ & $8.976\times 10^{-4} $ \\
$44$ & $3.735\times 10^{-7}$ & $6.949\times 10^{-3}$ & $5.458\times 10^{-4} $ \\
$45$ & $2.659\times 10^{-7}$ & $6.693\times 10^{-3}$ & $1.638\times 10^{-4} $ \\
$46$ & $1.987\times 10^{-7}$ & $6.452\times 10^{-3}$ & $3.271\times 10^{-4} $ \\
$47$ & $1.539\times 10^{-7}$ & $6.223\times 10^{-3}$ & $5.814\times 10^{-4} $ \\
$48$ & $1.224\times 10^{-7}$ & $6.006\times 10^{-3}$ & $6.424\times 10^{-4} $ \\
$49$ & $9.914\times 10^{-8}$ & $5.801\times 10^{-3}$ & $4.937\times 10^{-4} $ \\
$50$ & $8.134\times 10^{-8}$ & $5.606\times 10^{-3}$ & $1.857\times 10^{-4} $ \\
$51$ & $6.731\times 10^{-8}$ & $5.420\times 10^{-3}$ & $8.195\times 10^{-7} $ \\
$52$ & $5.598\times 10^{-8}$ & $5.244\times 10^{-3}$ & $2.353\times 10^{-6} $ \\
$53$ & $4.664\times 10^{-8}$ & $5.076\times 10^{-3}$ & $4.603\times 10^{-6} $ \\
$54$ & $3.883\times 10^{-8}$ & $4.917\times 10^{-3}$ & $5.806\times 10^{-6} $ \\
$55$ & $3.220\times 10^{-8}$ & $4.764\times 10^{-3}$ & $5.374\times 10^{-6} $ \\
$56$ & $2.652\times 10^{-8}$ & $4.619\times 10^{-3}$ & $4.509\times 10^{-6} $ \\
$57$ & $2.160\times 10^{-8}$ & $4.480\times 10^{-3}$ & $3.667\times 10^{-6} $ \\
$58$ & $1.732\times 10^{-8}$ & $4.348\times 10^{-3}$ & $2.614\times 10^{-6} $ \\
$59$ & $1.360\times 10^{-8}$ & $4.221\times 10^{-3}$ & $2.730\times 10^{-6} $ \\
$60$ & $1.036\times 10^{-8}$ & $4.100\times 10^{-3}$ & $3.020\times 10^{-6} $ \\
$61$ & $7.589\times 10^{-9}$ & $3.983\times 10^{-3}$ & $2.782\times 10^{-6} $ \\
$62$ & $5.291\times 10^{-9}$ & $3.872\times 10^{-3}$ & $2.102\times 10^{-6} $ \\
$63$ & $3.493\times 10^{-9}$ & $3.766\times 10^{-3}$ & $8.517\times 10^{-7} $ \\
$64$ & $2.197\times 10^{-9}$ & $3.663\times 10^{-3}$ & $9.238\times 10^{-7} $ \\
$65$ & $1.332\times 10^{-9}$ & $3.565\times 10^{-3}$ & $7.732\times 10^{-7} $ \\
$66$ & $7.805\times 10^{-10}$ & $3.471\times 10^{-3}$ & $2.168\times 10^{-7} $ \\
$67$ & $4.374\times 10^{-10}$ & $3.381\times 10^{-3}$ & $4.768\times 10^{-9} $ \\
$68$ & $2.378\times 10^{-10}$ & $3.294\times 10^{-3}$ & $5.772\times 10^{-9} $ \\
$69$ & $1.353\times 10^{-10}$ & $3.210\times 10^{-3}$ & $5.944\times 10^{-9} $ \\
\end{tabular}
\begin{tabular}{c|ccc}
$70$ & $8.533\times 10^{-11}$ & $3.129\times 10^{-3}$ & $5.588\times 10^{-9} $ \\
$71$ & $5.912 \times 10^{-11}$ & $3.052\times 10^{-3}$ & $5.053\times 10^{-9} $ \\
$72$ & $4.375 \times 10^{-11}$ & $2.977\times 10^{-3}$ & $4.449\times 10^{-9} $ \\
$73$ & $3.385 \times 10^{-11}$ & $2.905\times 10^{-3}$ & $3.745\times 10^{-9} $ \\
$74$ & $2.701 \times 10^{-11}$ & $2.836\times 10^{-3}$ & $2.851\times 10^{-9} $ \\
$75$ & $2.201 \times 10^{-11}$ & $2.769\times 10^{-3}$ & $1.632\times 10^{-9} $ \\
$76$ & $1.821 \times 10^{-11}$ & $2.704\times 10^{-3}$ & $1.420\times 10^{-9} $ \\
$77$ & $1.522 \times 10^{-11}$ & $2.642\times 10^{-3}$ & $2.227\times 10^{-9} $ \\
$78$ & $1.282 \times 10^{-11}$ & $2.582\times 10^{-3}$ & $1.900\times 10^{-9} $ \\
$79$ & $1.084 \times 10^{-11}$ & $2.524\times 10^{-3}$ & $3.832\times 10^{-10} $ \\
$80$ & $9.189 \times 10^{-12}$ & $2.468\times 10^{-3}$ & $2.785\times 10^{-10} $ \\
$81$ & $7.784 \times 10^{-12}$ & $2.413\times 10^{-3}$ & $1.923\times 10^{-10} $ \\
$82$ & $6.576 \times 10^{-12}$ & $2.361\times 10^{-3}$ & $6.659 \times 10^{-12}$ \\
$83$ & $5.526 \times 10^{-12}$ & $2.310\times 10^{-3}$ & $5.588 \times 10^{-12}$ \\
$84$ & $4.607 \times 10^{-12}$ & $2.261\times 10^{-3}$ & $4.657 \times 10^{-12}$ \\
$85$ & $3.794 \times 10^{-12}$ & $2.213\times 10^{-3}$ & $3.840 \times 10^{-12}$ \\
$86$ & $3.073 \times 10^{-12}$ & $2.167\times 10^{-3}$ & $3.117 \times 10^{-12}$ \\
$87$ & $2.429 \times 10^{-12}$ & $2.122\times 10^{-3}$ & $2.477 \times 10^{-12}$ \\
$88$ & $1.855 \times 10^{-12}$ & $2.079\times 10^{-3}$ & $1.911 \times 10^{-12}$ \\
$89$ & $1.343 \times 10^{-12}$ & $2.037\times 10^{-3}$ & $1.417 \times 10^{-12}$ \\
$90$ & $8.972 \times 10^{-13}$ & $1.996\times 10^{-3}$ & $1.001 \times 10^{-12}$ \\
$91$ & $5.314 \times 10^{-13}$ & $1.956\times 10^{-3}$ & $6.898 \times 10^{-13}$ \\
$92$ & $2.774 \times 10^{-13}$ & $1.918\times 10^{-3}$ & $5.099 \times 10^{-13}$ \\
$93$ & $1.404 \times 10^{-13}$ & $1.881\times 10^{-3}$ & $3.859 \times 10^{-13}$ \\
$94$ & $7.384 \times 10^{-14}$ & $1.844\times 10^{-3}$ & $1.464 \times 10^{-13}$ \\
$95$ & $3.873 \times 10^{-14}$ & $1.809\times 10^{-3}$ & $3.873 \times 10^{-14}$ \\
$96$ & $1.791 \times 10^{-14}$ & $1.775\times 10^{-3}$ & $1.791 \times 10^{-14}$ \\
$97$ & $4.351 \times 10^{-15}$ & $1.742\times 10^{-3}$ & $4.351 \times 10^{-15}$ \\
$98$ & $0$ & $1.710\times 10^{-3}$ & $0 $ \\
$99$ & $0$ & $1.678\times 10^{-3}$ & $0 $ \\
$100$ & $0$ & $1.648\times 10^{-3}$ & $0 $ \\
$101$ & $0$ & $1.618\times 10^{-3}$ & $0 $ \\
$102$ & $0$ & $1.589\times 10^{-3}$ & $0 $ \\
$103$ & $0$ & $1.561\times 10^{-3}$ & $0 $ \\
$104$ & $0$ & $1.534\times 10^{-3}$ & $0 $ \\
\hline
\hline
\end{tabular}
\caption{Numerical values for calculations in Fig.~\ref{fig:plot}. Transitions between the labeled numbers of cores for the mixed states are $\Tilde{r}=(1,2,5,10,18,30,98)$ and for the MNF identified pure states are $\Tilde{r}=(19,33,46,51,64,67,80,82,95,98)$.}
\end{table}

\end{document}